\documentclass[english]{article}
\usepackage{lmodern}

\usepackage[T1]{fontenc}
\usepackage[a4paper]{geometry}
\usepackage{dsfont}
\usepackage{color}
\usepackage{babel}
\usepackage{amsmath}
\usepackage{amsthm}
\usepackage{blindtext}
\usepackage{tikz-cd}
\usepackage{bm}
\usepackage{amsmath} 
\usepackage{enumitem}
\usepackage{booktabs}
\usepackage{multirow}
\usepackage{float}
\usepackage{subcaption}
\usepackage{graphicx}
\usepackage{amsthm,mathrsfs}
\usepackage{pdfpages}
\usepackage{lscape}
\usepackage{float}
\usepackage[linesnumbered,ruled,lined]{algorithm2e}
\usepackage{makecell}
\usepackage{amssymb}
\usepackage{stmaryrd}
\usepackage{graphicx}
\usepackage{setspace}
\usepackage{esint}
\usepackage{caption}

\usepackage[authoryear]{natbib}
\usepackage{float}
\usepackage{nicefrac}

\graphicspath{{plots/}}     

\usepackage{epstopdf}
\usepackage{url} 
\usepackage{caption}
\usepackage{multirow}
\usepackage{amssymb}
\usepackage{multicol}
\usepackage{booktabs}
\usepackage{tikz}
\usepackage{bbm}
\usepackage{bm}
\usetikzlibrary{positioning, bayesnet}
\usepackage{algorithm2e}
\RestyleAlgo{ruled}

\newtheorem{lemma}{Lemma}
\newtheorem{theorem}{Theorem}
\newtheorem{example}{Example}

\newtheorem{proposition}{Proposition}
\newtheorem{assumption}{Assumption}
\newtheorem{assumption*}{Assumption}[section]
\theoremstyle{remark}
\newtheorem{remark}{Remark}

\usepackage[textsize=tiny]{todonotes}

\usepackage[
    unicode=true,
    colorlinks=true,
    linkcolor=blue,
    citecolor=blue,
    urlcolor=blue
]{hyperref}

\def\ve{\varepsilon}
\def\vs{\varsigma}

\def\bY{\mathbf{Y}}
\def\bX{\mathbf{X}}
\def\bZ{\mathbf{Z}}
\def\bW{\mathbf{W}}
\def\bT{\mathbf{T}}
\def\bV{\mathbf{V}}
\def\bve{\bm{\varepsilon}}
\def\bU{\mathbf{U}}
\def\bxi{\bm{\xi}}
\def\bnu{\bm{\nu}}
\def\bmu{\bm{\mu}}
\def\bzeta{\bm{\zeta}}
\def\bmeta{\bm{\eta}}
\def\btheta{\bm{\theta}}
\def\bTheta{\bm{\Theta}}
\def\bPhi{\bm{\Phi}}
\def\bQ{\mathbf{Q}}
\def\bOmega{\mathbf{\Omega}}
\def\bzero{\mathbf{0}}
\def\bR{\mathbf{R}}
\def\bI{\mathbf{I}}

\def\bGamma{\mathbf{\Gamma}}
\def\bA{\mathbf{A}}

\def\PhiJ{\bPhi_{\bzeta}[\cdot,J]}
\def\hPhiJ{\hat{\bPhi}_{\bzeta}[\cdot,J]}
\def\bDelta{\mathbf{\Delta}}

\def\bone{\mathbf{1}}
\def\balpha{\bm{\alpha}}
\def\tbzeta{\widetilde{\bzeta}}
\def\tZ{\widetilde{Z}}

\def\btZ{\mathbf{\widetilde{Z}}}
\def\talpha{\widetilde{\alpha}}
\def\hbnu{\hat{\bnu}}
\def\tT{\widetilde{T}}
\def\bG{\mathbf{G}}
\def\bM{\mathbf{M}}
\def\bS{\mathbf{S}}

\def\bB{\mathbf{B}}
\def\bF{\mathbf{F}}
\def\bSigma{\bm{\Sigma}}
\def\bL{\mathbf{L}}
\def\bQ{\mathbf{Q}}
\def\tbSigma{\bm{\widetilde{\Sigma}}}

\def\bC{\mathbf{C}}
\def\tT{\widetilde{T}}
\def\tbPhi{\widetilde{\mathbf{\Phi}}}

\def\cG{\mathcal{G}}
\def\cE{\mathcal{E}}
\def\cV{\mathcal{V}}
\def\hcV{\hat{\mathcal{V}}}
\def\cN{\mathcal{N}}

\def\cA{\mathcal{A}}
\def\cO{\mathcal{O}}

\def\E{\mathbb{E}}
\def\P{\mathbb{P}}
\def\X{\mathbb{X}}
\def\Y{\mathbb{Y}}
\def\Z{\mathbb{Z}}
\def\hZ{\hat{\mathbb{Z}}}
\def\O{\mathbb{O}}
\def\R{\mathbb{R}}
\def\Ok{\mathbb{O}^{(k)}}
\def\bu{\mathbf{u}}

\def\hT{\hat{\mathbb{T}}}
\def\hS{\hat{\mathbb{S}}}

\def\TT{\mathbb{T}}
\def\SS{\mathbb{S}}
\def\Q{\mathbb{Q}}
\def\hQ{\hat{\mathbb{Q}}}

\DeclareMathOperator\rank{rank}
\DeclareMathOperator\cov{cov}
\DeclareMathOperator\var{var}
\DeclareMathOperator\pa{pa}
\DeclareMathOperator\an{an}
\DeclareMathOperator\de{de}
\DeclareMathOperator\leaf{leaf}
\DeclareMathOperator*\argmin{\arg\min}
\DeclareMathOperator\sgn{sgn}
\DeclareMathOperator\TP{TP}
\DeclareMathOperator\RE{RE}
\DeclareMathOperator\FP{FP}
\DeclareMathOperator\FN{FN}

\DeclareMathOperator\FDR{FDR}
\DeclareMathOperator\TPR{TPR}
\DeclareMathOperator\SHD{SHD}
\DeclareMathOperator\JI{JI}

\DeclareMathOperator\inter{in}
\DeclareMathOperator\ARG{ARG}

\DeclareMathOperator\TSLS{2SLS}

\newcommand\relph[1]{\mathrel{\phantom{#1}}}

\SetKw{Return}{Return:}

\makeatletter
\newcommand{\myfnsymbol}[1]{%
  \expandafter\@myfnsymbol\csname c@#1\endcsname
}
\newcommand{\@myfnsymbol}[1]{%
  \ifcase #1
  \or 1
  \or 2
  \or 3
  \or 4
  \or \TextOrMath{\textasteriskcentered}{*}
  \or \TextOrMath{\textdagger}{\dagger}
  \fi
}
\newcommand{\affiliationA}{\@myfnsymbol{1}}
\newcommand{\affiliationB}{\@myfnsymbol{2}}
\newcommand{\affiliationC}{\@myfnsymbol{3}}
\newcommand{\affiliationD}{\@myfnsymbol{4}}
\newcommand{\equalcontribution}{\@myfnsymbol{5}}
\newcommand{\correspondingA}{\@myfnsymbol{6}}
\makeatother

\makeatother

\begin{document}
	\title{Using Heteroscedasticity for Causal Discovery under Hidden Confounding}
	\date{}
\author{
    Jing Zou\textsuperscript{\affiliationA},
    Wei Li\textsuperscript{\affiliationB},
    Wei Lin\textsuperscript{\affiliationC}
}
\renewcommand{\thefootnote}{\myfnsymbol{footnote}}
\maketitle
\footnotetext[1]{School of Mathematical Sciences, Peking University,
   Beijing, 100871, China. Email: jingzou@stu.pku.edu.cn}%
\footnotetext[2]{Center for Applied Statistics and School of Statistics, Renmin University of China,
Beijing, 100872, China. Email: weilistat@ruc.edu.cn}%
\footnotetext[3]{School of Mathematical Sciences and Center for Statistical Science, Peking University, 
   Beijing, 100871, China. Email: weilin@math.pku.edu.cn}

    \begin{abstract}
The presence of hidden confounding in observational studies poses a fundamental challenge to the identification and estimation of causal structures encoded by directed acyclic graphs (DAGs). While instrumental variables (IVs) have been widely used to address hidden confounding, most existing methods require valid instruments, which may be unavailable in practice. In this paper, we develop a novel framework for causal discovery and inference in DAGs that accommodates both hidden confounding and invalid IVs by exploiting heteroscedasticity. We establish a novel linear characterization of leaf nodes through heteroscedasticity-induced relationships between cross-covariance and self-covariance vectors, enabling recursive recovery of the causal order without requiring any candidate IV to be valid. Given the identified order, we then sequentially construct surrogate IVs to identify the causal effects and directions, thereby recovering the entire DAG. We also propose a data-adaptive procedure for estimating the DAG and causal effects. We show that the proposed method consistently recovers the underlying DAG and yields asymptotically normal estimators of causal effects. Extensive simulation studies demonstrate substantial improvements over existing methods. We further illustrate the practical utility of the proposed method through an application to gene regulatory network analysis. 
        \end{abstract}
        

       \section{Introduction} \label{sec:intro}


Causal discovery, which aims to learn directed acyclic graphs (DAGs) that encode causal relationships among a set of variables, has become increasingly important in statistics and machine learning, with broad applications in areas such as systems biology \citep{sachs2005causal} and medical imaging \citep{castro2020causality}. While randomized experiments are the gold standard for causal discovery, they can be expensive, unethical, or even infeasible. Consequently, learning causal structures from observational data has become an active area of research (e.g., \cite{shimizu2006linear,buhlmann2014cam,mooij2016distinguishing}). A major challenge in observational studies is the presence of hidden confounding, which induces spurious associations among observed variables and may lead to incorrect orientation of causal directions, thereby compromising the reliability of causal discovery.

To address hidden confounding, instrumental variables (IVs) have been widely used in causal inference and causal discovery \citep{angrist1996identification,hemani2017orienting,chen2018two}. A valid IV must satisfy three key assumptions: (a) (relevance) it is associated with the treatment; (b) (independence) it is independent of unmeasured confounders; and (c) (exclusion restriction) it has no direct effect on the outcome. However, because the independence and exclusion restriction assumptions are generally untestable, candidate IVs may fail to satisfy these conditions in practice, leading to biased or inconsistent inference. A prominent example arises in Mendelian randomization (MR) studies in genetic epidemiology \citep{smith2004mendelian}, where genetic variants are used as IVs to infer causal relationships and construct gene regulatory networks. In such settings, the independence assumption may be violated due to linkage disequilibrium or population stratification \citep{lawlor2008mendelian}, while the exclusion assumption may fail because of unknown pleiotropic effects of genetic variants \citep{little2003mendelian,lawlor2008mendelian,burgess2017interpreting}. These challenges motivate the development of identification and estimation methods for DAGs under hidden confounding that remain robust to violations of IV assumptions, even in the absence of any valid IVs.

\subsection{Related work}\label{ssec:rela}


Various methods have been proposed for causal discovery in the presence of hidden confounding. A common strategy is to relax the goal of identifying a fully directed causal graph and instead recover a partial ancestral graph (PAG), which represents an equivalence class of causal structures compatible with hidden confounding \citep{colombo2012learning,bernstein2020ordering,akbari2021recursive}. To obtain more informative causal directions, \cite{frot2019robust} and \cite{shah2020right} introduced additional assumptions on the confounding mechanisms and established identification results for the completed partially directed acyclic graph (CPDAG). Another line of research exploits valid IVs to address hidden confounding in causal discovery \citep{hemani2017orienting,chen2018two}. For example, building on the classical two-stage least squares framework with valid IVs, \cite{chen2018two} proposed a two-stage penalized least squares estimator for learning causal graphs under linear structural equation models. In contrast, we consider causal discovery and causal effect estimation for DAGs while accounting for both hidden confounding and invalid IVs.


In classical causal inference settings where the causal direction is known, considerable effort has been devoted to handling invalid IVs. One line of research assumes that only a small proportion of candidate IVs are invalid. For example, \cite{kang2016instrumental} and \cite{windmeijer2019use} established identification of causal effects under the majority rule, which requires more than half of the candidate IVs to be valid. Relaxing this requirement, \cite{hartwig2017robust} and \cite{guo2018confidence} introduced the more general plurality rule, under which the number of valid IVs exceeds the size of the largest group of invalid IVs sharing the same ratio between their associations with the outcome and the treatment.  \cite{guo2018confidence} further developed the two-stage hard thresholding (TSHT) procedure with voting, which achieves consistent estimation in both low- and high-dimensional linear models. Clearly, all of these methods require a sufficient number of valid IVs. 
In contrast, \cite{bowden2015mendelian} and \cite{kolesar2015identification} allowed all candidate IVs to violate the exclusion restriction, provided that the direct effects of IVs on the outcome are asymptotically orthogonal to their effects on the treatment. 
But as pointed out by \cite{kang2016instrumental}, this orthogonality condition may be difficult to justify in MR settings because it requires the pleiotropic effects of genetic variants to be unrelated to their effects on the treatment, a restriction that generally lacks biological justification.

Another line of work identifies causal effects by leveraging heteroscedasticity while allowing all instruments to violate the exclusion restriction \citep{lewbel2012using,tchetgen2021genius,sun2022selective,ye2024genius}. Unlike the orthogonality condition of \cite{bowden2015mendelian,kolesar2015identification}, these methods do not rely on assumptions linking the mechanisms through which instruments affect the treatment and the outcome. 
\cite{lewbel2012using} exploited heteroscedasticity in the treatment model to identify causal effects under linear models and developed G-estimation procedures even when no valid IV is available. \cite{tchetgen2021genius} further generalized this identification strategy to a broader class of MR settings and referred to this framework as G-estimation under no interaction with unmeasured selection (GENIUS). Building on the GENIUS framework, \cite{sun2022selective} proposed a multiply robust estimator that incorporates generic machine learning methods for nuisance parameter estimation, while \cite{ye2024genius} developed an extension that is robust to many weak invalid instruments. In a related line of work, \cite{liu2023mendelian} leveraged heteroscedasticity in the outcome model rather than the treatment model for identification and proposed efficient likelihood-based estimators. Despite their  differences, all of these methods assume that the causal direction is known a priori and therefore focus on causal effect estimation rather than causal discovery. Consequently, they cannot be directly applied to learning causal structures in DAGs, where both the graph structure and causal directions must be inferred from the data.

To infer causal directions between two variables in the presence of invalid IVs, existing methods have relied on standard identification strategies, such as the covariance heterogeneity condition and the plurality rule, or combinations thereof \citep{lewbel2012using,xue2020inferring,li2026doubly}. However, it remains unclear how these approaches can be extended to multivariate causal networks. Among the few attempts to learn DAGs with invalid IVs under hidden confounding, \cite{chen2024discovery} established identification results for Gaussian DAGs by assuming that the majority rule holds for each primary variable. \cite{zou2025semiparametric} identified DAGs in partially linear models by constructing surrogate IVs, assuming that the candidate instruments are mutually independent and that at least one valid IV is available.
While theoretically appealing, these assumptions may be restrictive in practice, and valid IVs may not always be available. We therefore seek to develop a causal discovery framework under alternative conditions that remains applicable even when no valid IV exists.

\subsection{Our contributions}\label{ssec:contr}

In this paper, we develop a novel framework for causal discovery and inference in DAGs that leverages covariance heterogeneity to accommodate both hidden confounding and invalid IVs. Our main contributions are summarized as follows.

First, we establish a new identification theory for DAGs under hidden confounding by exploiting covariance heterogeneity. Unlike existing causal discovery methods that require a majority or plurality of instruments to be valid, the proposed framework remains effective even when no valid instrument is available. The proposed identification strategy decomposes DAG identification into two sequential tasks. We first recover the causal order through a recursive leaf-identification procedure. Specifically, by exploiting the linear relationships among covariance vectors induced by covariance heterogeneity, we establish a novel characterization of leaf nodes through cross-covariance vectors, which enables recursive identification and removal of leaf nodes to recover the causal order. Conditioning on the recovered causal order, we then construct surrogate IVs sequentially to identify the causal effects and directions, thereby recovering the entire DAG.

Second, we develop a data-adaptive estimation procedure, termed the Covariance Heterogeneity based Approach to Recovering Graphical Effects (CHARGE), to implement the proposed identification strategy. A key component of CHARGE is a recursive leaf-estimation procedure that reformulates leaf identification as a sequence of goodness-of-fit tests. We develop an F-type test statistic together with a theoretically motivated threshold that simultaneously accounts for estimation uncertainty and multiple testing, without relying on repeated significance testing based on $p$-values. The estimated causal order is subsequently incorporated as structural constraints into a G-estimation procedure for causal effect estimation, guaranteeing acyclicity of the estimated graph while substantially reducing the parameter space for estimation. We further establish the consistency of CHARGE for both causal discovery and causal effect estimation, together with the asymptotic normality of the causal effect estimators. Simulation studies and a real-data application to gene regulatory network analysis demonstrate the  finite-sample performance and practical utility of CHARGE.


\subsection{Organization}\label{ssec:org}

The rest of this paper is organized as follows. Section \ref{sec:model} introduces the causal graphical model. Section \ref{sec:iden} presents identification conditions and establishes identifiability of causal directions and effects. Section \ref{sec:est} describes the proposed method for causal discovery and effect estimation and provides theoretical justification for it. We then investigate the finite-sample performance of the proposed method via simulations in Section \ref{sec:sim}, and apply the method to a gene regulatory network analysis in Section \ref{sec:app}. We finally conclude with a discussion in Section \ref{sec:dis} and relegate proofs and extensions to the Supplementary Material \citep{zou2026supplement}.

The following notation will be used throughout the paper. For an index set $s\subseteq\{1,\ldots,K\}$ and a $K$-dimensional vector $\bnu=(\nu_1,\ldots,\nu_K)^T$, let $\bnu_s=(\nu_i)_{i\in s}$ and $\bnu_{-s}=(\nu_i)_{i\notin s}$. Given vectors $\bnu_1,\ldots,\bnu_L$ of the same dimension, let $\mathrm{span}(\{\bnu_1,\ldots,\bnu_L\})$ denote the linear subspace spanned by these vectors. 
For a matrix $\bW$, let $\bW_i$ denote its $i$th column and $W_{ij}$ its $(i,j)$th entry. Let $\bI_d$ denote the $d\times d$ identity matrix. 
Let $\bX$ denote population variables and $\X$ their corresponding realizations. We write $\X^{(k)}$ for the $k$th observation, $\X_i$ for the $i$th feature, and $\X_i^{(k)}$ for the  $i$th feature of the $k$th observation.
For a vector $\bnu\in\R^K$ and a matrix $\bW\in\R^{K\times L}$, let $P_{\bW}\bnu$ denote the linear projection of $\bnu$ onto the column space of $\bW$, and let $P_{\bW}^{\perp}\bnu=(\bI_K-P_{\bW})\bnu$ denote the corresponding residual vector. Similarly, for random vectors $\bX\in \R^q$ and $\bY\in\R^p$, let $\bX\perp\!\!\!\perp \bY$ denote independence, $P_\bX \bY=\cov(\bY,\bX)\var(\bX)^{-1}\bX$ and $P^\perp_\bX \bY=(\bI_p-P_\bX) \bY$ denote the associated linear projection and residual, respectively. Double subscripts are interpreted according to context. Specifically, $W_{ij}$ denotes the $(i,j)$th entry of the matrix $\bW$, and a bold symbol with double subscripts, such as $\bzeta_{ij}$, denotes a vector indexed by the pair $(i,j)$. 


\section{Causal graphical model}\label{sec:model}

Let  $\bY=(Y_1,\dots,Y_p)^T$ denote the $p$ endogenous primary variables and $\bX=(X_1,\dots,X_q)^T$ the $q$ auxiliary variables, both of which have finite variance.
Suppose that there are $n$ independent and identically distributed observations $\{(\Y^{(k)T},\X^{(k)T})^T\}_{k=1}^n$ from $(\bY^T,\bX^T)^T$, yielding the observed data matrices $\Y=(\Y^{(1)T},\dots,\Y^{(n)T})^T$ and $\X=(\X^{(1)T},\dots,\X^{(n)T})^T$. We  assume that the covariance matrix of $\bX$ is invertible. Since $\bX$ is introduced only to provide candidate IVs, the dependence structure among its components is not of primary interest and is therefore left unspecified.

Consider the following linear structural equation model (SEM):
\begin{equation}\label{eq:SEM}
  \bY=\bV^T \bY+\bW^T \bX+\bve,
\end{equation}
where $\bV\in\mathbb{R}^{p\times p}$ and $\bW\in\mathbb{R}^{q\times p}$ are unknown coefficient matrices, and the error vector $\bve=(\ve_1,\ldots,\ve_p)^T$ satisfies $\E(\bve)=\bzero$ and $\cov(\bX,\bve)=\bzero$. The coefficient matrix $\bV$, which is our primary object of interest, encodes the direct causal relationships among the primary variables. Specifically, $V_{ij}\neq0$ indicates that $Y_i$ has a direct causal effect on $Y_j$, denoted by $Y_i\rightarrow Y_j$. Likewise, $W_{\ell j}\neq0$ indicates that the auxiliary variable $X_\ell$ has a direct causal effect on $Y_j$, denoted by $X_\ell\rightarrow Y_j$. Consistent with the classical IV framework, we assume that there are no causal effects from $\bY$ to $\bX$.
To explicitly represent hidden confounding, one possible way is to write each error term as
\begin{equation}\label{eq:add-error}
  \ve_i=g_i(\bU)+\xi_i,\qquad i=1,\ldots,p,
\end{equation}
where $\bU$ denotes the latent confounders, $g_i(\bU)$ is an arbitrary measurable function satisfying $\E\{g_i(\bU)\}=0$, and $\xi_i$ is an idiosyncratic noise term. Because the latent confounders may simultaneously affect multiple components of $\bY$, the error term $\ve_i$ is generally correlated with $\bY_{-\{i\}}$, thereby inducing endogeneity that poses the primary challenge for identifying the causal coefficient matrix $\bV$. Although the main text focuses on the linear SEM in \eqref{eq:SEM}, the proposed framework can be extended to partially linear structural equation models. The corresponding identification results are given in Section  S3 of the Supplementary Material \citep{zou2026supplement}.

The proposed model is paired with a DAG $\cG=(\cV;\cE_1,\cE_2)$, which consists of a vertex set $\cV=\{Y_1,\dots,Y_p\}\cup\{X_1,\dots,X_q\}$, an edge set $\cE_1=\{(i,j):Y_i\to Y_j\}$ representing the edges among $\bY$, and an edge set $\cE_2=\{(\ell,j):X_\ell \to Y_j\}$ representing the edges from $\bX$ to $\bY$. Our interest lies in recovering the causal structure among the primary variables $\bY$, encoded by the edge set $\cE_1$. Considering the edges in $\cE_1$, we introduce some terminology as follows. If a directed edge $Y_i\to Y_j$ exists in $\cE_1$, we say $Y_i$ is a parent of $Y_j$ and $Y_j$ is a child of $Y_i$. A vertex $Y_i$ with no child in $\cG$ is called a leaf of $\cG$, and the set of all leaves is denoted by $\leaf(\cG)$. If a directed path $Y_i\to \dots \to Y_j$ exists, then we say $Y_i$ is an ancestor of $Y_j$ and $Y_j$ is a descendant of $Y_i$. 
The topological height $h_\cG(i)$ of a vertex $Y_i$ is the maximal length of a directed path from $Y_i$ to a leaf of $\cG$. Clearly, a leaf node has height zero, and we have $0\le h_\cG(i)\le H_\cG\le p-1$ for all $i=1,\dots,p$, where $H_\cG:=\max_{1\le i\le p}h_\cG(i)$ is the length of the longest directed path in $\cE_1$. Let $\cV(h)=\{i:h_\cG(i)= h\}$ for $0\le h\le H_\cG$ denote the indices of primary variables at topological height $h$, and $\cV^+(h)=\bigcup_{h^\prime>h}\cV(h^\prime)$ for $-1 \le h\le H_\cG$ denote the indices of those with height larger than $h$. It follows that $\emptyset\equiv \cV^+(H_\cG)\subset \cV^+(H_\cG-1)\subset \cdots \subset \cV^+(-1)\equiv \{1,\dots,p\}$ and $\cV^+(h_\cG(i))$ contains all the ancestors of $Y_i$ without containing any descendant of $Y_i$. Therefore, the topological height essentially determines a certain causal order among $\bY$. For each $h=-1,\dots, H_\cG-1$, we define 
\[
\cG(h)=(\bY_{\cV^+(h)}\cup \bX;\cE_1(h),\cE_2(h))
\]
as the subgraph of $\cG$ induced by $\bY_{\cV^+(h)}$ and $\bX$, where the edge sets $\cE_1(h)=\{(i,j)\in \cE_1:i,j\in \cV^+(h)\}$ and $\cE_2(h)=\{(\ell,j)\in \cE_2: j\in \cV^+(h)\}$. {Intuitively, $\cG(h)$ is the remaining subgraph obtained from $\cG$ by removing all primary variables with topological height at most $h$ and all edges connected to them. }
We provide an example in Section S1 of the Supplementary Material \citep{zou2026supplement} for further illustration.

The additive linear constant-effects model of \cite{holland1988causal}, extended to allow violations of the exclusion restriction \citep{small2007sensitivity}, has been widely adopted in the literature on invalid IVs. This modeling framework underlies a broad range of methods in classical causal inference \citep[e.g.,][]{kolesar2015identification,kang2016instrumental,windmeijer2019use,guo2018confidence} and has also been commonly used for causal discovery with potentially invalid IVs \citep[e.g.,][]{xue2020inferring,li2023inference,chen2024discovery,li2026doubly}. The SEM in \eqref{eq:SEM} falls within this well-established class of linear IV models. Despite its linear structure, however, recovering the underlying DAG remains highly challenging.
One major challenge is the presence of hidden confounding, which induces endogeneity among the primary variables and renders standard regression-based methods inconsistent. While the auxiliary variables $\bX$ may serve as IVs to mitigate confounding bias, the unknown coefficient matrix $\bW$ makes it impossible to determine a priori which candidate IVs satisfy the exclusion restriction \citep{li2023inference,chen2024discovery}. Furthermore, even if the valid IVs were known for every primary variable, recovering the entire DAG would still be difficult. Existing IV-based procedures, such as the two-stage least squares approach of \cite{chen2018two}, estimate causal effects locally and may produce cyclic graphs under finite samples. In contrast, globally searching over the space of DAGs, as in score-based methods \citep{chickering2002optimal,ramsey2017million}, is computationally prohibitive because the number of candidate DAGs grows super-exponentially with the number of variables \citep{mokhtarian2025recursive}. To address these challenges, we exploit covariance heterogeneity to develop a novel two-step framework for DAG identification and estimation. Specifically, the proposed framework first identifies the causal order of the primary variables and then recovers the causal effects and graph structure.

\section{Causal identification theory}\label{sec:iden}

In this section, we establish the identification theory for the causal directions and effects under hidden confounding by leveraging covariance heterogeneity. Section \ref{ssec:ass} introduces a new set of identification assumptions and summarizes the main identification results. We then elaborate on the identification strategy in the following two subsections. In Section \ref{ssec:order}, we establish the identification of the causal order in $\cG$. Building on this result, we study the identification of direct causal effects and consequently the full DAG in Section \ref{ssec:eff}. 

\subsection{Identification conditions and main results}\label{ssec:ass}

In this subsection, we introduce our identification assumptions and present the main identification results. We begin with the following conditions.


\begin{assumption}\label{ass:uncorr}
    For all $i,j=1,\dots, p$ and $i\ne j$, $\cov(\bX,\ve_i\ve_j)=\bzero$.
\end{assumption}

 Assumption~\ref{ass:uncorr} requires that the auxiliary variables $\bX$ are uncorrelated with the cross-products of  distinct structural errors.
For DAGs with two primary variables, Assumption~\ref{ass:uncorr} reduces to a key orthogonality condition imposed in \cite{lewbel2012using} for identifying the causal effect between an exposure and an outcome in the presence of hidden confounding. Here, we extend this condition to DAGs with multiple primary variables.

Under the error decomposition in \eqref{eq:add-error}, Assumption~\ref{ass:uncorr} is implied by the following stronger but more easily interpretable conditions: (i) $\bX\perp\!\!\!\perp\bU$; and (ii) the idiosyncratic noise satisfies
$\E(\xi_i\mid \bX,\bU,\bxi_{-\{i\}})=0$ for $i=1,\dots,p$.
Indeed, under these conditions,
 $\cov(\bX, g_i(\bU)\xi_j)=\bzero$ and $\cov(\bX, \xi_i\xi_j)=\bzero$ for all $i,j=1,\dots,p$ and $i\ne j$. Therefore, $\cov(\bX, \ve_i\ve_j)=\cov(\bX,g_i(\bU)g_j(\bU)+g_i(\bU)\xi_j+g_j(\bU)\xi_i+\xi_i\xi_j)=\bzero$, which establishes Assumption~\ref{ass:uncorr}. 


 The next assumption imposes a heteroscedastic covariance restriction on model \eqref{eq:SEM}. For each $i=1,\ldots,p$, let  $\bmeta_i:=\cov(\bX,\ve_i^2)$ denote the covariance between $\bX$ and the squared structural error $\ve_i^2$, and define the $q\times p$ matrix $\bPhi_{\bmeta}:=(\bmeta_1,\dots,\bmeta_p)$. A nonzero $\bmeta_i$ implies that $\ve_i$ is heteroscedastic with respect to $\bX$. We then require the matrix $\bPhi_{\bmeta}$ to have full column rank. 

\begin{assumption}\label{ass:hete}
  The  matrix $\bPhi_{\bmeta}$ is of full column rank, i.e., $\rank(\bmeta_1,\dots,\bmeta_p)=p$. 
\end{assumption}

Assumption~\ref{ass:hete} requires structural errors to exhibit  distinct heteroscedasticity patterns, as characterized by the covariance vectors $\bmeta_i$. In particular, it implies that $q\ge p$ and $\bmeta_i\neq\bzero$ for every $i=1,\ldots,p$. Assumption~\ref{ass:hete} is hereditary: if it holds for $\cG$, then it also holds for every subgraph $\cG(h)$, since each submatrix $(\bmeta_i)_{i\in\cV^+(h)}$ of $\bPhi_{\bmeta}$ retains full column rank.

When $p=2$, Assumption~\ref{ass:hete} reduces to the covariance heterogeneity condition proposed by \cite{lewbel2012using}. Heteroscedasticity has been widely exploited as a source of identification in causal inference without conventional exclusion restrictions \citep[e.g.,][]{lewbel2012using,tchetgen2021genius,sun2022selective}. More recently, \citet{li2026doubly} combined heteroscedasticity with an enhanced plurality rule to identify causal direction between two variables in the presence of hidden confounding, allowing for potentially bidirectional relationships.
Although \cite{li2026doubly} successfully address bivariate problems, extending heteroscedasticity based identification to general DAGs is substantially more challenging because the goal is no longer to determine the causal relationship between a pair of variables, but to recover an entire DAG. Our identification strategy differs fundamentally from existing bivariate methods. Specifically, we first recover the causal order using the heteroscedasticity structure and then identify the causal coefficients and graph structure conditional on that order.
To our knowledge, this is the first work to exploit heteroscedasticity for identifying general DAGs under hidden confounding. Below, we provide an example under which Assumption~\ref{ass:hete} is satisfied.

\begin{figure}[t]
    \centering
    
    \begin{subfigure}{0.18\textwidth}
        \centering
        \includegraphics[width=\linewidth]{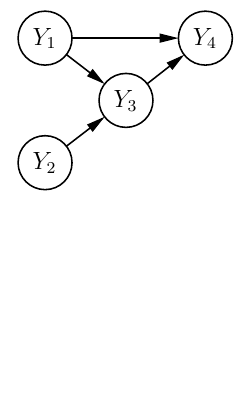}
        \caption{The true $\cE_1$.}
        \label{fig:true_dag}
    \end{subfigure}
    \hfill  
    \begin{subfigure}{0.8\textwidth}
        \centering
        \includegraphics[width=\linewidth]{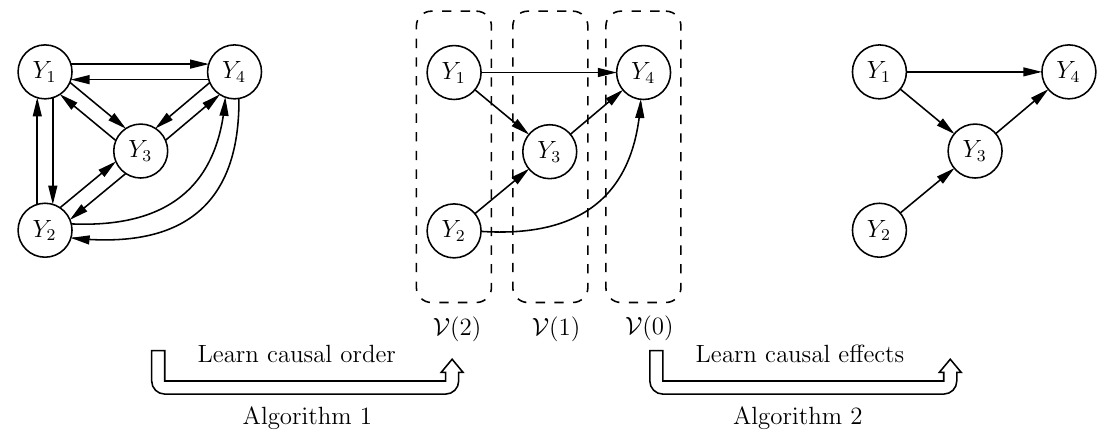}
        \caption{The two-step pipeline for identifying $\cE_1$ in Figure \ref{fig:true_dag}.}
        \label{fig:pipeline}
    \end{subfigure}
\caption{
Illustration of the proposed two-step identification procedure.
(a) shows the true edge set among the primary variables and 
(b) illustrates the two-step identification pipeline.
A directed edge from $Y_i$ to $Y_j$ indicates that $Y_i$ remains a potential parent of $Y_j$ at the current stage.
The left, middle, and right panels in (b) correspond to the initial search space, the candidate edges after causal order identification, and the final DAG after identifying the nonzero causal effects, respectively.
}
    \label{fig:summarize}
\end{figure}

\begin{example}
Suppose that $q=p=4$ and consider the DAG  in Figure~\ref{fig:true_dag}, where the auxiliary variables $\bX$ and latent confounders $\bU$ are omitted for simplicity. Let the  errors in \eqref{eq:add-error} satisfy $\xi_i=\vs_iX_i$, where $\vs_i\sim\cN(0,1)$ independently for $i=1,\ldots,4$, and assume that $\vs_i\perp\!\!\!\perp(\bX,\bU)$. Assume further that $X_i$'s are pairwise independent, $\E(\bX)=\bzero$, and $\bX\perp\!\!\!\perp\bU$. Under these conditions, the  matrix $\bPhi_{\bmeta}$ is diagonal with diagonal entries $\E(X_1^3),\ldots,\E(X_4^3)$. Therefore, if $\E(X_i^3)\neq0$ for all $i=1,\ldots,4$, then $\rank(\bPhi_{\bmeta})=4$, and Assumption~\ref{ass:hete} is satisfied.
\end{example}

The disturbance $\vs_i$ in the above example may be interpreted as unobserved genetic, biological, or environmental factors that modify the effect of $X_i$ on $Y_i$, thereby inducing latent effect heterogeneity. Such heterogeneity is common in many biomedical applications and is particularly natural in MR studies, where the effects of genetic variants often vary across individuals because of unmeasured environmental or biological factors. Indeed, gene--environment interactions have frequently been invoked as one possible explanation for the missing heritability observed in genetic association studies \citep{manolio2009finding}. From this perspective, latent effect heterogeneity provides a plausible mechanism under which Assumption~\ref{ass:hete} is satisfied. This interpretation is closely related to approaches that exploit gene--environment interactions for IV analysis \citep{spiller2019detecting,spiller2022interaction}. Unlike these methods, however, our framework does not require the interacting environmental factors to be observed explicitly.


Beyond its practical interpretation, Assumption \ref{ass:hete} is also testable given Assumption \ref{ass:uncorr}. Define $\bZ=(Z_1,\dots,Z_p)^T$ as the residuals obtained by linearly projecting $\bY$ onto $\bX$, 
\[
\bZ:=P^\perp_\bX \bY=(\bI_p-\bV^T)^{-1}\bve.
\]
Let $\bzeta_{ij}:=\cov(\bX,Z_iZ_j)$ denote the covariance of $\bX$ and $Z_iZ_j$ for $i,j=1,\dots,p$, and let $\bPhi_{\bzeta}:=(\bzeta_{11},\bzeta_{22},\dots,\bzeta_{pp})$ denote the $q\times p$ matrix with columns given by the vectors $\bzeta_{ii}$'s. Since  $(\bY^T,\bX^T)^T$ is observed,  $\bPhi_{\bzeta}$ is estimable from the data. The following proposition establishes the relationship between $\bPhi_{\bmeta}$ and $\bPhi_{\bzeta}$.

\begin{proposition}\label{prop:test}
  Under Assumption \ref{ass:uncorr}, the column spaces of $\bPhi_{\bmeta}$ and $\bPhi_{\bzeta}$ are the same, and thus $\rank(\bPhi_{\bmeta})=p$ if and only if $\rank(\bPhi_{\bzeta})=p$. 
\end{proposition}

Proposition \ref{prop:test} shows that, although $\bPhi_{\bmeta}$ cannot be directly obtained from the observed data due to the unknown coefficients $\bV$ and $\bW$, Assumption \ref{ass:hete} can nevertheless be assessed by checking the rank of the estimable matrix $\bPhi_{\bzeta}$ under Assumption \ref{ass:uncorr}. In practice, one may apply existing rank testing procedures, such as those proposed by \cite{cragg1997inferring} and \cite{chen2019improved}, to the estimator $\hat{\bPhi}_{\bzeta}$. Alternatively, one may test whether the determinant of $\bPhi_{\bzeta}^T \bPhi_{\bzeta}$ is significantly different from zero. 


Equipped with these assumptions, we derive the identifiability of the causal coefficients and the underlying graph structure. The main results are summarized in the following theorem, and the detailed identification procedure is presented in the subsequent two subsections.

\begin{theorem}\label{thm:iden}
  Under Assumptions \ref{ass:uncorr} and \ref{ass:hete}, the coefficients $(\bV,\bW)$ in model \eqref{eq:SEM} and the corresponding edge sets $(\cE_1,\cE_2)$ are identifiable. 
\end{theorem}

Notably, Theorem \ref{thm:iden} imposes no restrictions on the coefficients $\bW$, implying that any auxiliary variable $X_\ell$ may have direct causal effects on multiple primary variables $\bY$. Consequently, all candidate IVs may violate the exclusion restriction, allowing the possibility that no valid IVs are available. In contrast, existing IV-based methods for causal discovery in DAGs require at least some valid IVs, \citep[e.g., ][]{oates2016estimating,chen2018two,chen2024discovery}, regardless of whether they account for hidden confounding. This motivates the development of identification and estimation methods under alternative assumptions that do not rely on the existence of valid IVs.

The rest of this section details the identification strategies in Theorem \ref{thm:iden}. We first establish the causal order among the variables in $\bY$ in Section \ref{ssec:order}, and then recover the causal effects and directions in Section \ref{ssec:eff}. Figure \ref{fig:summarize} provides a graphical illustration of this two-step pipeline. Similar order-based frameworks for DAG identification under hidden confounding have been studied in prior literature. For example, assuming Gaussian distribution of errors and sublinear growth of causal effects, \cite{li2024nonlinear} reconstructed the topological order by iteratively applying normality tests. In cases with instruments, \cite{chen2024discovery} and \cite{wang2024causal} leveraged the exclusion conditions of valid IVs and the local faithfulness of interventions to recover the causal order. While these identification conditions relate to causal order in a straightforward and intuitive way, such approaches do not apply to our setting, where we impose no distributional assumptions on the errors and do not require the existence of valid IVs. Moreover, because the covariance vectors $\bmeta_i$
 are not  estimable from the observed data, Assumption~\ref{ass:hete} cannot be directly exploited to recover the causal order.

\subsection{Identification of causal order}\label{ssec:order}

In this subsection, we establish the identification of the causal order among primary variables using covariance heterogeneity. In particular, we consider the causal order determined by the topological height within the DAG. The following proposition provides insights into the connection between covariances $\{\bzeta_{ij}\}_{i,j=1}^p$ and the causal order.

\begin{proposition}\label{prop:leaf}
  Under Assumptions \ref{ass:uncorr} and \ref{ass:hete}, the leaves of $\cG$ are identified by
  \begin{equation}\label{eq:leaf}
        \leaf(\cG)=\{i:\bzeta_{ij}\in \mathrm{span}(\{\bzeta_{kk}:k\ne i\})\ \text{for all}\ j\ne i\}.   
  \end{equation}
\end{proposition}

Proposition~\ref{prop:leaf} provides an observable characterization of leaf nodes using the covariance vectors $\{\bzeta_{ij}\}_{i,j=1}^p$. Rather than recovering the entire DAG directly, it identifies one layer of the graph at a time. Specifically, a node is a leaf if every cross-covariance vector associated with that node can be represented as a linear combination of the self-covariance vectors of the remaining nodes. This characterization forms the basis for recursively recovering the causal order.

The intuition behind Proposition~\ref{prop:leaf} is that heterogeneity behaves like a signal that propagates only along directed causal paths. Under the linear SEM~\eqref{eq:SEM}, the covariance vector $\bmeta_i$ characterizes the heterogeneity associated with the structural error of $Y_i$, and this heterogeneity propagates only from ancestors to descendants. Although the vectors $\bmeta_i$ are not directly observable, the covariance vectors $\bzeta_{ij}$, constructed from the observed variables $Y_i$ and $Y_j$, inherit the heterogeneity transmitted to these variables and therefore follow the same propagation pattern. Consequently, if $Y_i$ is a leaf, its heterogeneity cannot propagate to any other primary variable. Therefore, every cross-covariance vector $\bzeta_{ij}$ associated with $Y_i$ contains no heterogeneity unique to $Y_i$ and can be represented by the self-covariance vectors $\{\bzeta_{kk}:k\neq i\}$ of the remaining nodes. In contrast, a non-leaf node necessarily transmits part of its heterogeneity to at least one child. As a result, the corresponding cross-covariance vectors retain a component unique to $Y_i$ that cannot be represented by the remaining self-covariance vectors. This distinction gives rise to the characterization in Proposition~\ref{prop:leaf}.  Below, we focus on nodes $Y_2$ and $Y_4$ in Figure \ref{fig:true_dag} for further illustration.


\begin{example}
Consider the DAG $\cG$ with $\cE_1$ shown in Figure \ref{fig:true_dag}. We illustrate the distinction between leaf and non-leaf nodes using $Y_4$ and $Y_2$, respectively, starting with the leaf node $Y_4$. Since $Y_4$ is a leaf, its heterogeneity cannot propagate to any other primary variable. Indeed, under Assumption \ref{ass:uncorr},
\[
\begin{aligned}
  \bzeta_{41}&=(V_{14}+V_{13}V_{34})\bmeta_1,\\
  \bzeta_{42}&=V_{23}V_{34}\bmeta_2,\\
  \bzeta_{43}&=(V_{14}+V_{13}V_{34})V_{13}\bmeta_1+V_{23}^2V_{34}\bmeta_2+V_{34}\bmeta_3.
\end{aligned}
\]
Obviously, none of these cross-covariance vectors contains the heterogeneity component $\bmeta_4$ and $\bzeta_{4i}\in\mathrm{span} (\{\bmeta_k:k\ne 4\})$ for all $i\ne 4$. Meanwhile, the self-covariance vectors are
\begin{equation}\label{eq:self_cov}
\begin{aligned}
&\bzeta_{11}=\bmeta_1, \\
&\bzeta_{22}=\bmeta_2,\\
&\bzeta_{33}=V_{13}^2\bmeta_1+V_{23}^2\bmeta_2+\bmeta_3, \\
&\bzeta_{44}=(V_{14}+V_{13}V_{34})^2\bmeta_1+(V_{23}V_{34})^2\bmeta_2+V_{34}^2\bmeta_3+\bmeta_4. 
\end{aligned}
\end{equation}
It is straightforward to verify that $\mathrm{span}(\{\bzeta_{kk}:k\ne 4\})=\mathrm{span}(\{\bmeta_k:k\ne 4\})$. Therefore, $\bzeta_{4i}\in \mathrm{span} (\{\bzeta_{kk}:k\ne 4\})$ for all $i\ne 4$, which is exactly the identification criterion of a leaf node in Proposition \ref{prop:leaf}.

We next turn to the non-leaf node $Y_2$, which has a child node $Y_3$. Unlike the leaf node $Y_4$, the heterogeneity of $Y_2$ propagates to $Y_3$ along the edge $Y_2\to Y_3$. Accordingly, under Assumption \ref{ass:uncorr}, $\bzeta_{23}=V_{23}\bzeta_{22}=V_{23}\bmeta_2$, which shows that the cross-covariance vector $\bzeta_{23}$ directly contains the heterogeneity unique to $Y_2$. In contrast, as shown in \eqref{eq:self_cov}, although the self-covariance vectors $\{\bzeta_{kk}:k\ne 2\}$ may contain $\bmeta_2$, this component always appears in linear combinations with other heterogeneity components $\{\bmeta_k:k\neq2\}$ and cannot be isolated. Indeed, by the linear independence of $\{\bmeta_k\}$ under Assumption \ref{ass:hete}, any linear combination of $\{\bzeta_{kk}:k\ne 2\}$ representing $\bmeta_2$  would require the coefficient of $\bzeta_{44}$ to be zero to eliminate $\bmeta_4$, and then the coefficient of $\bzeta_{33}$ to be zero to eliminate $\bmeta_3$, leaving no $\bmeta_2$ component. Since $V_{23}\ne 0$, it follows that $\bzeta_{23}\notin\mathrm{span}\{\bzeta_{kk}:k\ne 2\}$, and hence Proposition \ref{prop:leaf} correctly identifies $Y_2$ as a non-leaf node.
\end{example}

After identifying the leaf nodes, namely the vertices with topological height zero, $\cV(0)$, we remove them from the graph to obtain the induced subgraph $\cG(0)$. Since leaf nodes have no children, removing them from the graph does not alter the structural equations governing the remaining primary variables in model~\eqref{eq:SEM}. Consequently, Assumptions~\ref{ass:uncorr} and~\ref{ass:hete} continue to hold for the subgraph $\cG(0)$, allowing Proposition~\ref{prop:leaf} to be applied recursively. Applying the proposition to $\cG(0)$ identifies the leaves of the subgraph, which coincide with the vertices of topological height one, $\cV(1)$. Repeating this procedure recursively identifies the vertex sets $\cV(h)$ and $\cV^{+}(h)$ for $h=0,\ldots,H_{\cG}$, thereby recovering the complete causal order.  This recursive identification procedure is summarized in Algorithm \ref{alg:iden_order}. 

\begin{algorithm}[t]
  \caption{Recursive procedure for learning the causal order of $\cG$ at the population level.}\label{alg:iden_order}
  \KwIn{Random variables $\bY$ and $\bX$}

  $h\gets -1$, $p\gets \mathrm{dim}(\bY)$, $\cV^+(-1)\gets \{1,\dots,p\}$
  
   $\bZ\gets P^\perp_\bX\bY$

  \For{$i=1$ to $p$}{
    \For{$j=i$ to $p$}{
      $\bzeta_{ij}\gets \cov(\bX,Z_iZ_j)$

      $\bzeta_{ji}\gets\bzeta_{ij}$
    }
  }

  \While{$\cV^+(h)\ne \emptyset$}{
    $\cV(h+1)\gets \bigl\{i\in \cV^+(h):\bzeta_{ij}\in \mathrm{span}(\{\bzeta_{kk}:k\ne i\ and\ k\in \cV^+(h)\})\ for\ all\ j\ne i\ and\ j\in \cV^+(h)\ \bigr\}$

    $\cV^+(h+1)\gets \cV^+(h)\setminus \cV(h+1)$

    $h\gets h+1$
  }

  $H_\cG\gets h$

  \Return{$\{\cV(h)\}_{h=0,\dots, H_\cG}$, $\{\cV^+(h)\}_{h=0,\dots, H_\cG}$, and $H_\cG$}
        
\end{algorithm}

Algorithm~\ref{alg:iden_order} first computes the covariance vectors $\{\bzeta_{ij}\}$ and then recursively identifies and removes the leaf nodes of the current working subgraph. At each iteration (lines 9--13), Proposition~\ref{prop:leaf} is applied to the current subgraph $\cG(h)$ to identify its leaf nodes $\cV(h+1)$, which are subsequently removed to form the next subgraph. By the recursive argument above, Proposition~\ref{prop:leaf} remains applicable throughout the procedure. Consequently, the topological order of the entire DAG is recovered, yielding the following result.

\begin{proposition}\label{prop:order}
Under Assumptions~\ref{ass:uncorr} and~\ref{ass:hete}, the topological height of each primary variable is identifiable.
\end{proposition}

Once the causal order has been identified, recovering the entire DAG reduces to identifying the parent set of each primary variable. Indeed, for any node $Y_i$, the set $\cV^+(h_{\cG}(i))$ contains all of its ancestors while excluding all of its descendants. Therefore, it suffices to determine which variables in $\cV^+(h_{\cG}(i))$ are the direct parents of $Y_i$, as established in the next subsection.


\subsection{Identification of causal effects and directions}\label{ssec:eff}

In this subsection, we establish a sequential procedure for identifying the causal effects and directions by recovering the coefficient matrices in the SEM \eqref{eq:SEM}. Once $\bV$ is identified, the coefficient matrix $\bW$ is immediately identifiable because $\bX$ is uncorrelated with the error term $\bve$ in \eqref{eq:SEM}. We therefore begin by establishing the identifiability of $\bV$.

Given the identified causal order, the set $\cV^+(h)$ contains all parents of vertices in $\cV(h)$ while excluding their descendants, whereas vertices in $\cV(H_\cG)$ have no parents. Consequently, identifying $\bV$ reduces to estimating the direct causal effects from variables in $\cV^+(h)$ to those in $\cV(h)$ for $h=0,\dots,H_\cG-1$. Specifically, let $j\in\cV(h)$ for some $h=0,\dots,H_\cG-1$. Assumption~\ref{ass:uncorr} implies
$\cov(\bX,\ve_i\ve_j)=\E\{(\bX-\E(\bX))\ve_i\ve_j\}=\bzero$ for all $i\in \cV^+(h)$.
Moreover, since
$Z_j=P^\perp_\bX Y_j=\bV_j^T \bZ+\ve_j$,
it follows that, for all $i\in\cV^+(h)$ and $j\in \cV(h)$,
\begin{equation}\label{eq:uncorr}
\E\Bigl\{
(\bX-\E(\bX))\ve_i
\Bigl(
Z_j-\sum_{k\in\cV^+(h)}V_{kj}Z_k
\Bigr)
\Bigr\}
=
\bzero.
\end{equation}
To identify the coefficient vector $\bV_j$, the moment condition \eqref{eq:uncorr} naturally motivates the construction of IVs satisfying the required orthogonality condition. We therefore define the surrogate instruments
\[
\bT(h)
:=
\bigl(
(\bX-\E(\bX))\ve_i
\bigr)_{i\in\cV^+(h)}
\in
\mathbb{R}^{q|\cV^+(h)|},
\qquad
h=0,\dots,H_\cG-1.
\]
To serve as valid instruments, they must also satisfy the relevance condition. Section S2.4 of the Supplementary Material \citep{zou2026supplement} shows that the relevance of these surrogate instruments is determined by the covariance heterogeneity encoded in $\bmeta_i$. The following proposition establishes that the surrogate instruments satisfy the rank condition for multiple valid IVs \citep{schmidt1976econometrics}, ensuring that $\bV_j$ is the unique solution to \eqref{eq:uncorr}.

\begin{proposition}\label{prop:rank}
Suppose that Assumptions~\ref{ass:uncorr} and~\ref{ass:hete} hold. Then, for all $h=0,\dots,H_\cG-1$, the surrogate instruments $\bT(h)$ satisfy the rank condition
\[
\rank\!\left\{
\E\!\left(
\bT(h)\bZ_{\cV^+(h)}^T
\right)
\right\}
=
|\cV^+(h)|.
\]
\end{proposition}

However, since the structural errors $\ve_i$ are unobserved, the surrogate instruments $\bT(h)$ cannot be constructed directly, preventing Proposition~\ref{prop:rank} from being applied to identify the causal effects $\bV_j$ for $j\in\cV(h)$. To overcome this difficulty, we develop a sequential top-down identification strategy.

We begin with the simplest case, where $h=H_\cG-1$ and $j\in\cV(H_\cG-1)$. In this case, the parents of $Y_j$ are restricted to $\cV^+(H_\cG-1)=\cV(H_\cG)$. Since vertices in $\cV(H_\cG)$ have no parents, $\bV_i=\mathbf 0$ and hence
$Z_i=\bV_i^T\bZ+\ve_i=\ve_i$ for $
i\in\cV(H_\cG)$.
Therefore, the structural errors are directly observable through $Z_i$, and the surrogate instruments reduce to
\[
\bT(H_\cG-1)
=
\bigl((\bX-\E(\bX))Z_i\bigr)_{i\in\cV(H_\cG)}.
\]
Proposition~\ref{prop:rank} then guarantees that $\bV_j$ is identifiable for all $j\in\cV(H_\cG-1)$.

\begin{algorithm}[t]
  \caption{Sequential procedure for learning causal effects and directions at the population level.}\label{alg:iden_dag}
  \KwIn{Random variables $\bY$ and $\bX$, causal order $\{\cV(h)\}_{0\le h\le H_\cG}$, $\{\cV^+(h)\}_{0\le h\le H_\cG}$, and $H_\cG$}

   $\bZ\gets P^\perp_\bX\bY$

  $\bV\gets\mathbf{NA}$

  \For{$j\in \cV(H_\cG)$}{
    $\bV_j\gets \bzero$
  }

  \For{$h=H_\cG-1$ to $0$}{
    $\bT(h)\gets\bigl((\bX-\E(\bX))(Z_i-\bV_i^T \bZ)\bigr)_{i\in\cV^+(h)}$

    \For{$j\in \cV(h)$}{

      $(V_{ij})_{i\notin \cV^+(h)}\gets \bzero$

      $(V_{ij})_{i\in \cV^+(h)}\gets\ solution\ to\ \E\Bigl(\bT(h)\Bigl(Z_j-\sum_{i\in \cV^+(h)}V_{ij}Z_i\Bigr)\Bigr)=\bzero$
    }
  }

  $\bW\gets \var(\bX)^{-1}\cov(\bX,(\bI_p-\bV^T)\bY)$

  $\cE_1\gets \{(i,j):V_{ij}\ne 0\}$
  
  $\cE_2\gets \{(\ell,j):W_{\ell j}\ne 0\}$

  \Return{$\bV$, $\bW$, $\cE_1$, and $\cE_2$}
\end{algorithm}

Next consider $h=H_\cG-2$. Since the coefficient vectors $\bV_i$ for $i\in\cV^+(H_\cG-2)$ have already been identified in the previous step, the structural errors can be recovered as
$\ve_i
=
Z_i-\bV_i^T\bZ$ for $
i\in\cV^+(H_\cG-2)$,
which again yields observable surrogate instruments,
\[
\bT(H_\cG-2)
=
\bigl((\bX-\E(\bX))(Z_i-\bV_i^T\bZ)\bigr)_{i\in\cV^+(H_\cG-2)}.
\]
Applying Proposition~\ref{prop:rank} identifies $\bV_j$ for all $j\in\cV(H_\cG-2)$. Repeating this argument recursively identifies the entire coefficient matrix $\bV$, and hence the edge set $\cE_1$. This sequential identification procedure is summarized in Algorithm~\ref{alg:iden_dag}. In particular, line 2 initializes $\bV$ as $\mathbf{NA}$ to indicate that the corresponding coefficients are initially unidentified.

Algorithm~\ref{alg:iden_dag} learns the causal effects in a top-down manner according to the identified causal order. At each iteration (lines~6--12), it constructs observable surrogate instruments $\bT(h)$ using the previously identified coefficients $\bV_i$ for $i\in\cV^+(h)$, and applies Proposition~\ref{prop:rank} to identify the coefficient vectors $\bV_j$ for $j\in\cV(h)$. Once $\bV$ has been identified, $\bW$ is recovered in line~13 as $\var(\bX)^{-1}\cov(\bX,(\bI_p-\bV^T)\bY)$,
because $(\bI_p-\bV^T)\bY=\bW^T\bX+\bve$ and $\cov(\bX,\bve)=\bzero$. Hence, the entire DAG with edges $(\cE_1,\cE_2)$ is identified. We conclude the results for identifying the causal effects and directions in Proposition \ref{prop:eff}.

\begin{proposition}\label{prop:eff}
  Suppose that Assumptions \ref{ass:uncorr} and \ref{ass:hete} hold. Then the coefficients $\bV$ and $\bW$ are identifiable by the unique solution to
  \begin{equation}\label{eq:iden_eff}
  \begin{aligned}
      & \E\{\bX(\bY-\bV^T\bY-\bW^T\bX)^T\}=\bzero,\\
      & \E\{(\bX-\E(\bX))(Y_i-\bV_i^T\bY-\bW_i^T\bX)(Y_j-\bV_j^T\bY-\bW_j^T\bX)\}=\bzero,~~ \text{for}~ i\ne j, \\
      & \text{subject to}~ V_{ij}=0,~~ \text{for}~ h_\cG(i)\le h_\cG(j). 
  \end{aligned}
\end{equation}
  Subsequently, the edges in $\cG$ are identified as 
  \[
  \cE_1= \{(i,j):V_{ij}\ne 0\}~~ \text{and}~~ \cE_2= \{(\ell,j):W_{\ell j}\ne 0\}.
  \]
\end{proposition}

Combining Propositions~\ref{prop:order} and~\ref{prop:eff}, the preceding results establish a complete identification theory for DAGs under covariance heterogeneity. As summarized in Figure~\ref{fig:summarize}, the proposed identification strategy consists of two stages. First, the causal order is recovered by recursively identifying and removing the leaf nodes of the working subgraph. Given the identified order, the causal effects and directions are then sequentially identified by recursively constructing observable surrogate IVs. Section~\ref{sec:est} builds on these identification results to develop finite-sample procedures for causal discovery and effect estimation.

\section{Estimation methodology}\label{sec:est}

In this section, we present the Covariance Heterogeneity based Approach to Recovering Graphical Effects (CHARGE), a data-adaptive procedure for estimating causal effects and learning causal directions. We focus on the low-dimensional setting where $p$ and $q$ are fixed. Extensions to high-dimensional settings, including $n\ll p$ and $p\le n\ll q$, together with the corresponding theoretical analysis, are presented in Section~S6 of the Supplementary Material \citep{zou2026supplement}.
Following the two-step identification strategy developed in the previous section, CHARGE also proceeds in two stages. Section~\ref{ssec:est_order} presents the algorithm for learning the causal order, followed by Section~\ref{ssec:est_eff}, which develops the estimation procedure for the causal effects and directions. The theoretical guarantees of CHARGE are established in Section~\ref{ssec:theo}.

\subsection{Learning causal order}\label{ssec:est_order}

In this subsection, we develop the empirical counterpart of Algorithm~\ref{alg:iden_order} for learning the causal order from the observed data $(\Y,\X)$. To obtain a finite-sample analogue of Proposition~\ref{prop:leaf}, we first establish the following auxiliary result.

\begin{proposition}\label{prop:full}
Under Assumption~\ref{ass:uncorr},
\[
\bzeta_{ij}\in \mathrm{span}\bigl(\{\bzeta_{kk}:k=1,\dots,p\}\bigr),
\quad \text{for all $
i\neq j$ and $i,j=1,\dots,p$.}
\]
\end{proposition}

Propositions~\ref{prop:full} and~\ref{prop:leaf} together transform leaf identification into a goodness-of-fit testing problem. The key observation is that Proposition~\ref{prop:full} guarantees a full linear representation of every cross-covariance vector $\bzeta_{ij}$, whereas Proposition~\ref{prop:leaf} shows that, when $Y_i$ is a leaf node, the same vector admits a reduced representation obtained by excluding the self-covariance vector $\bzeta_{ii}$. Specifically, Proposition~\ref{prop:full} implies that, for any $j\neq i$,
\[
\bzeta_{ij}
=
\sum_{k=1}^{p}
\gamma_k^{(i,j)}
\bzeta_{kk},
\]
for some coefficients $\gamma_k^{(i,j)}$. Moreover, Proposition~\ref{prop:test} establishes that $\{\bzeta_{kk}:k=1,\dots,p\}$ are linearly independent under Assumption~\ref{ass:hete}, ensuring the uniqueness of the above representation. Thus, for a leaf node $Y_i$, comparison with the reduced representation guaranteed by Proposition~\ref{prop:leaf} yields $\gamma_i^{(i,j)}=0$, and hence 
\[
\bzeta_{ij}
=
\sum_{k\neq i}
\gamma_k^{(i,j)}
\bzeta_{kk},
\qquad
j\neq i.
\] 
Therefore, testing whether $Y_i$ is a leaf node is equivalent to testing
\begin{equation}\label{eq:eq_test}
\begin{aligned}
&H_0:\;
\bzeta_{ij}\in
\mathrm{span}\bigl(\{\bzeta_{kk}:k\neq i\}\bigr),
\quad
\text{for all }j\neq i,
\\
&\hspace{2cm}\text{versus}
\\
&H_1:\;
\bzeta_{ij}\in
\mathrm{span}\bigl(\{\bzeta_{kk}:k=1,\dots,p\}\bigr)
\setminus
\mathrm{span}\bigl(\{\bzeta_{kk}:k\neq i\}\bigr),
\quad
\text{for some }j\neq i.
\end{aligned}
\end{equation}

The hypotheses in \eqref{eq:eq_test} compare two nested linear models. The full model uses all self-covariance vectors $\{\bzeta_{kk}\}_{k=1,\ldots,p}$ to represent $\bzeta_{ij}$, whereas the reduced model excludes $\bzeta_{ii}$. Hence, determining whether $Y_i$ is a leaf amounts to assessing whether removing $\bzeta_{ii}$ leads to a significant loss of fit. This establishes a goodness-of-fit testing problem for nested linear models.
If the null hypothesis in \eqref{eq:eq_test} is not rejected, we classify $Y_i$ as a leaf node of $\cG$. Since Assumptions~\ref{ass:uncorr} and~\ref{ass:hete} remain valid for every working subgraph $\cG(h)$ with $h=0,\ldots,H_\cG-1$, the same testing procedure can be applied recursively to each working subgraph. This recursive goodness-of-fit procedure provides the basis for the finite-sample algorithm for learning the causal order.

Equipped with the above characterization, we now present Algorithm~\ref{alg:est_order} for learning the causal order from finite samples. The algorithm first computes the empirical covariance vectors $\{\hat{\bzeta}_{ij}\}$ in lines~2--9 and then recursively estimates the causal order in lines~10--23 by repeatedly identifying the leaf nodes of the current working subgraph.

\begin{algorithm}[t]
  \caption{Empirical method for learning causal order.}\label{alg:est_order}
  \KwIn{Data matrices $\Y$ and $\X$}

  $h\gets -1$, $p\gets \mathrm{dim}_2(\Y)$, $\hcV^+(-1)\gets \{1,\dots,p\}$
  
   $\hZ\gets (\bI_n-\X(\X^T \X)^{-1}\X^T)\Y$ 

   $\hat{\bmu}\gets \frac{1}{n}\sum_{k=1}^n \X^{(k)}$

  \For{$i=1$ to $p$}{
    \For{$j=i$ to $p$}{
      $\hat{\bzeta}_{ij}\gets \frac{1}{n}\sum_{k=1}^n (\X^{(k)}-\hat{\bmu})\hZ^{(k)}_{i}\hZ^{(k)}_{j}$

      $\hat{\bzeta}_{ji}\gets\hat{\bzeta}_{ij}$
    }
  }

  \While{$\hcV^+(h)\ne \emptyset$}{
    $\hcV(h+1)\gets \emptyset$

    $\hat{\bPhi}_{\bzeta}(h)\gets (\hat{\bzeta}_{kk})_{k\in \hcV^+(h)}$

    \For{$i\in \hcV^+(h)$}{
        $\hat{\bR}_i(h)\gets (\hat{\bzeta}_{kk})_{k\ne i,k\in \hcV^+(h)}$

        \begin{equation}\label{eq:F_modi}
          \hat{F}_i(h)\gets \max_{j\ne i,j\in \hcV^+(h)}\frac{\hat{\bzeta}_{ij}^T\Bigl(P_{\hat{\bPhi}_{\bzeta}(h)}-P_{\hat{\bR}_i(h)}\Bigr)\hat{\bzeta}_{ij}}{\hat{\bzeta}_{ij}^T\Bigl(\bI_q-P_{\hat{\bPhi}_{\bzeta}(h)}\Bigr)\hat{\bzeta}_{ij}+C_0/n}
        \end{equation}

        \If{$\hat{F}_i(h)<\sqrt{2.01\log n}$}{
            $\hcV(h+1)\gets \hcV(h+1)\cup \{i\}$

            $\hat{h}_\cG(i)\gets h+1$
        }
    }
    $\hcV^+(h+1)\gets \hcV^+(h)\setminus \hcV(h+1)$

    $h\gets h+1$
  }

  $\hat{H}_\cG\gets h$

  \Return{$\{\hat{h}_\cG(i)\}_{1\le i\le p}$, $\{\hcV(h)\}_{0\le h\le \hat{H}_\cG}$, $\{\hcV^+(h)\}_{0\le h\le \hat{H}_\cG}$, and $\hat{H}_\cG$}
\end{algorithm}

To determine whether a node is a leaf, the algorithm tests the goodness-of-fit criterion in \eqref{eq:eq_test} using the statistic $\hat F_i(h)$ defined in \eqref{eq:F_modi}, which is motivated by the classical F-statistic for nested linear models. Since the null hypothesis requires the reduced representation to hold simultaneously for every $j\neq i$ in the current working subgraph, the statistic is defined as the maximum over all such $j$. Ignoring this maximization, the numerator measures the additional lack of fit caused by removing $\bzeta_{ii}$ from the full representation, whereas the denominator normalizes this quantity to account for sampling variability. A vanishing stabilization term $C_0/n$, with any constant $C_0>0$, is further added to the denominator to avoid division by zero while retaining favorable asymptotic properties. 

If $\hat F_i(h)$ falls below the threshold $\sqrt{2.01\log n}$, the null hypothesis is not rejected and $Y_i$ is classified as a leaf node, as implemented in lines~16--19. The constant $2.01$ follows from a Gaussian tail bound, whereas the factor $\sqrt{\log n}$ accounts for the multiple testing involved in simultaneously examining all candidate leaf nodes. After identifying the leaf nodes, the algorithm removes them from the current working subgraph (lines~21--22) and repeats the procedure until the graph becomes empty, thereby producing the estimated causal order.

By leveraging the estimated causal order, we not only guarantee that the estimated graph is acyclic, but also substantially reduce the parameter space and computational complexity. This estimated causal order serves as the foundation for the sequential estimation of the causal effects and directions developed in the next subsection.

\subsection{Estimating causal effects and directions}\label{ssec:est_eff}

In this subsection, we develop an estimator for the causal effects and directions based on the estimated causal order obtained from Algorithm~\ref{alg:est_order}. Building upon the moment conditions established in Proposition~\ref{prop:eff}, we estimate the causal parameters via the generalized method of moments (GMM).

Let $\bmu=\E(\bX)$ denote the expectation of $\bX$, let $\btheta=(\bmu,\bV,\bW)$ denote the unknown parameter tuple,
and let $\Ok=(\Y^{(k)T},\X^{(k)T})^T$ denote the $k$th observation. Define the following vector-valued moment functions
\begin{equation}\label{eq:moments}
\begin{aligned}
&\bQ_1(\btheta,\Ok)=\X^{(k)}-\bmu,\\
&\bQ_2(\btheta,\Ok)=\X^{(k)}\bigl(\Y_i^{(k)}-\bV_i^T\Y^{(k)}-\bW_i^T\X^{(k)}\bigr),~~ i=1,\dots,p,\\
&\bQ_3(\btheta,\Ok)=\bigl(\X^{(k)}-\bmu\bigr)\bigl(\Y_i^{(k)}-\bV_i^T\Y^{(k)}-\bW_i^T\X^{(k)}\bigr)\bigl(\Y_j^{(k)}-\bV_j^T\Y^{(k)}-\bW_j^T\X^{(k)}\bigr),~~ i\neq j,
\end{aligned}
\end{equation}
and let $\bQ(\btheta,\Ok)$ denote the vector obtained by stacking the above moment functions. Proposition~\ref{prop:eff} implies that the true parameter $\btheta$ is the unique element of
\[
\bTheta=
\Bigl\{
(\bmu,\bV,\bW):
V_{ij}=0,\ \text{for all }h_\cG(i)\le h_\cG(j)
\Bigr\},
\]
satisfying
\[
\E\{\bQ(\btheta,\Ok)\}=\bzero.
\]

Replacing the unknown topological heights by their estimates from Algorithm~\ref{alg:est_order}, we estimate $\btheta$ by the GMM estimator
\begin{equation}\label{eq:GMM}
\hat{\btheta}
=
\argmin_{\btheta\in\hat{\bTheta}}
\left\{
\sum_{k=1}^{n}
\bQ(\btheta,\Ok)
\right\}^{\!T}
\bOmega_n
\left\{
\sum_{k=1}^{n}
\bQ(\btheta,\Ok)
\right\},
\end{equation}
where
\[
\hat{\bTheta}
=
\Bigl\{
(\bmu,\bV,\bW):
V_{ij}=0,\ \text{for all }\hat{h}_\cG(i)\le\hat{h}_\cG(j)
\Bigr\},
\]
and $\bOmega_n$ is a positive definite weighting matrix.

The estimated causal order restricts the parameter space by setting $\hat V_{ij}=0$ whenever $\hat h_\cG(i)\le\hat h_\cG(j)$, while the remaining parameters are estimated by solving the optimization problem in \eqref{eq:GMM}. The choice of the weighting matrix $\bOmega_n$ determines the asymptotic efficiency of the GMM estimator. As in classical GMM, semiparametric efficiency can be achieved using either the two-step or the iterated GMM procedure of \cite{hansen1982large}. When the moment conditions in \eqref{eq:moments} are weak, inference can instead be based on GMM procedures that are robust to weak identification, such as those proposed by \cite{staiger1997instrumental} and \cite{stock2000gmm}. Since weak identification is not the focus of this paper, we do not pursue this direction further and refer the reader to \cite{stock2002survey} for a comprehensive review.

After estimating the coefficient matrices $\bV$ and $\bW$, we recover the edge sets $\cE_1$ and $\cE_2$ using the hard-thresholding principle \citep{donoho1994ideal,donoho1995noising}. Hard thresholding has been widely used for support recovery in high-dimensional estimation and has also been adopted in causal inference, for example by \cite{guo2018confidence} to identify strong and valid IVs. Specifically, an edge is included if the corresponding estimated coefficient is sufficiently far from zero. Since $\hat{V}_{ij}=0$ whenever $\hat{h}_\cG(i)\le \hat{h}_\cG(j)$, it suffices to detect edges among the remaining candidate pairs. The resulting estimators of $\cE_1$ and $\cE_2$ are
\begin{equation}\label{eq:thresh}
\begin{aligned}
  &\hat{\cE}_1=\Biggl\{(i,j):\hat{h}_\cG(i)>\hat{h}_\cG(j),|\hat{V}_{ij}|\ge \sqrt{\widehat{\sigma^2_{\bV}[i,j]}}\sqrt{\frac{2.01\log n}{n}}\Biggr\},\\
  &\hat{\cE}_2=\Biggl\{(\ell,j):|\hat{W}_{\ell j}|\ge \sqrt{\widehat{\sigma^2_{\bW}[\ell,j]}}\sqrt{\frac{2.01\log n}{n}}\Biggr\},
\end{aligned}
\end{equation}
where $\widehat{\sigma^2_{\bV}[i,j]}$ and $\widehat{\sigma^2_{\bW}[\ell,j]}$ are consistent estimators of the asymptotic variances of $\hat{V}_{ij}$ and $\hat{W}_{\ell j}$, respectively.
The threshold in \eqref{eq:thresh} consists of two components. Taking $\hat{\cE}_1$ as an example, the term $\sqrt{\widehat{\sigma^2_{\bV}[i,j]}/n}$ corresponds to the standard error of $\hat{V}_{ij}$ and accounts for the estimation uncertainty, whereas the factor $\sqrt{2.01\log n}$ adjusts for the multiple comparisons involved in simultaneously testing whether the coefficients are nonzero. The constant $2.01$ arises from a Gaussian tail bound. Consequently, the proposed hard-thresholding rule simultaneously accounts for estimation variability and multiplicity, yielding a consistent estimator of the edge sets.

In addition to the GMM estimator, we also develop a sequential estimation procedure following Algorithm~\ref{alg:iden_dag}. This approach is computationally simpler but typically less efficient than the GMM estimator. Since it can be formulated within the GMM framework with a particular weighting matrix $\bOmega_n$, we omit the details here and refer the reader to Section~S5 of the Supplementary Material \citep{zou2026supplement}. In practice, this sequential procedure may also be useful for providing a consistent initial value for the efficient GMM estimator. 

To summarize, the proposed method consists of three components: a recursive procedure for learning the causal order (Algorithm~\ref{alg:est_order}), a GMM estimator \eqref{eq:GMM} for estimating the causal effects, and a hard-thresholding step \eqref{eq:thresh} for recovering the DAG. The theoretical properties of the proposed procedure are established in the next subsection.

\subsection{Theoretical guarantees}\label{ssec:theo}

In this subsection, we establish the theoretical guarantees for the proposed procedure, including causal order learning, causal effect estimation, and causal discovery. We begin by establishing the consistency of Algorithm~\ref{alg:est_order}, which provides the foundation for the subsequent estimation and graph recovery results.

\begin{lemma}[Consistent causal order]\label{lem:con_order}
Suppose that Assumptions~\ref{ass:uncorr} and~\ref{ass:hete} hold. Then
\[
\lim_{n\to\infty}
\P\bigl(
\hat{h}_\cG(i)=h_\cG(i),
\ \text{for all }i=1,\dots,p
\bigr)
=1.
\]
\end{lemma}

Lemma~\ref{lem:con_order} justifies replacing the unknown causal order by its estimator in the subsequent GMM procedure. In particular, the estimated parameter space in \eqref{eq:GMM} satisfies
\[
\lim_{n\to\infty}\P(\hat{\bTheta}=\bTheta)=1,
\]
so that the zero constraints imposed by the estimated causal order are asymptotically correct. Consequently, the uncertainty arising from causal order learning becomes asymptotically negligible for estimating the causal coefficients. The following theorem establishes the asymptotic distribution of the resulting GMM estimators.

\begin{theorem}[Asymptotic normality]\label{thm:asy}
Suppose that Assumptions~\ref{ass:uncorr} and~\ref{ass:hete} hold. Then, under some regularity conditions in Section~S4.3 of the Supplementary Material \citep{zou2026supplement},
\[
\begin{aligned}
\sqrt{n}(\hat{V}_{ij}-V_{ij})
&\to
\cN(0,\sigma_{\bV}^{2}[i,j]),
\qquad
\text{for all }h_\cG(i)>h_\cG(j),\\
\sqrt{n}(\hat{W}_{\ell j}-W_{\ell j})
&\to
\cN(0,\sigma_{\bW}^{2}[\ell,j]),
\qquad
\ell=1,\dots,q,\;
j=1,\dots,p,
\end{aligned}
\]
where the explicit forms of
$\sigma_{\bV}^{2}[i,j]$
and
$\sigma_{\bW}^{2}[\ell,j]$
are given in Section~S4.3 of the Supplementary Material \citep{zou2026supplement}.
\end{theorem}

Theorem~\ref{thm:asy} provides valid statistical inference for all nonzero causal coefficients that remain after imposing the estimated causal order. Combined with the consistency of Algorithm~\ref{alg:est_order}, it ensures that estimating the causal order does not compromise the asymptotic validity of the subsequent GMM estimator.

\begin{theorem}[Consistent causal discovery]\label{thm:con_graph}
Suppose that Assumptions~\ref{ass:uncorr} and~\ref{ass:hete} hold. Then
\[
\lim_{n\to\infty}
\P\bigl(
\hat{\cE}_1=\cE_1
\ \text{and}\
\hat{\cE}_2=\cE_2
\bigr)
=1.
\]
\end{theorem}

Theorem~\ref{thm:con_graph} establishes the consistency of the proposed hard-thresholding procedure for recovering the edge sets. Together with Lemma~\ref{lem:con_order} and Theorem~\ref{thm:asy}, it provides a complete theoretical guarantee for the proposed two-stage framework, demonstrating that the recursive causal order learning, GMM estimation, and final graph recovery are all asymptotically valid.

\section{Simulation studies}\label{sec:sim}

In this section, we conduct simulation studies to evaluate the finite-sample performance of CHARGE. The weighting matrix $\bOmega_n$ in \eqref{eq:GMM} is specified by the sequential procedure outlined in Section S5 of the Supplementary Material \citep{zou2026supplement}, and the stabilization constant in \eqref{eq:F_modi} is set to $C_0=10^{-6}$. For causal discovery, we compare our method with LiNGAM \citep{shimizu2006linear}, RFCI \citep{colombo2012learning}, and LRpS-GES \citep{frot2019robust}. LiNGAM is included as a representative method for linear SEMs, while RFCI and LRpS-GES are representative methods for causal discovery under hidden confounding, targeting a PAG and a CPDAG, respectively. For causal effect estimation, since neither RFCI nor LRpS-GES estimates direct causal effects, we restrict the comparison to LiNGAM.

We consider three types of DAGs with hidden confounding: random graphs, hub graphs, and chain graphs. Specifically, define the adjacency matrix $\bA\in\R^{p\times p}$ of the DAG $\cG=(\cV;\cE_1,\cE_2)$ as $A_{ij}=1$ if $(i,j)\in \cE_1$ and $0$ otherwise. For random graphs, each off-diagonal entry in the upper triangle of the adjacency matrix $\{A_{ij}\}_{i<j}$ is independently generated from $\text{Bernoulli}\{2/(3p)\}$, while the other entries are set to $0$. For hub graphs, the adjacency matrix is defined by $A_{1j}=1$ for $j=2,\dots, p$ with all the other entries being $0$. For chain graphs, the adjacency matrix has entries $A_{j,j+1}=1$ for $j=1,\dots, p-1$ and zeros elsewhere. We set the dimension of primary variables to $p=10$ and the dimension of auxiliary variables to $q=2p=20$. 
Consider the following data generation model
\[
\bY=\bV^T \bY+\bW^T \bX+ \bGamma^T \bU+\bxi, 
\]
where $\bU$ is the unobserved confounder of dimension $r=3$ and $\bxi$ is the noise term. Given the adjacency matrix $\bA$, we set the causal effect $V_{ij}\sim\mathrm{Uniform}\{(-1.2,-0.8)\cup (0.8,1.2)\}$ if $A_{ij}\ne 0$ and $0$ otherwise.  To evaluate the performance of the proposed method when no valid IVs are available, we set $W_{\ell j}=3$ for all $\ell=1,\dots,q$ and $j=1,\dots,p$. This implies that every $X_\ell$ has causal effects on all primary variables and thus violates the exclusion restriction condition, resulting in no valid IVs. For the confounding effects $\bGamma$, we sample each entry $\Gamma_{kj}$ from $\mathrm{Uniform}\{(-1,-0.5)\cup(0.5,1)\}$. We generate the auxiliary variables $\bX$ by independently drawing each $X_\ell$ from a three-level $(1,2,3)$ discrete variable with probabilities $(0.6,0.2,0.2)$. The unobserved confounder is generated as $\bU\sim \cN_r(\bzero,\bI_r)$, and the noise is generated as $\xi_i=\vs_i(X_i+X_{i+p})$ for $i=1,\dots,p$, where $\vs_i$ is independently drawn from $\cN(0,1)$ and is independent of $\bX$ and $\bU$. Under these conditions, $\bve=\bGamma^T\bU+\bxi$ satisfies $\cov(\bX,\ve_i\ve_j)=0$ for all $i\ne j$, thereby verifying Assumption \ref{ass:uncorr}. Besides, only the $i$th and $(i+p)$th elements of $\bmeta_i$ are nonzero for $i=1,\dots,p$, which implies that Assumption \ref{ass:hete} holds.

 \begin{figure}[htbp]
    \centering
    \includegraphics[width=1\textwidth]{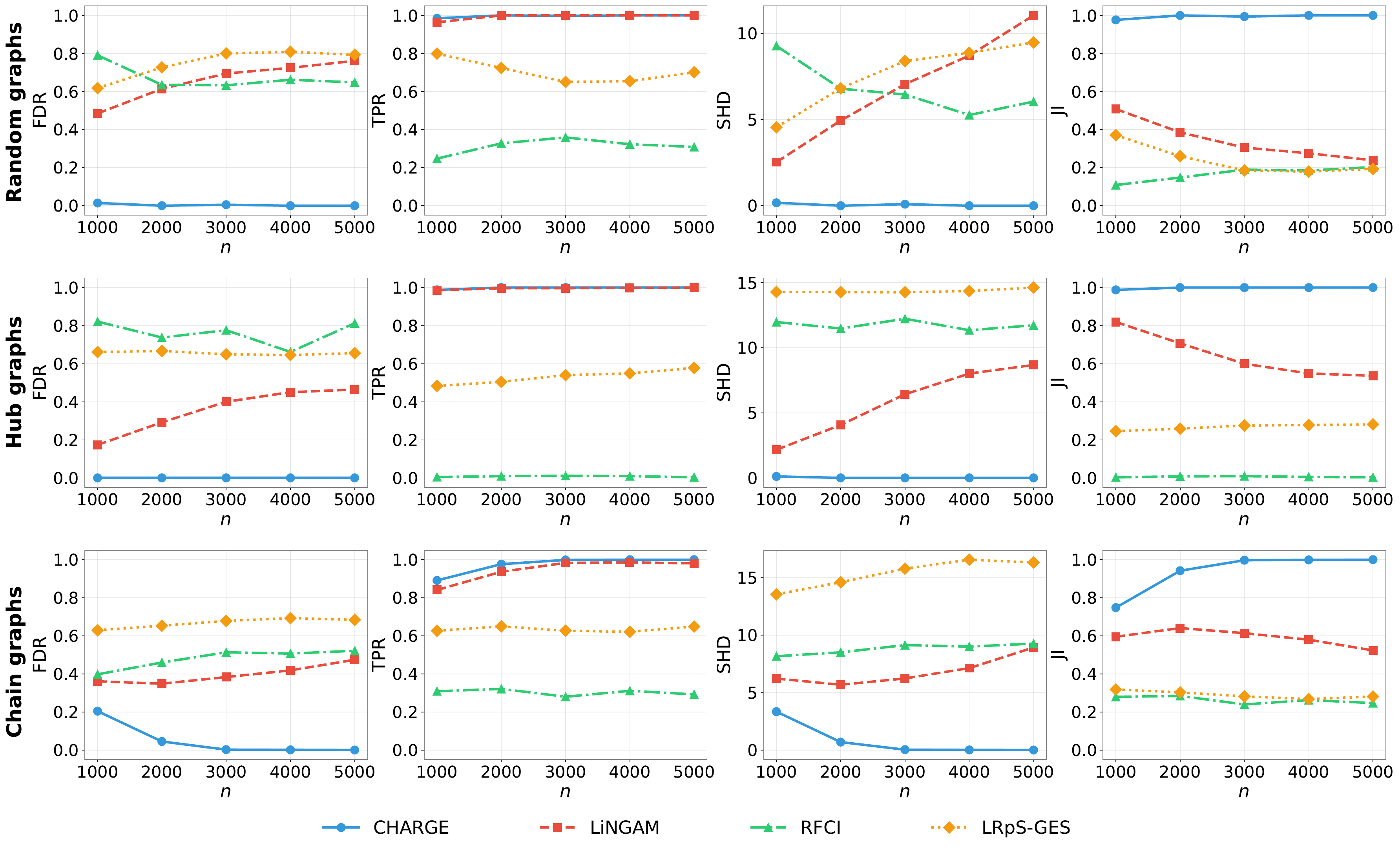}
    \caption{Comparisons between the proposed (CHARGE) and existing methods in recovering $\cE_1$ across different graphs, evaluated by the average false discovery rate (FDR), true positive rate (TPR), structural Hamming distance (SHD), and Jaccard index (JI).}
    \label{fig:sim_str}
\end{figure}

As our  interest is the causal graph among the primary variables $\bY$, we vary the sample size $n$ to compare the performance of different methods in recovering the edge set $\cE_1$ and the causal coefficient matrix $\bV$. For evaluation of the causal discovery, we consider four commonly-used graph metrics: the false discovery rate (FDR), the true positive rate (TPR), the structural Hamming distance (SHD), and the Jaccard index (JI). To compute these metrics, let $\TP$, $\RE$, $\FP$, and $\FN$ denote the numbers of the estimated edges with correct directions, those with reverse directions, those not in the true DAG, and the missing edges from the true DAG, respectively. Then $\FDR=(\RE+\FP)/(\TP+\RE+\FP)$, $\TPR=\TP/(\TP+\FN)$, $\SHD=\FP+\FN+\RE$, and $\JI=\TP/(\TP+\SHD)$. 
Accordingly, smaller values of FDR and SHD and larger values of TPR and JI indicate better graph recovery performance. 
Since RFCI returns a PAG and LRpS-GES returns a CPDAG, both of which may contain undirected edges, we evaluate these methods favorably by assigning the correct orientation whenever an undirected edge is consistent with the true DAG, following \cite{zheng2018dags}.
 To evaluate causal effect estimation, we compute the entrywise $\ell_2$-loss of $\hat{\bV}$ as $(\sum_{i,j=1}^p (\hat{V}_{ij}-V_{ij})^2)^{1/2}$. For each simulation setting, all reported results are averaged over
 $100$ experiments.

The causal discovery results are summarized in Figure~\ref{fig:sim_str}. Overall, CHARGE consistently achieves the best performance across all simulation settings in terms of FDR, TPR, SHD, and JI. Specifically, both RFCI and LRpS-GES perform poorly in the presence of substantial hidden confounding. LiNGAM achieves a TPR comparable to that of CHARGE, but performs considerably worse on the remaining evaluation metrics. In contrast, CHARGE attains nearly perfect graph recovery in all settings except for chain graphs with small sample sizes. Moreover, its performance on chain graphs improves steadily as the sample size increases, providing empirical support for the consistency of CHARGE in recovering  $\cE_1$.

\begin{figure}[ht]
    \centering
    \includegraphics[width=1\textwidth]{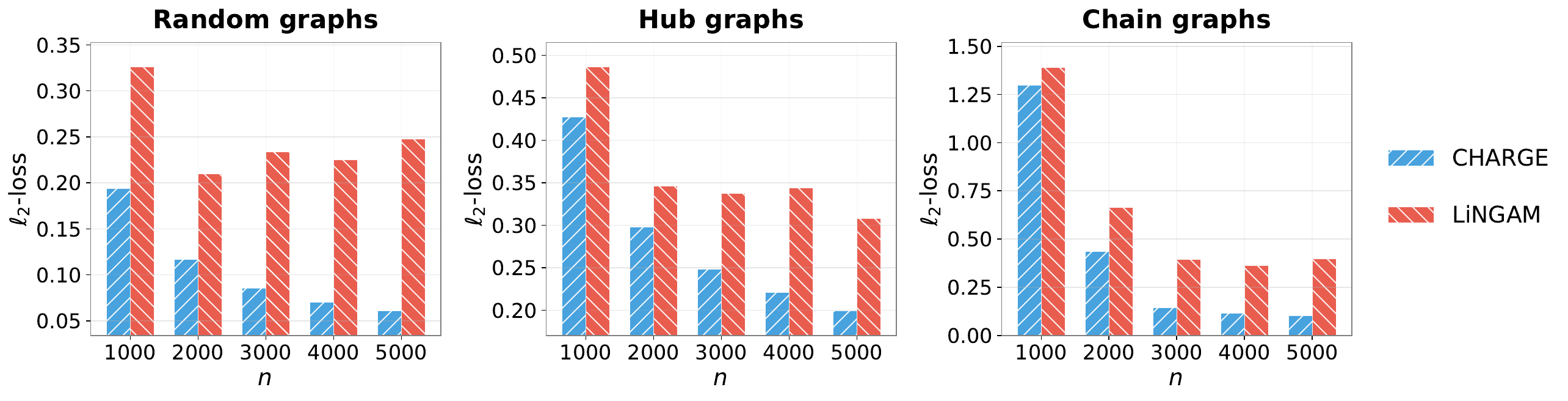}
    \caption{Comparisons between the proposed method (CHARGE) and LiNGAM in recovering $\bV$ across different graphs, evaluated by the average entrywise $\ell_2$-loss.}
    \label{fig:sim_para}
\end{figure}

Figure~\ref{fig:sim_para} presents the simulation results for causal effect estimation. CHARGE outperforms LiNGAM across the different graph structures. Although the entrywise $\ell_2$-loss of LiNGAM decreases as the sample size increases for hub and chain graphs, its performance remains poor for random graphs. In contrast, the estimation error of CHARGE decreases as the sample size increases across all graph types. For chain graphs, the  error of CHARGE declines markedly as $n$ increases from $1000$ to $2000$. This pattern is consistent with the causal discovery results for chain graphs shown in Figure~\ref{fig:sim_str}, where the accuracy of CHARGE also improves substantially over the same range of sample sizes. These findings provide further empirical support for the consistency of CHARGE in estimating $\bV$.

\newpage

\section{Application}\label{sec:app}

To illustrate the applicability of the proposed method, this section presents an analysis of the mouse obesity dataset described by \cite{wang2006genetic}. In this study, F2 mice were fed a high-fat Western diet from 8 to 24 weeks of age, and their body weights were recorded. Genome-wide SNPs were genotyped, and liver tissue gene expression was profiled. We use the preprocessed data provided by \cite{lin2015regularization} to learn a gene regulatory network associated with obesity. Specifically, we focus on genes involved in fatty acid metabolism, which are grouped into Module 221 (\url{https://robotics.stanford.edu/~erans/cancer/modules/module_221.html}).

In our analysis, we select $p=20$ genes from Module 221 without missing values as the primary variables $\bY$. These genes are functionally related through their roles in fatty acid beta-oxidation, fatty acid degradation, and lipid metabolism \citep{koo2013nonalcoholic,rosen2014we}. Following \cite{chen2024discovery}, we further select the two marginally most correlated SNPs for each gene, yielding $q=40$ auxiliary variables $\bX$. To investigate how gene regulatory mechanisms differ with obesity status, we divide the mice into lean and obese groups according to the median body weight, resulting in sample sizes of 143 and 144, respectively. We then apply the proposed method to learn the gene regulatory networks for the two groups.

\begin{figure}[t]
    \centering
    \begin{subfigure}[b]{0.49\textwidth}
        \includegraphics[width=\textwidth]{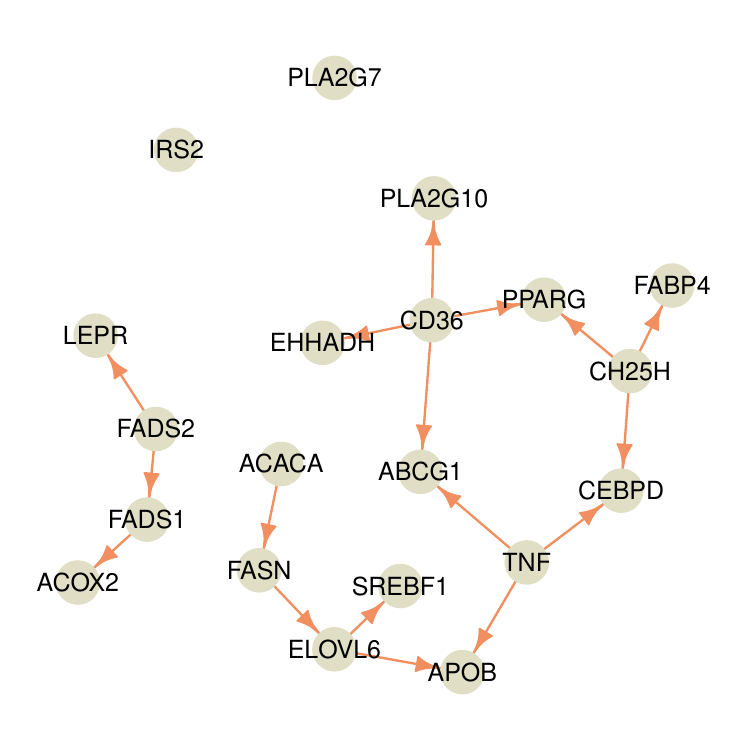}
        \caption{Lean mice}
        \label{fig:exp_collinear_pair}
    \end{subfigure}
    \begin{subfigure}[b]{0.49\textwidth}
        \includegraphics[width=\textwidth]{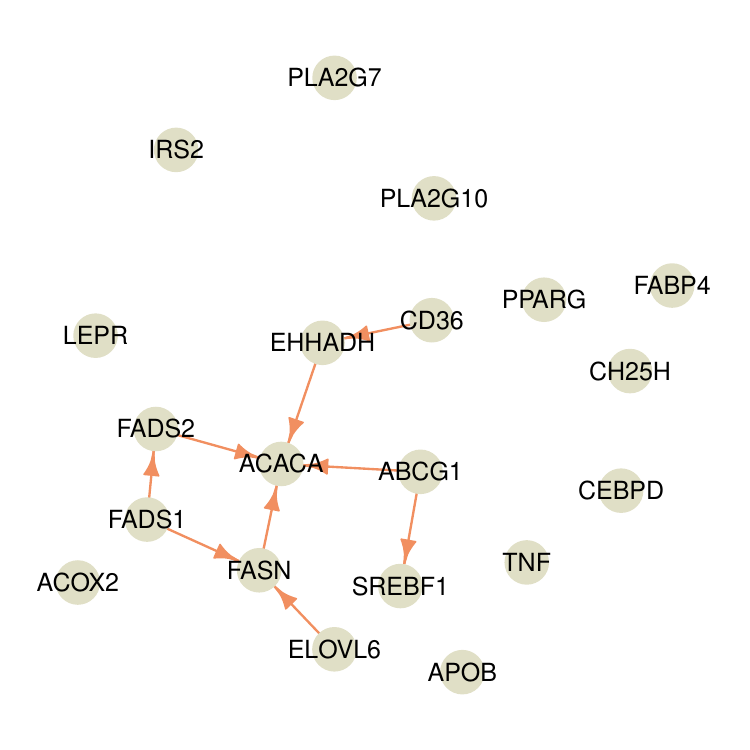}
        \caption{Obese mice}
        \label{fig:exp_collinear_cos}
    \end{subfigure}
    \caption{Estimated gene regulatory networks for the (a) lean and (b) obese mice.} 
    \label{fig:real}
\end{figure}

The estimated DAGs for lean and obese mice are shown in Figure \ref{fig:real}, where some of our discoveries are consistent with existing biological evidence. Compared with the obese group, the lean group exhibits substantially more outgoing edges from TNF, suggesting a more central regulatory role of TNF under the lean condition. Indeed, TNF encodes a multifunctional cytokine TNF-$\alpha$, which is known to contribute to metabolic dysregulation by influencing both adipose tissue function and its ability to store excess fuel \citep{hotamisligil1999role,cawthorn2008tnf}. 
Among the outgoing edges from TNF in the lean group, the edge TNF $\to$ CEBPD
 may support the previous finding that TNF-$\alpha$ regulates CEBPD in hepatocytes directly \citep{wu2011ccaat} or through a post-transcriptional mechanism \citep{yin1996tumor}. The edge TNF $\to$ APOB also differs between lean and obese groups, consistent with the evidence that TNF-$\alpha$ regulates APOB expression in rat hepatocytes, thereby affecting the release of very-low-density lipoprotein (VLDL), which is critical for hepatic lipid metabolism and systemic energy homeostasis \citep{bartolome2007upregulation,song2017crebh}. We further observe that the lean group includes paths from CH25H to PPARG, FABP4, and CEBPD, whereas these paths are absent in the obese group. Interestingly, the edges  CH25H $\to$ PPARG and CH25H $\to$ FABP4 seem to support the conclusion that CH25H regulates these target genes through its product 25-HC, which ultimately inhibits adipogenic differentiation and mitigates obesity \citep{wang2021cholesterol,moseti202025}. In addition, the edges associated with ACACA exhibit notable differences between the two groups, which aligns with the finding that ACACA plays a pivotal role in regulating fatty acid metabolism and biological development \citep{abu2005mutant,mao2006liver}. 

\section{Discussion}\label{sec:dis}

This paper develops a new framework for causal discovery and inference in DAGs under hidden confounding and invalid IVs by leveraging covariance heterogeneity. Unlike existing IV-based causal discovery methods, the proposed framework does not require the existence of any valid instrument. Our identification strategy proceeds in two stages: it first recursively recovers the causal order using covariance heterogeneity and then constructs surrogate IVs to identify the causal effects, leading to identification of the entire DAG. Building on these identification results, we develop CHARGE, a data-adaptive estimation procedure for learning both causal directions and causal effects.  We show that the proposed method can consistently recover the underlying causal relations and effects. Simulations and a real-world data application further illustrate its finite-sample performance and practical utility.

Our work may be extended or improved in several directions. First, while the proposed framework is built upon a linear SEM, its extension to more flexible models, particularly those capable of capturing nonlinear causal relations among primary variables, remains to be explored. Second, in addition to covariance heterogeneity, it would be worth considering whether higher-order moment conditions can provide additional information for identification. We plan to pursue these and other related issues in future research.

	\bibliographystyle{apalike}
\bibliography{references}

\newpage
\appendix
\begin{center}
{\Large\bf Supplement to ``Using Heteroscedasticity for Causal Discovery Under Hidden Confounding''}
\end{center}

This is the supplementary material for \cite{zou2026using}. Section \ref{sec:exp} includes an example to illustrate the terminology in DAGs; Section \ref{sec:proof_iden} provides the proofs for the proposed identification theory; Section \ref{sec:plm} outlines the extension of the identification to partially linear models; Section \ref{sec:proof_est} shows the proofs for estimation guarantees; Section \ref{sec:seq} presents a sequential estimation procedure for causal effect estimation; Section \ref{sec:high} investigates the extension of our method to high-dimensional settings.

\section{An example of DAG terminology}\label{sec:exp}

In this section, we provide an example to illustrate the terminology in DAGs introduced in Section \ref{sec:model}. Consider a DAG with $\cE_1$ as shown in Figure \ref{fig:true_dag}. The vertex $Y_2$ is a parent of $Y_3$ and an ancestor of $Y_4$, because there is an edge $Y_2\to Y_3$ and a directed path $Y_2\to Y_3\to Y_4$. The topological height for each node is $h_\cG(1)=h_\cG(2)=2$, $h_\cG(3)=1$, and $h_\cG(4)=0$. Thus, we have $H_\cG=2$, $\cV^+(-1)=\{1,2,3,4\}$, $\cV^+(0)=\{1,2,3\}$, $\cV^+(1)=\{1,2\}$, and $\cV^+(2)=\emptyset$. The subgraph $\cG(0)$ induced by $\bY_{\cV^+(0)}\cup \bX=\{Y_1,Y_2,Y_3\}\cup \bX$ corresponds to edge sets $\cE_1(0)=\{(1,3),(2,3)\}$. 

\section{Proofs for Section \ref{sec:iden}}\label{sec:proof_iden}

In this section, we provide the proofs for Section \ref{sec:iden}. Since the proof of Theorem \ref{thm:iden} depends on the technical foundations developed in Sections \ref{ssec:order} and \ref{ssec:eff}, we present its proof until these have been established. We begin with the following lemmas, which are useful for the subsequent proofs. Let $\pa_\cG(i)$, $\an_\cG(i)$, and $\de_\cG(i)$ denote the index set of parents, ancestors, and descendants of $Y_i$ in the edge set $\cE_1$ of $\cG$, respectively. Let $\bone(\cdot)$ denote the indicator function, $\an_\cG^+(i)=\an_\cG(i)\cup \{i\}$, and define $\de^+_\cG(i)=\de_\cG(i)\cup \{i\}$ and $\pa_\cG^+(i)=\pa_\cG(i)\cup \{i\}$ for $i=1,\dots, p$. If $Y_i\to Y_j$ in $\cE_1$ and there is no other directed path from $Y_i$ to $Y_j$, then $Y_i$ is called an unmediated parent of $Y_j$. 

\begin{lemma}\label{lem:Zrepre}
  For each $i=1,\dots,p$, there exists a set of coefficients $\{\alpha_{k i}\}_{k \in \an_\cG^+(i)}$ such that 
  \begin{equation}\label{eq:Zrepre}
      Z_i=\sum_{k \in \an_\cG^+(i)} \alpha_{k i}\ve_k, 
  \end{equation}
  where $\alpha_{ii}=1$ and $\alpha_{k i}=V_{k i}\ne 0$ if $Y_k$ is an unmediated parent of $Y_i$.
\end{lemma}

\begin{proof}
  The proof proceeds by mathematical induction. Due to the definition of $\bZ$, it follows that 
  \begin{equation}\label{eq:bZ}
      Z_i=P_\bX^\perp Y_i=\bV_i^T \bZ+\ve_i. 
  \end{equation}
  For $i\in \cV(H_\cG)$, since $\bV_i=\bzero$, \eqref{eq:bZ} implies that $Z_i=\ve_i$, which is consistent with Lemma \ref{lem:Zrepre}. Suppose that \eqref{eq:Zrepre} holds for all $i\in \cV^+(h)$ with some $0\le h<H_\cG$, we next show that it also holds for all $i\in \cV(h)$. According to \eqref{eq:bZ}, for any $i\in\cV(h)$, we have  
  \[
  \begin{aligned}
        Z_i&=\sum_{j\in \pa_\cG(i)}V_{ji}Z_j+\ve_i\\
        &\stackrel{\text{(i)}}{=}\sum_{j\in \pa_\cG(i)}V_{ji}\sum_{k\in \an_\cG^+(j)} \alpha_{k j}\ve_{k}+\ve_i\\ \label{eq:bZ_line2}
        &\stackrel{\text{(ii)}}{=}\sum_{k \in \an_\cG(i)}\sum_{j\in \pa_\cG(i) \cap\de_\cG^+(k)}V_{ji}\alpha_{kj}\ve_k+\ve_i\\
        &\stackrel{\text{(iii)}}{=}\sum_{k \in \an_\cG(i)} \alpha_{ki}\ve_k+\ve_i,
        \end{aligned}
\]
  where the equality (i) holds because $\pa_\cG(i)\subseteq \cV^+(h)$ and thus \eqref{eq:Zrepre} is supposed to hold, equality (ii) follows from $\an_\cG(i)=\cup_{j\in\pa_\cG(i)}\an_\cG^+(j)$, and the equality (iii) holds by taking $\alpha_{ki}=\sum_{j\in \pa_\cG(i) \cap \de_\cG^+(k)} V_{ji}\alpha_{kj}$. If $Y_k$ is an unmediated parent of $Y_i$, then $\alpha_{ki}=\sum_{j\in \pa_\cG(i) \cap \de_\cG^+(k)} V_{ji}\alpha_{kj}=V_{ki}\alpha_{kk}=V_{ki}$, where we have used $\alpha_{kk}=1$ since $k\in \cV^+(h)$. Therefore, \eqref{eq:Zrepre} holds for all $i\in \cV(h)$. By induction, \eqref{eq:Zrepre} holds for all $i=1,\dots, p$, which leads to Lemma \ref{lem:Zrepre}. 
\end{proof}

\begin{lemma}\label{lem:zeta_repre}
  Suppose that Assumption \ref{ass:uncorr} holds. Then for all $i\ne j$ and $h_\cG(i)\ge h_\cG(j)$, $\bzeta_{ij}$ can be expressed as  
  \[
    \bzeta_{ij}=\sum_{k\in \an_\cG(i)\cap \an_\cG(j)}\alpha_{ki}\alpha_{kj}\bmeta_k+\bone(i\in \an_\cG(j))\alpha_{ij} \bmeta_i. 
  \]
  For all $i=1,\dots, p$, $\bzeta_{ii}$ can be expressed as 
  \begin{equation}\label{eq:zeta_ii_repre}
    \bzeta_{ii}=\sum_{k \in \an_\cG(i)} \alpha_{ki}^2\bmeta_k+\bmeta_i.
  \end{equation}
  The coefficients $\alpha_{..}$ are as defined in Lemma \ref{lem:Zrepre}.
\end{lemma}

\begin{proof}
  Under Assumption \ref{ass:uncorr}, according to Lemma \ref{lem:Zrepre}, for all $i\ne j$ and $h_\cG(i)\ge h_\cG(j)$, 
  \[
  \begin{aligned}
        \bzeta_{ij}&=\cov(\bX,Z_iZ_j)\\
        &=\cov\Biggl\{\bX,\Biggl(\sum_{k\in \an_\cG(i)}\alpha_{ki}\ve_k+\ve_i\Biggr)\Biggl(\sum_{\ell\in \an_\cG(j)}\alpha_{\ell j}\ve_\ell+\ve_j\Biggr)\Biggr\}\\
        &=\sum_{k \in \an_\cG(i)\cap \an_\cG(j)}\alpha_{k i}\alpha_{k j}\bmeta_k+\bone(i\in \an_\cG(j))\alpha_{ij} \bmeta_{i},
    \end{aligned}
  \]
  where the last equality holds since $\cov(\bX,\ve_k\ve_\ell)=0$ for all $k\ne \ell$. For all $i=1,\dots,p$, we have 
    \[
    \begin{aligned}
            \bzeta_{ii}&=\cov(\bX,Z_i^2)\\
            &=\cov\Biggl[\bX,\Biggl\{\sum_{k \in \an_\cG^+(i)} \alpha_{ki}\ve_k\Biggr\}^2\Biggr]\\
            &=\cov\Biggl(\bX,\sum_{k \in \an_\cG(i)} \alpha_{ki}^2\ve_k^2+\ve_i^2\Biggr)\\
            &=\sum_{k \in \an_\cG(i)} \alpha_{ki}^2\bmeta_k+\bmeta_i.
    \end{aligned}
    \]
    Combining these results leads to Lemma \ref{lem:zeta_repre}. 
\end{proof}

\begin{lemma}\label{lem:eta_repre}
  Suppose that Assumption \ref{ass:uncorr} holds. Then for each $i=1,\dots,p$, there exists a set of coefficients $\{\beta_{ki}\}_{k\in \an_\cG^+(i)}$, such that 
  \begin{equation}\label{eq:eta_repre}
    \bmeta_i=\sum_{k\in \an_\cG^+(i)}\beta_{ki}\bzeta_{kk},
  \end{equation}
  where $\beta_{ii}=1$. 
\end{lemma}

\begin{proof}
  The proof proceeds by mathematical induction. For $i\in \cV(H_\cG)$, since $\an_\cG(i)=\emptyset$, it follows from \eqref{eq:zeta_ii_repre} that $\bmeta_i=\bzeta_{ii}$, which is consistent with Lemma \ref{lem:eta_repre}. Suppose that \eqref{eq:eta_repre} holds for all $i\in \cV^+(h)$ with some $0\le h<H_\cG$, we next show that it also holds for all $i\in \cV(h)$. According to \eqref{eq:zeta_ii_repre}, for any $i\in\cV(h)$, we have 
  \[
  \begin{aligned}
            \bmeta_i&=\bzeta_{ii}-\sum_{k\in \an_\cG(i)}\alpha_{ki}^2\bmeta_k\\
            &=\bzeta_{ii}-\sum_{k\in \an_\cG(i)}\alpha_{ki}^2 \Biggl(\bzeta_{kk}+\sum_{\ell\in \an_\cG(k)}\beta_{\ell k}\bzeta_{\ell\ell}\Biggr)\\
            &=\sum_{k\in \an_\cG(i)}\beta_{ki} \bzeta_{kk}+\bzeta_{ii},
        \end{aligned}
  \]
  where the second equality holds since $\an_\cG(i)\subseteq \cV^+(h)$ thereby \eqref{eq:eta_repre} is supposed to hold, and the last equality holds by taking $\beta_{ki}=-\alpha_{ki}^2-\sum_{\ell:\ell\in \an_\cG(i)\cap \de_\cG(k)}\beta_{k \ell}\alpha_{\ell i}^2$ for $k\in \an_\cG(i)$. Therefore, \eqref{eq:eta_repre} holds for all $i\in \cV(h)$. By induction, \eqref{eq:eta_repre} holds for all $i=1,\dots, p$, which leads to Lemma \ref{lem:eta_repre}. 
\end{proof}

\subsection{Proof of Proposition \ref{prop:test}}

\begin{proof}
Suppose that Assumption \ref{ass:uncorr} holds.  Lemma \ref{lem:zeta_repre} implies that $\bzeta_{ii}\in \mathrm{span}(\{\bmeta_k:k=1,\dots,p\})$ for all $i=1,\dots, p$. Hence, $\mathrm{span}(\{\bzeta_{kk}:k=1,\dots,p\})\subseteq \mathrm{span}(\{\bmeta_k:k=1,\dots,p\})$. Lemma \ref{lem:eta_repre} in turn gives $\bmeta_i\in \mathrm{span}(\{\bzeta_{kk}:k=1,\dots,p\})$ for all $i=1,\dots, p$, which implies $\mathrm{span}(\{\bmeta_k:k=1,\dots,p\}) \subseteq \mathrm{span}(\{\bzeta_{kk}:k=1,\dots,p\})$. Combining these two parts leads to Proposition \ref{prop:test}. 
\end{proof}

\subsection{Proof of Proposition \ref{prop:leaf}}\label{ssec:prf_repre}

We begin with the following lemma, which is useful for proving Proposition \ref{prop:leaf}. 

\begin{lemma}\label{lem:repre}
    Suppose that Assumption \ref{ass:uncorr} holds. Then for any $i\ne j$ and $i\notin \an_\cG(j)$, $\bzeta_{ij}\in \mathrm{span}(\{\bzeta_{kk}:k\ne i\})$. If suppose further that Assumption \ref{ass:hete} holds, then for any unmediated parent $Y_i$ of $Y_j$, $\bzeta_{ij}\notin \mathrm{span}(\{\bzeta_{kk}:k\ne i\})$. 
\end{lemma}

\begin{proof}[Proof of Lemma \ref{lem:repre}]
We begin by proving the first part of Lemma \ref{lem:repre}. That is, under Assumption \ref{ass:uncorr}, for any $i\ne j$ and $i\notin \an_\cG(j)$, we have $\bzeta_{ij}\in \mathrm{span}(\{\bzeta_{kk}:k\ne i\})$. If $i\ne j$ and $h_\cG(i)\ge h_\cG(j)$, then by Lemmas \ref{lem:zeta_repre} and \ref{lem:eta_repre}, 
\[
\begin{aligned}
\bzeta_{ij}&=\sum_{k\in \an_\cG(i)\cap \an_\cG(j)}\alpha_{ki}\alpha_{kj}\bmeta_k\\
&=\sum_{k\in \an_\cG(i)\cap \an_\cG(j)}\alpha_{ki}\alpha_{kj}\sum_{\ell\in \an_\cG^+(k)}\beta_{\ell k} \bzeta_{\ell \ell}. 
\end{aligned}
\]
Since for any $k\in \an_\cG(i)\cap \an_\cG(j)$ and $\ell\in \an_\cG^+(k)$, we have $\ell\in \an_\cG(i)\cap \an_\cG(j)$ and thus $\ell \ne i$. Therefore, the first part of Lemma \ref{lem:repre} holds for any $h_\cG(i)\ge h_\cG(j)$. For $i\ne j$ and $h_\cG(j)\ge h_\cG(i)$, using Lemmas \ref{lem:zeta_repre} and \ref{lem:eta_repre} yields 
\[
\begin{aligned}
\bzeta_{ji}&=\sum_{k\in \an_\cG(i)\cap \an_\cG(j)}\alpha_{ki}\alpha_{kj}\bmeta_k+\bone(j\in \an_\cG(i))\alpha_{ji}\bmeta_j\\
&=\sum_{k\in \an_\cG(i)\cap \an_\cG(j)}\alpha_{ki}\alpha_{kj}\sum_{\ell\in \an^+_\cG(k)}\beta_{\ell k} \bzeta_{\ell \ell}+\bone(j\in \an_\cG(i))\alpha_{ji}\sum_{k\in \an_\cG^+(j)}\beta_{kj}\bzeta_{kk}. 
\end{aligned}
\]
Note that for any $k\in \an_\cG(i)\cap \an_\cG(j)$ and $\ell\in \an^+_\cG(k)$, we have $\ell\in \an_\cG(i)\cap \an_\cG(j)$ and thus $\ell \ne i$. Also, for any $k\in \an_\cG^+(j)$, we have $k\ne i$. The proof of the first part of Lemma \ref{lem:repre} is completed by combining these two parts. 

For the second part of Lemma \ref{lem:repre}, consider the case where there exists some $Y_i$ that is an unmediated parent of $Y_j$, but $\bzeta_{ij}\in \mathrm{span}(\{\bzeta_{kk}:k\ne i\})$. Using Lemmas \ref{lem:zeta_repre} and \ref{lem:eta_repre} for this $\bzeta_{ij}$ leads to 
\[
  \begin{aligned}
        \bzeta_{ij}=&\sum_{k\in \an_\cG(i)} \alpha_{ki}\alpha_{kj}\bmeta_k+\alpha_{ij}\bmeta_i\\
    =& \sum_{k\in \an_\cG(i)} \alpha_{ki}\alpha_{kj}\sum_{\ell\in \an_\cG^+(k)}\beta_{\ell k}\bzeta_{\ell \ell}+\alpha_{ij}\bmeta_i,  
\end{aligned}
\]
where $\alpha_{ij}=V_{ij}\ne 0$. Note that for any $k\in \an_\cG(i)$ and $\ell\in\an^+_\cG(k)$, we have $\ell\ne i$. From $\bzeta_{ij}\in \mathrm{span}(\{\bzeta_{kk}:k\ne i\})$, it follows that $\bmeta_i\in \mathrm{span}(\{\bzeta_{kk}:k\ne i\})$. In contrast, by Lemma \ref{lem:eta_repre}, the coefficient of $\bzeta_{ii}$ in the linear representation of $\bmeta_i$ is $1$, as shown in \eqref{eq:eta_repre}. This implies that the linear representation of $\bmeta_i$ in terms of $\{\bzeta_{kk}\}_{k=1,\dots,p}$ is not unique, which contradicts Proposition \ref{prop:test} that $\rank(\{\bzeta_{kk}:k=1,\dots,p\})=p$ under Assumption \ref{ass:hete}. Therefore, we complete the proof of Lemma \ref{lem:repre}. 
\end{proof}

\begin{proof}[Proof of Proposition \ref{prop:leaf}]
  For any leaf node $Y_i$ of $\cG$ and $j\ne i$, $i\notin \an_\cG(j)$. Therefore, for any $i\in\leaf(\cG)$ and $j\ne i$, we have $\bzeta_{ij}\in \mathrm{span}(\{\bzeta_{kk}:k\ne i\})$ according to Lemma \ref{lem:repre}. This shows that $\leaf(\cG)\subseteq\{i:\bzeta_{ij}\in \mathrm{span}(\{\bzeta_{kk}:k\ne i\})\ for\ all\ j\ne i\}$. Conversely, for any $i\notin \leaf(\cG)$, there exists some $Y_j$ whose unmediated parents include $Y_i$. Lemma \ref{lem:repre} shows that $\bzeta_{ij}\notin \mathrm{span}(\{\bzeta_{kk}:k\ne i\})$, and thus $\{i:\bzeta_{ij}\in \mathrm{span}(\{\bzeta_{kk}:k\ne i\})\ for\ all\ j\ne i\}\subseteq \leaf(\cG)$. Combining these pieces yields Proposition \ref{prop:leaf}. 
\end{proof}

\subsection{Proof of Proposition \ref{prop:order}}

\begin{proof}
The proof proceeds by mathematical induction. Using Proposition \ref{prop:leaf}, it follows that $\leaf(\cG)$, which is exactly $\cV(0)$, is identifiable under Assumptions \ref{ass:uncorr} and \ref{ass:hete}. Suppose that $\cV(0),\dots, \cV(h)$ is identifiable for some $h=0,\dots, H_\cG-1$. It follows that the set $\cV^+(h)=\bigl(\cup_{\ell=0}^{h}\cV(\ell)\bigr)^c$ is identifiable. We then prove that $\cV(h+1)$ is identifiable. Since $\cV(h+1)=\leaf(\cG(h))$, where $\cG(h)$ is the subgraph induced by the identified vertices $\bY_{\cV^+(h)}\cup \bX$, it suffices to prove the identifiability of $\leaf(\cG(h))$. Note that for all $i,j\in \cV^+(h)$ and $i\ne j$, we have $\cov(\bX,\ve_i\ve_j)=\bzero$ under Assumption \ref{ass:uncorr}. Also, $\rank(\{\bmeta_i:i\in \cV^+(h)\})=|\cV^+(h)|$ under Assumption \ref{ass:hete}. Therefore, Assumptions \ref{ass:uncorr} and \ref{ass:hete} remain valid in the subgraph $\cG(h)$. Using Proposition \ref{prop:leaf}, we can then identify the leaves of $\cG(h)$ as 
\[
\leaf(\cG(h))=\{i\in \cV^+(h):\bzeta_{ij}\in \mathrm{span}(\{\bzeta_{kk}:k\in \cV^+(h)\ and\ k\ne i\})\ for\ all\ j\in \cV^+(h)\ and\ j\ne i\}.
\]
By induction, the leaf nodes of every subgraph $\cG(h)$ for $h=-1,\dots,H_\cG-1$ are identifiable. Since $\cV(h)=\leaf(\cG(h-1))$ for $h=0,\dots, H_\cG$, we conclude that $\{\cV(h)\}_{h=0,\dots,H_\cG}$ is identifiable, which leads to Proposition \ref{prop:order}. 
\end{proof}

\subsection{Proof of Proposition \ref{prop:rank}}\label{ssec:prf_prop_rank}

\begin{proof}
  Without loss of generality, we rearrange the indices of the primary variables such that $\cV(H_\cG)=\{1,2,\dots,i_{H_\cG}\}$, $\cV(H_\cG-1)=\{i_{H_\cG}+1,\dots, i_{H_\cG-1}\}$, ..., and $\cV(0)=\{i_1+1,\dots, i_0\}$ where $i_0=p$. Under this condition, for any $i>j$, we have $h_\cG(i)\le h_\cG(j)$ and thus $i\notin \an_\cG(j)$. To compute $\E\{\bT(h)\bZ^T_{\cV^+(h)}\}=\cov(\bT(h),\bZ_{\cV^+(h)})$ for a given $h=0,\dots, H_\cG-1$, we calculate each of their elements separately. Let $\bT_i(h)=(\bX-\E(\bX))\ve_i$ denote the $i$th instrument for $i\in\cV^+(h)$. Then, for any $i,j\in \cV^+(h)$ and $i> j$, we have 
  \[
  \begin{aligned}
  \cov(\bT_i(h),Z_j)&=\cov\bigl((\bX-\E(\bX))\ve_i,Z_j\bigr)\\
  &=\cov\Biggl((\bX-\E(\bX))\ve_i,\sum_{k\in \an_\cG^+(j)}\alpha_{kj}\ve_k\Biggr)\\
  &=\sum_{k\in \an_\cG^+(j)}\alpha_{kj}\cov\bigl((\bX-\E(\bX))\ve_i,\ve_k\bigr)\\
  &=\sum_{k\in \an_\cG^+(j)}\alpha_{kj}\E\bigl((\bX-\E(\bX))\ve_i\ve_k\bigr)\\
  &=\sum_{k\in \an_\cG^+(j)}\alpha_{kj}\cov(\bX,\ve_i\ve_k)\\
  &=\bzero,
  \end{aligned}
  \]
  where the second line holds by implementing Lemma \ref{lem:Zrepre} and the last line holds due to $i\notin \an^+_\cG(j)$ and Assumption \ref{ass:uncorr}. Similarly, for any $i\in \cV^+(h)$, using Lemma \ref{lem:Zrepre} and Assumption \ref{ass:uncorr} yields
  \[
    \begin{aligned}
  \cov(\bT_i(h),Z_i)&=\cov\bigl((\bX-\E(\bX))\ve_i,Z_i\bigr)\\
  &=\cov\Biggl((\bX-\E(\bX))\ve_i,\sum_{k\in \an_\cG^+(i)}\alpha_{ki}\ve_k\Biggr)\\
  &=\sum_{k\in \an_\cG^+(i)}\alpha_{ki}\cov\bigl((\bX-\E(\bX))\ve_i,\ve_k\bigr)\\
  &=\sum_{k\in \an_\cG^+(i)}\alpha_{ki}\E\bigl((\bX-\E(\bX))\ve_i\ve_k\bigr)\\
  &=\sum_{k\in \an_\cG(i)}\alpha_{ki}\cov(\bX,\ve_i\ve_k)+\cov(\bX,\ve_i^2)\\
  &=\bmeta_i.
  \end{aligned}
  \]
  Therefore, the matrix $\E\{\bT(h)\bZ_{\cV^+(h)}^T\}=\cov(\bT(h),\bZ_{\cV^+(h)})$ takes the form 
  \begin{equation}\label{eq:tria}
  \E\{\bT(h)\bZ_{\cV^+(h)}^T\}=\begin{bmatrix}
 \bmeta_1 & * & * & *\\
 0 & \bmeta_2 & * & *\\
 \dots& \dots & \dots &\dots\\
 0 & 0 & \dots &\bmeta_{i_{h+1}}
\end{bmatrix},
\end{equation}
which is a block-wise upper triangular matrix. Since the covariances $\bmeta_i$'s are heterogeneous due to Assumption \ref{ass:hete}, we have $\bmeta_i\ne \bzero$ for $i=1,\dots,p$. Thus, Proposition \ref{prop:rank} holds. 
\end{proof}

\begin{remark}
As shown in \eqref{eq:tria}, the matrix $\E\{\bT(h)\bZ_{\cV^+(h)}^T\}$ can be rearranged into a block upper triangular matrix with diagonal elements $\bmeta_i$ for $i\in \cV^+(h)$. The rank condition in Proposition \ref{prop:rank} then reduces to checking whether any $\bmeta_i$ equals zero for $i=1,\dots, p$.  Therefore, the strength of the reconstructed instruments after controlling $\bX$ is proportional to the heterogeneity of the covariances $\bmeta_i$'s, which corresponds to the degree of heteroscedasticity in Assumption \ref{ass:hete}. 
\end{remark}

\subsection{Proof of Proposition \ref{prop:eff}}\label{ssec:prf_eff}

\begin{proof}
  The proof proceeds by mathematical induction. Note that the topological height is identifiable by Proposition \ref{prop:order}. Since $\var(\bX)$ is invertible, and $\bV_i=\bzero$ for $i\in \cV(H_\cG)$, we can then identify $\bW_i$ by $\cov(\bX,Y_i-\bW_i^T\bX)=0$ as $\bW_i=\var(\bX)^{-1}\cov(\bX, Y_i)$. 

Suppose that $\bV_i$ and $\bW_i$ are identifiable for all $i\in \cV^+(h)$ with some $h=0,\dots, H_\cG-1$. We then show that $\bV_j$ and $\bW_j$ for all $j\in \cV(h)$ are identifiable by \eqref{eq:iden_eff}. Since $\bV_i$ and $\bW_i$ are identifiable for all $i\in \cV^+(h)$, the moment conditions in \eqref{eq:iden_eff} for $\bV_j$ and $\bW_j$ with $j\in \cV(h)$ become
 \begin{equation}\label{eq:prf_eff_i}
 \begin{aligned}
  &\E\Biggl\{\bX\Biggl(Y_j-\sum_{i\in \cV^+(h)}V_{ij}Y_i-\bW_j^T\bX\Biggr)\Biggr\}=\bzero,\\
      & \E\Biggl\{\bT(h)\Biggl(Y_j-\sum_{i\in \cV^+(h)}V_{ij}Y_i-\bW_j^T\bX\Biggr)\Biggr\}=\bzero, \\
      & subject\ to\ V_{kj}=0,\ for\ k\notin \cV^+(h),
 \end{aligned}
\end{equation}
where $\bT(h)$ is constructed without using unidentified parameters as 
\[
  \bT(h):=\bigl(\{\bX-\E(\bX)\}\ve_i\bigr)_{i\in \cV^+(h)}=\bigl(\{\bX-\E(\bX)\}(Y_i-\bV_i^T\bY-\bW_i^T\bX)\bigr)_{i\in \cV^+(h)}.
\]
Using Proposition \ref{prop:rank} and the fact that $\var(\bX)$ is invertible, it follows that the true value of $(\bV_j,\bW_j)$ is the unique solution to \eqref{eq:prf_eff_i} given $\bV_i$ and $\bW_i$ for $i\in \cV^+(h)$. Therefore, $\bV_j$ and $\bW_j$ are identifiable by \eqref{eq:iden_eff} for all $j\in \cV(h)$. By induction, $(\bV,\bW)$ is the unique solution to \eqref{eq:iden_eff}, and is therefore identifiable. Consequently, the edge sets are identifiable as $\cE_1= \{(i,j):V_{ij}\ne 0\}$ and $\cE_2= \{(\ell,j):W_{\ell j}\ne 0\}$ due to the definition. 
\end{proof}

\subsection{Proof of Theorem \ref{thm:iden}}

\begin{proof}
  Under Assumptions \ref{ass:uncorr} and \ref{ass:hete}, the topological order $h_\cG(i)$ for $i=1,\dots, p$ is identifiable using Proposition \ref{prop:order}. Based on the identified topological order, we can identify the causal effects $(\bV, \bW)$ by the unique solution of \eqref{eq:iden_eff} using Proposition \ref{prop:eff}. It then follows that edges $(\cE_1,\cE_2)$ in $\cG$ can be identified as $\cE_1= \{(i,j):V_{ij}\ne 0\}$ and $\cE_2= \{(\ell,j):W_{\ell j}\ne 0\}$ by definition. Therefore, the coefficients $(\bV,\bW)$ in model \eqref{eq:SEM} and the corresponding edge sets $(\cE_1,\cE_2)$ are identifiable under Assumptions \ref{ass:uncorr} and \ref{ass:hete}. 
\end{proof}

\section{Extension of identification theory to partially linear models}\label{sec:plm}

In this section, we present the extension of our identification theory to partially linear models (PLMs). Consider a DAG $\cG=(\cV;\cE_1,\cE_2)$ paired with the following PLM 
\begin{equation}\label{eq:plm}
    Y_i=\bV_i^T \bY+f_i(\bX)+\ve_i,\ i=1,\dots, p,
\end{equation}
where $f_i$ is any arbitrary measurable function and the error $\ve_i$ satisfies $\E(\ve_i\mid \bX)=0$ for $i=1,\dots,p$.  For the identification of $\cE_2$, we additionally require the causal minimality condition \citep{pearl2009causality,li2024nonlinear}. Let the set $\inter_\cG(i):=\{\ell:(\ell,i)\in \cE_2\}$ denote the indices of auxiliary variables that have direct causal effects on $Y_i$, and $\ARG(f_i)$ denote the minimal argument set $s\subseteq \inter_\cG(i)$ such that the value of $f_i$ only depends on $\bX_s$. The causal minimality requires $\inter_\cG(i)=\ARG(f_i)$ for $i=1,\dots, p$. This condition is a form of causal faithfulness \citep{spirtes2001causation} such that the problem of nonlinear causal discovery is well-defined \citep{li2024nonlinear}. Under the causal minimality condition, we define the directed edges by $Y_i\to Y_j$ whenever $V_{ij}\ne 0$ and $X_\ell \to Y_j$ whenever $\ell\in\ARG(f_j)$. This gives edge sets $\cE_1=\{(i,j):Y_i\to Y_j\}$ and $\cE_2=\{(\ell,j):X_\ell\to Y_j\}$. Define $\bmeta_i=\cov(\bX,\ve_i^2)$ as in the linear case. Let $\pa_\cG(i)$, $\an_\cG(i)$, and $\de_\cG(i)$ denote the index set of parents, ancestors, and descendants of $Y_i$ in the edge set $\cE_1$ of $\cG$, respectively. The following result holds under the causal minimality condition.

\begin{theorem}
   Under Assumptions \ref{ass:uncorr} and \ref{ass:hete}, the coefficient $\bV$ in \eqref{eq:plm} and the edge sets $(\cE_1,\cE_2)$ are identifiable. 
\end{theorem}

\begin{proof}
   Define $\btZ=\bY-\E(\bY\mid \bX)$ and with a slight abuse of notation, redefine $\bzeta_{ij}:=\cov(\bX,\tZ_i\tZ_j)$ for $i,j=1\dots,p$. Since $\E(\ve_i\mid \bX)=0$, we have $\btZ=\bV^T\btZ+\bve$. Therefore, following the same strategy in Lemma \ref{lem:Zrepre}, we have
    \[
    \tZ_i=\sum_{\ell \in \an_\cG(i)\cup \{i\}} \talpha_{\ell i}\ve_\ell,
    \]
    with $\talpha_{ii}=1$ and $\talpha_{\ell i}=V_{\ell i}\ne 0$ if $Y_\ell$ is an unmediated parent of $Y_i$. Similarly, since $\btZ=\bV^T\btZ+\bve$, we can obtain the same results in Lemmas \ref{lem:zeta_repre} and \ref{lem:eta_repre} under Assumption \ref{ass:uncorr}. Therefore, we can still use the results in Proposition \ref{prop:leaf} under Assumptions \ref{ass:uncorr} and \ref{ass:hete}, and identify the leaves of $\cG$ by 
    \[
    \leaf(\cG)=\{i:\bzeta_{ij}\in \mathrm{span}(\{\bzeta_{kk}:k\ne i\})\ for\ all\ j\ne i\}.   
    \]
    By using the same argument in the proof of Proposition \ref{prop:order}, it can be shown that the causal order of $\cG$ is identifiable. 
    
    It then suffices to establish the identifiability of $\bV$ and $f_i(\cdot)$. Since $\E(\bve\mid \bX)=\bzero$, $f_i(\bX)=\E(Y_i-\bV_i^T \bY\mid \bX)$ is identifiable once $\bV_i$ is identified. We then focus on the identification of $\bV$ after controlling $\bX$. Given $j\in \cV(h)$ for some $h=0,\dots, H_{\cG-1}$, Assumption \ref{ass:uncorr} implies that for all $i\in \cV^+(h)$, 
\[
  \begin{aligned}
    \cov(\bX,\ve_i\ve_j)&=\E\{(\bX-\E(\bX))\ve_i\ve_j\}\\
    &=\E\Bigl\{(\bX-\E(\bX))\ve_i\Bigl(\tZ_j-\sum\nolimits_{k\in \cV^+(h)}V_{kj}\tZ_k\Bigr)\Bigr\}\\
&=\bzero,
  \end{aligned}
\]
where the second line holds due to $\tZ_j=\bV_j^T\btZ+\ve_j$. Therefore, we can still construct the instruments in line with the linear SEM \eqref{eq:SEM} as
\[
\bT(h)=\bigl((\bX-\E(\bX))\ve_i\bigr)_{i\in\cV^+(h)}\in \R^{q|\cV^+(h)|},\ h=0,\dots, H_\cG-1.
\]
We now show that $\bT(h)$ satisfies the rank condition analogous to Proposition \ref{prop:rank} under \eqref{eq:plm}. Following the same arrangement in Section \ref{ssec:prf_prop_rank}, we assume that $\cV(H_\cG)=\{1,2,\dots,i_{H_\cG}\}$, $\cV(H_\cG-1)=\{i_{H_\cG}+1,\dots, i_{H_\cG-1}\}$, ..., and $\cV(0)=\{i_1+1,\dots, i_0\}$ where $i_0=p$. To compute $\E\{\bT(h)\btZ^T_{\cV^+(h)}\}=\cov(\bT(h),\btZ_{\cV^+(h)})$ for a given $h=0,\dots, H_\cG-1$, we calculate each of their elements separately. Let $\bT_i(h)=(\bX-\E(\bX))\ve_i$ denote its $i$th instrument. Then, for any $i,j\in \cV^+(h)$ and $i> j$, we have 
  \[
  \begin{aligned}
  \cov(\bT_i(h),\tZ_j)&=\cov\bigl((\bX-\E(\bX))\ve_i,\tZ_j\bigr)\\
  &=\cov\Biggl((\bX-\E(\bX))\ve_i,\sum_{k\in \an_\cG^+(j)}\talpha_{kj}\ve_k\Biggr)\\
  &=\sum_{k\in \an_\cG^+(j)}\talpha_{kj}\cov\bigl((\bX-\E(\bX))\ve_i,\ve_k\bigr)\\
  &=\sum_{k\in \an_\cG^+(j)}\talpha_{kj}\E\bigl((\bX-\E(\bX))\ve_i\ve_k\bigr)\\
  &=\sum_{k\in \an_\cG^+(j)}\talpha_{kj}\cov(\bX,\ve_i\ve_k)\\
  &=\bzero,
  \end{aligned}
  \]
  where the last line holds due to $i\notin \an^+_\cG(j)$ and Assumption \ref{ass:uncorr}. Similarly, for any $i\in \cV^+(h)$,
  \[
    \begin{aligned}
  \cov(\bT_i(h),\tZ_i)&=\cov\bigl((\bX-\E(\bX))\ve_i,\tZ_i\bigr)\\
  &=\cov\Biggl((\bX-\E(\bX))\ve_i,\sum_{k\in \an_\cG^+(i)}\talpha_{ki}\ve_k\Biggr)\\
  &=\sum_{k\in \an_\cG^+(i)}\talpha_{ki}\cov\bigl((\bX-\E(\bX))\ve_i,\ve_k\bigr)\\
  &=\sum_{k\in \an_\cG^+(i)}\talpha_{ki}\E\bigl((\bX-\E(\bX))\ve_i\ve_k\bigr)\\
  &=\sum_{k\in \an_\cG(i)}\talpha_{ki}\cov(\bX,\ve_i\ve_k)+\cov(\bX,\ve_i^2)\\
  &=\bmeta_i.
  \end{aligned}
  \]
Therefore, the matrix $\E\{\bT(h)\btZ_{\cV^+(h)}^T\}=\cov(\bT(h),\btZ_{\cV^+(h)})$ takes the form 
  \[
  \E\{\bT(h)\btZ_{\cV^+(h)}^T\}=\begin{bmatrix}
 \bmeta_1 & * & * & *\\
 0 & \bmeta_2 & * & *\\
 \dots& \dots & \dots &\dots\\
 0 & 0 & \dots &\bmeta_{i_{h+1}}
\end{bmatrix}.
\]
Under Assumption \ref{ass:hete}, $\bT(h)$ satisfies $\rank\bigl(\E(\bT(h)\btZ_{\cV^+(h)}^T)\bigr)=|\cV^+(h)|$ and thus is valid. Following the same induction strategy in Section \ref{ssec:prf_eff}, we can identify $\bV$ and $f_i(\cdot)$ for $i=1,\dots, p$ by the unique solution to
\[
  \begin{aligned}
      & \E\{Y_i-\bV_i^T\bY-f_i(\bX)\mid \bX\}=\bzero,\\
      & \E[\{\bX-\E(\bX)\}\{Y_i-\bV_i^T\bY-f_i(\bX)\}\{Y_j-\bV_j^T\bY-f_j(\bX)\}]=\bzero,\ for\ i\ne j, \\
      & subject\ to\ V_{ij}=0,\ for\ h_\cG(i)\le h_\cG(j). 
  \end{aligned}
\]
Consequently, the edge sets are identifiable as $\cE_1= \{(i,j):V_{ij}\ne 0\}$ and $\cE_2= \{(\ell,j):\ell\in \ARG(f_j)\ and\ j=1,\dots,p\}$. 
\end{proof}

\section{Proofs for Section \ref{sec:est}}\label{sec:proof_est}

In this section, we provide the proofs for Section \ref{sec:est}. For the DAG $\cG=(\cV;\cE_1,\cE_2)$, let $\pa_\cG(i)$, $\an_\cG(i)$, and $\de_\cG(i)$ denote the index set of parents, ancestors, and descendants of $Y_i$ in the edge set $\cE_1$, respectively. For a matrix $\bA$, let $\sigma_{min}(\bA)$ denote the minimal singular value of $\bA$. For two positive sequences $a_n$ and $b_n$, we use $a_n\asymp b_n$ to denote $C\le a_n/b_n\le C^\prime$ for some constants $C,C^\prime>0$.  

\subsection{Proof of Proposition \ref{prop:full}}
\begin{proof}
Since $\bzeta_{ij}=\bzeta_{ji}$, we assume $h_\cG(i)\ge h_\cG(j)$ without loss of generality. Then under Assumption \ref{ass:uncorr}, it follows from Lemmas~\ref{lem:zeta_repre} and \ref{lem:eta_repre} that, for all $i\ne j$,
\begin{equation}\label{eq:zeta_ij_repre}
\begin{aligned}
\bzeta_{ij}&=\sum_{k\in \an_\cG(i)\cap \an_\cG(j)}\alpha_{ki}\alpha_{kj}\bmeta_k+\bone(i\in \an_\cG(j))\alpha_{ij}\bmeta_i\\
&=\sum_{k\in \an_\cG(i)\cap \an_\cG(j)}\alpha_{ki}\alpha_{kj}\sum_{\ell\in \an^+_\cG(k)}\beta_{\ell k} \bzeta_{\ell \ell}+\bone(i\in \an_\cG(j))\alpha_{ij}\sum_{k\in \an_\cG^+(i)}\beta_{ki}\bzeta_{kk}. 
\end{aligned}
\end{equation}
Therefore, $\bzeta_{ij}\in \mathrm{span}(\{\bzeta_{kk}:k=1,\dots,p\})$ for all $i\ne j$ and $i,j=1,\dots,p$. 
\end{proof}

\subsection{Proof of Lemma \ref{lem:con_order}}

\begin{proof}

It is equivalent to showing that
\[
\lim_{n\to\infty}\P\bigl\{\hat{\cV}(h)=\cV(h),\ h=0,\ldots,H_{\cG}\bigr\}=1.
\]
We prove this result by mathematical induction. Define, for$i\neq j$,
\[
\hat F_{ij}(-1):=\frac{\|P_{\hat{\bR}_i(-1)}^\perp\hat{\bzeta}_{ij}\|_2^2-\|P_{\hat{\bPhi}_{\bzeta}(-1)}^\perp\hat{\bzeta}_{ij}\|_2^2}{\|P_{\hat{\bPhi}_{\bzeta}(-1)}^\perp\hat{\bzeta}_{ij}\|_2^2+C_0/n
},
\]
where $C_0>0$ is any fixed constant, so that
\[
\hat F_i(-1)=\max_{j\neq i}\hat F_{ij}(-1).
\]
We first prove that
\[
\lim_{n\to\infty}\P\{\hat{\cV}(0)=\cV(0)\}=1.
\]
It suffices to show that for every $i\in\leaf(\cG)$,
\[
\lim_{n\to\infty}\P\left\{\hat F_i(-1)<\sqrt{2.01\log n}\right\}=1,
\]
and for every $i\notin\leaf(\cG)$,
\[
\lim_{n\to\infty}\P\left\{\hat F_i(-1)\geq\sqrt{2.01\log n}\right\}=1.
\]

Consider the full-model residual $P_{\hat{\bPhi}_{\bzeta}(-1)}^\perp\hat{\bzeta}_{ij}$. Under the low-dimensional setting, $p$ and $q$ are fixed. By standard least-squares theory and the central limit theorem, for all $i,j=1,\ldots,p$ we have
\[
\|\hat{\bzeta}_{ij}-\bzeta_{ij}\|_2=O_p(n^{-1/2}).
\]
Consequently,
\[
\|\hat{\bPhi}_{\bzeta}(-1)-\bPhi_{\bzeta}(-1)\|_2\leq\|\hat{\bPhi}_{\bzeta}(-1)-\bPhi_{\bzeta}(-1)\|_F=O_p(n^{-1/2}).
\]

Under Assumption \ref{ass:hete}, $\bPhi_{\bzeta}(-1)$ has full column rank and its minimum singular value is bounded away from zero. Therefore, with probability tending to one, $\hat{\bPhi}_{\bzeta}(-1)$ also has full column rank and $\sigma_{\min}\{\hat{\bPhi}_{\bzeta}(-1)\}\geq\sigma_{\min}\{\bPhi_{\bzeta}(-1)\}/2>0$. It follows from the continuity of the projection matrix on the set of
full-column-rank matrices that
\[
\|P_{\hat{\bPhi}_{\bzeta}(-1)}-P_{\bPhi_{\bzeta}(-1)}\|_2=O_p(n^{-1/2}).
\]

By Proposition \ref{prop:full}, for every $i\neq j$, $P_{\bPhi_{\bzeta}(-1)}^\perp\bzeta_{ij}=0$. Hence,
\begin{align*}
\|P_{\hat{\bPhi}_{\bzeta}(-1)}^\perp\hat{\bzeta}_{ij}\|_2&=\|P_{\hat{\bPhi}_{\bzeta}(-1)}^\perp\hat{\bzeta}_{ij}-P_{\bPhi_{\bzeta}(-1)}^\perp\bzeta_{ij}\|_2\\
&\leq\|P_{\hat{\bPhi}_{\bzeta}(-1)}^\perp(\hat{\bzeta}_{ij}-\bzeta_{ij})\|_2+\|(P_{\hat{\bPhi}_{\bzeta}(-1)}^\perp-P_{\bPhi_{\bzeta}(-1)}^\perp)\bzeta_{ij}\|_2\\
&\leq\|\hat{\bzeta}_{ij}-\bzeta_{ij}\|_2+\|P_{\hat{\bPhi}_{\bzeta}(-1)}-P_{\bPhi_{\bzeta}(-1)}\|_2\|\bzeta_{ij}\|_2\\
&=O_p(n^{-1/2}).
\end{align*}
Therefore,
\begin{equation}
\label{eq:full_residual_rate}
\|P_{\hat{\bPhi}_{\bzeta}(-1)}^\perp\hat{\bzeta}_{ij}\|_2^2=O_p(n^{-1}).
\end{equation}

We next study the reduced-model residual. Since $\bR_i(-1)$ consists of a subset of the columns of $\bPhi_{\bzeta}(-1)$, Assumption~\ref{ass:hete} also implies that $\bR_i(-1)$ has full column rank. Moreover, $\|\hat{\bR}_i(-1)-\bR_i(-1)\|_2=O_p(n^{-1/2})$, and by the same projection perturbation argument, $\|P_{\hat{\bR}_i(-1)}-P_{\bR_i(-1)}\|_2=O_p(n^{-1/2})$. Thus, for every $i\neq j$,
\begin{equation}
\label{eq:reduced_residual_rate}
\|P_{\hat{\bR}_i(-1)}^\perp\hat{\bzeta}_{ij}-P_{\bR_i(-1)}^\perp\bzeta_{ij}\|_2\leq\|\hat{\bzeta}_{ij}-\bzeta_{ij}\|_2+\|P_{\hat{\bR}_i(-1)}-P_{\bR_i(-1)}\|_2\|\bzeta_{ij}\|_2=O_p(n^{-1/2}).
\end{equation}

Consider a non-leaf node $i\notin\leaf(\cG)$. By Proposition \ref{prop:leaf}, there exists some $j\neq i$ such that $\|P_{\bR_i(-1)}^\perp\bzeta_{ij}\|_2^2>0$. For such a pair $(i,j)$, let $c_{ij}:=\|P_{\bR_i(-1)}^\perp\bzeta_{ij}\|_2^2>0$. Using \eqref{eq:reduced_residual_rate} and the inequality
\[
\bigl|\|\hat{\bnu}\|_2^2-\|\bnu\|_2^2\bigr|\leq\|\hat{\bnu}-\bnu\|_2\bigl(\|\hat{\bnu}-\bnu\|_2+2\|\bnu\|_2\bigr),
\]
we obtain
\[
\left|\|P_{\hat{\bR}_i(-1)}^\perp\hat{\bzeta}_{ij}\|_2^2-\|P_{\bR_i(-1)}^\perp\bzeta_{ij}\|_2^2\right|=O_p(n^{-1/2}).
\]
Hence,
\begin{equation}\label{eq:reduced_residual_rate_non_leaf}
\|P_{\hat{\bR}_i(-1)}^\perp\hat{\bzeta}_{ij}\|_2^2=c_{ij}+O_p(n^{-1/2})=c_{ij}+o_p(1).
\end{equation}

Let
\[
N_{ij,n}:=\|P_{\hat{\bR}_i(-1)}^\perp\hat{\bzeta}_{ij}\|_2^2-\|P_{\hat{\bPhi}_{\bzeta}(-1)}^\perp\hat{\bzeta}_{ij}\|_2^2
\]
and
\[
D_{ij,n}:=\|P_{\hat{\bPhi}_{\bzeta}(-1)}^\perp\hat{\bzeta}_{ij}\|_2^2+\frac{C_0}{n}.
\]
By \eqref{eq:full_residual_rate} and \eqref{eq:reduced_residual_rate_non_leaf}, $N_{ij,n}=c_{ij}+o_p(1)$ and $D_{ij,n}=O_p(n^{-1})$, where the latter rate holds since $C_0>0$ is fixed. Therefore,
\[
N_{ij,n}-\sqrt{2.01\log n}D_{ij,n}=c_{ij}+o_p(1).
\]
Noting $F_{ij}(-1)=N_{ij,n}/D_{ij,n}$, we have
\[
\P\left\{\hat F_{ij}(-1)>\sqrt{2.01\log n}\right\}\to 1.
\]
Since $\hat F_i(-1)=\max_{j\neq i}\hat F_{ij}(-1)\geq\hat F_{ij}(-1)$, it follows that for every $i\notin\leaf(\cG)$
\[
\lim_{n\to\infty}\P\left\{\hat F_i(-1)\geq\sqrt{2.01\log n}\right\}=1. 
\]

We next consider a leaf node $i\in\leaf(\cG)$. By Proposition \ref{prop:leaf}, for every $j\neq i$, $P_{\bR_i(-1)}^\perp\bzeta_{ij}=0$. By \eqref{eq:reduced_residual_rate}, we have
\[
N_{ij,n}\leq\|P_{\hat{\bR}_i(-1)}^\perp\hat{\bzeta}_{ij}\|_2^2=O_p(n^{-1}).
\]
Moreover,
\[
D_{ij,n}=\|P_{\hat{\bPhi}_{\bzeta}(-1)}^\perp\hat{\bzeta}_{ij}\|_2^2+\frac{C_0}{n}\geq\frac{C_0}{n}.
\]
Therefore,
\[
\hat F_{ij}(-1)=\frac{N_{ij,n}}{D_{ij,n}}\leq\frac{n}{C_0}N_{ij,n}=O_p(1).
\]
Since $p$ is fixed,
\[
\hat F_i(-1)=\max_{j\neq i}\hat F_{ij}(-1)=O_p(1)=o_p(\sqrt{\log n}).
\]
Thus, for all $i\in\leaf(\cG)$,
\[
\lim_{n\to\infty}
\P\left\{
\hat F_i(-1)
<
\sqrt{2.01\log n}
\right\}
=
1.
\]

Combining the two cases gives
\[
\lim_{n\to\infty}
\P\{\hat{\cV}(0)=\cV(0)\}
=1,
\]
and consequently
\[
\lim_{n\to\infty}
\P\{\hat{\cV}^{+}(0)=\cV^{+}(0)\}
=1.
\]

We next proceed by induction. Suppose that, for some
$h=0,\ldots,H_{\cG}-1$,
\[
\lim_{n\to\infty}
\P\bigl\{
\hat{\cV}(k)=\cV(k),
\ k=0,\ldots,h
\bigr\}
=1.
\]
Define $\cA_h:=\bigl\{\hat{\cV}(k)=\cV(k),\ k=0,\ldots,h\bigr\}$.
On the event $\cA_h$, we have $\hat{\cV}^{+}(h)=\cV^{+}(h)$, so the working subgraph used in lines 12--21 of Algorithm \ref{alg:est_order} coincides with the true subgraph
$\cG(h)$. The preceding argument, including the stabilization term
$C_0/n$, can therefore be applied to $\cG(h)$. Since Assumptions
\ref{ass:uncorr} and \ref{ass:hete} continue to hold for this subgraph, we have $\P\{\hat{\cV}(h+1)=\cV(h+1)\mid\cA_h\}\to 1$. By the induction hypothesis,
$\P(\cA_h)\to1$. Therefore,
\begin{align*}
\P\left\{
\hat{\cV}(h+1)=\cV(h+1)
\right\}
&\geq
\P\left\{
\hat{\cV}(h+1)=\cV(h+1)
\,\middle|\,
\cA_h
\right\}
\P(\cA_h)
\to 1.
\end{align*}
Thus, by induction,
\[
\lim_{n\to\infty}
\P\bigl\{
\hat{\cV}(h)=\cV(h),
\ h=0,\ldots,H_{\cG}
\bigr\}
=1.
\]
This completes the proof.

\end{proof}

\subsection{Proof of Theorem \ref{thm:asy}}\label{ssec:prf_thm_asy}
\begin{proof}

Define the event $\cA:=\{\hat{h}_\cG(i)=h_\cG(i),\ for\ all\ i=1,\dots,p\}$. Lemma \ref{lem:con_order} demonstrates that $\lim_{n\to \infty}\P(\cA)=1$. Therefore, the asymptotic properties of $\hat{\bV}$ and $\hat{\bW}$ can be analyzed conditional on the event  $\cA$, where $\hat{\bTheta}=\bTheta$. The proof of the asymptotic normality of $\hat{\bV}$ and $\hat{\bW}$ relies on the theory of the generalized method of moments (GMM). For further illustration, here we let $\btheta$ denote the vector whose elements consist of $\mu_k$ for $k=1,\dots,q$, $V_{ij}$ for $h_\cG(i)>h_\cG(j)$, and $W_{\ell j}$ for $\ell=1,\dots,q$ and $j=1,\dots, p$. Let $\bTheta$ denote the corresponding parameter space for $\btheta\in\R^d$ where $d=q+\sum_{h=0}^{H_\cG-1}|\cV^+(h)|\cdot|\cV(h)|+qp$. Let $\cO$ denote the sample space of $\Ok$. We impose the following regularity conditions \citep{hansen1982large}. 

\begin{assumption}[Regularity conditions on $\bQ(\btheta,\Ok)$] 
  The function $\bQ(\btheta,\Ok)$ satisfies that (i)  it is continuous on $\bTheta$ for each $\Ok\in \cO$; (ii) $\E(\bQ(\btheta,\Ok))$ exists and is finite for every $\btheta\in\bTheta$ ; (iii) $\E(\bQ(\btheta,\Ok))$ is continuous on $\bTheta$. 
\end{assumption}

\begin{assumption}[Regularity conditions on $\partial \bQ(\btheta,\Ok)/\partial \btheta^T$]
   (i) The derivative matrix $\partial \bQ(\btheta,\Ok)/\partial \btheta^T$ exists and is continuous on $\bTheta$ for each $\Ok\in \cO$; (ii) the true $\btheta$ is an interior point of $\bTheta$; (iii) $\E(\partial \bQ(\btheta,\Ok)/\partial \btheta^T)$ exists and is finite. 
\end{assumption}

\begin{assumption}[Compactness of $\bTheta$]
  $\bTheta$ is a compact set.
\end{assumption}

\begin{assumption}[Domination of $\bQ(\btheta,\Ok)$]
$\E(\sup_{\btheta\in \bTheta}\|\bQ(\btheta,\Ok)\|_2)<\infty$. 
\end{assumption}

\begin{assumption}[Properties of the variances of the sample moment]
(i) For the true $\btheta$, $\E\{\bQ(\btheta,\Ok)\bQ(\btheta,\Ok)^T\}$ exists and is finite; (ii) for the true $\btheta$, $\bS=\lim_{n\to+\infty}\ \mathrm{var}\big(n^{-1/2} \sum_{k=1}^n \bQ(\btheta,\Ok)\big)$ exists and is a ﬁnite valued positive deﬁnite matrix.
\end{assumption}

\begin{assumption}[Continuity of $\E\{\partial \bQ(\btheta,\Ok)/\partial \btheta^T\}$]
    $\E\{\partial \bQ(\btheta,\Ok)/\partial \btheta^T\}$ is continuous on some neighborhood $N_\ve$ of the true $\btheta$.
\end{assumption}

\begin{assumption}[Uniform Convergence of $\bG_n(\btheta):=n^{-1}\sum_{k=1}^n\partial \bQ(\btheta,\Ok)/\partial \btheta^T$] As $n\to\infty$, 
    \[
    \operatorname{sup}_{\btheta\in N_{\ve}}\left\|\bG_{n}(\btheta)-\E\big\{\partial \bQ(\btheta,\Ok)/\partial \btheta^T\big\}\right\|_2\xrightarrow[]{p}0.
    \]
\end{assumption}

\begin{assumption}[Properties of the weighting matrix]\label{ass:regu}
  $\bOmega_n$ is a positive semi-definite matrix which converges in probability to a positive definite matrix of constants $\bOmega$. 
\end{assumption}

Under Assumptions \ref{ass:uncorr} to \ref{ass:regu}, according to Theorem 3.1 of \cite{hansen1982large}, we have 
\[
\sqrt{n}(\hat{\btheta}-\btheta)\to \cN_d(\bzero,\bM\bS\bM^T),
\]
where $\bS=\var( \bQ(\btheta,\Ok))$ and $\bM=(\bG_0^T\bOmega \bG_0)^{-1}\bG_0^T \bOmega$ with $\bG_0=\E\big\{\partial \bQ(\btheta,\Ok)/\partial \btheta^T\big\}$. The proof of Theorem \ref{thm:asy} is finished by simply taking the corresponding entries in $\bM\bS\bM^T$ to obtain $\sigma_\bV[i,j]$ and $\sigma_\bW[\ell, j]$ as specified in the theorem. 
\end{proof}

\begin{remark}
Although the specific form of $\bG_0$ can be obtained in closed form, its expression can be complicated. To get consistent estimates of $\sigma_\bV[i,j]$ and $\sigma_\bW[\ell, j]$, alternative methods, such as bootstrap, can be adopted in practice. 
\end{remark}

\subsection{Proof of Theorem \ref{thm:con_graph}}
\begin{proof}
  Lemma~\ref{lem:con_order} shows that $\lim_{n\to \infty}\P(\cA)=1$ where $\cA:=\{\hat{h}_\cG(i)=h_\cG(i),\ for\ all\ i=1,\dots,p\}$. Following an argument analogous to that in the proof of Theorem~\ref{thm:asy}, we analyze the convergence of $\hat{\cE}_1$ and $\hat{\cE}_2$ conditional on the event $\cA$. We first focus on $\hat{\cE}_1$, the proof for $\hat{\cE}_2$ is similar. On the event $\cA$, we have 
\[
\begin{aligned}
  \hat{\cE}_1&=\Biggl\{(i,j):\hat{h}_\cG(i)>\hat{h}_\cG(j),|\hat{V}_{ij}|\ge \sqrt{\widehat{\sigma^2_{\bV}[i,j]}}\sqrt{\frac{2.01\log n}{n}}\Biggr\}\\
&=\Biggl\{(i,j):h_\cG(i)>h_\cG(j),\frac{\sqrt{n}|\hat{V}_{ij}|}{\sqrt{\widehat{\sigma^2_{\bV}[i,j]}}}\ge \sqrt{2.01\log n}\Biggr\}. 
\end{aligned}
\]
Since $\widehat{\sigma^2_{\bV}[i,j]}$ is consistent for $\sigma^2_{\bV}[i,j]$, using Theorem \ref{thm:asy} yields
\[
\frac{\sqrt{n}(\hat{V}_{ij}-V_{ij})}{\sqrt{\widehat{\sigma^2_{\bV}[i,j]}}}\to \cN(0,1),\ for\ h_\cG(i)>h_\cG(j).
\]
Thus, 
\[
\lim_{n\to\infty}\P\Bigg(\frac{\sqrt{n}|\hat{V}_{ij}|}{\sqrt{\widehat{\sigma^2_{\bV}[i,j]}}}<\sqrt{2.01\log n},\ for\ all\ V_{ij}=0\ and\ h_\cG(i)> h_\cG(j)\Bigg)=1, 
\]
which leads to $\lim_{n\to \infty}\P(\hat{\cE}_1\subseteq\cE_1)=1$. In contrast, if $V_{ij}\ne 0$, then 
\[
\lim_{n\to\infty}\P\Bigg(\frac{\sqrt{n}|\hat{V}_{ij}|}{\sqrt{\widehat{\sigma^2_{\bV}[i,j]}}}\ge\sqrt{2.01\log n}\Bigg)\ge\lim_{n\to\infty}\P\Bigg(\frac{\sqrt{n}|V_{ij}|}{\sqrt{\widehat{\sigma^2_{\bV}[i,j]}}}-\frac{\sqrt{n}|\hat{V}_{ij}-V_{ij}|}{\sqrt{\widehat{\sigma^2_{\bV}[i,j]}}}\ge\sqrt{2.01\log n}\Bigg)=1.
\]
Since $p=O(1)$ in the low-dimensional setting, 
\[
\lim_{n\to\infty}\P\Bigg(\frac{\sqrt{n}|\hat{V}_{ij}|}{\sqrt{\widehat{\sigma^2_{\bV}[i,j]}}}\ge\sqrt{2.01\log n},\ for\ all\ V_{ij}\ne 0\Bigg)=1, 
\]
which implies that $\lim_{n\to \infty}\P(\cE_1\subseteq\hat{\cE}_1)=1$. Therefore, $\lim_{n\to \infty}\P(\hat{\cE}_1=\cE_1)=1$. The proof is finished by repeating the same analysis for $\hat{\cE}_2$.  
\end{proof}

\section{A sequential method for estimating causal effects and directions in low-dimensional settings}\label{sec:seq}

In this section, we provide a sequential method for estimating causal effects and directions in low-dimensional settings, which follows the top-down manner as in Algorithm \ref{alg:iden_dag}. The whole procedure appears in Algorithm \ref{alg:est_order_seq}. For vectors $\bnu_1$ and $\bnu_2$ of the same dimension, let $\bnu_1 \odot \bnu_2$ denote the elementwise product. 

\begin{algorithm}[htbp]
  \caption{Sequential methods for estimating the causal effects and directions.}\label{alg:est_order_seq}
  \KwIn{Data matrices $\Y$ and $\X$, causal order $\{\hat{h}_\cG(i)\}_{i=1,\dots, p}$, $\{\hcV(h)\}_{h=0,\dots, \hat{H}_\cG}$, $\{\hcV^+(h)\}_{h=0,\dots, \hat{H}_\cG}$, and $\hat{H}_\cG$}

   $\hat{\bmu}\gets \frac{1}{n}\sum_{k=1}^n \X^{(k)}$

  $\hat{\bV}\gets \mathbf{NA}$, $\hat{\bW}\gets \mathbf{NA}$

  \For{$j\in \hat{\cV}(\hat{H}_\cG)$}{
    $\hat{\bV}_j\gets \bzero$

    $\hat{\bW}_j\gets (\X^T\X)^{-1}\X^T\Y_j$
  }

  \For{$h=\hat{H}_\cG-1$ to $0$}{
    $\hT^{(k)}(h)\gets\bigl((\X^{(k)}-\hat{\bmu})\odot(\Y_i^{(k)}-\hat{\bV}_i^T \Y^{(k)}-\hat{\bW}_i^T \X^{(k)}) \bigr)_{i\in \hat{\cV}^+(h)}$ for $k=1,\dots, n$

    \For{$j\in \hat{\cV}(h)$}{

      $(\hat{V}_{ij})_{i\notin \hat{\cV}^+(h)}\gets \bzero$
       
      \begin{equation}\label{eq:2sls}
      \bigl((\hat{V}_{ij})_{i\in \hat{\cV}^+(h)}, \hat{\bW}_j\bigr)\gets \TSLS(\Y_j\sim (\Y_i)_{i\in \hat{\cV}^+(h)}+\X\mid \hT(h)+\X)
      \end{equation}
    }
  }

  Compute consistent asymptotic variance estimators $\widehat{\sigma^2_{\bV[i,j]}}$ for $\hat V_{ij}$, where $\hat h_{\cG}(i)> \hat h_{\cG}(j)$

  $\hat{\cE}_1=\Biggl\{(i,j):\hat{h}_\cG(i)>\hat{h}_\cG(j),|\hat{V}_{ij}|\ge \sqrt{\widehat{\sigma^2_{\bV}[i,j]}}\sqrt{\frac{2.01\log n}{n}}\Biggr\}$

  Compute consistent asymptotic variance estimators $\widehat{\sigma^2_{\bW[\ell,j]}}$ for $\hat W_{\ell j}$, where $\ell=1,\dots,q$ and $j=1,\dots,p$

  $\hat{\cE}_2=\Biggl\{(\ell,j):|\hat{W}_{\ell j}|\ge \sqrt{\widehat{\sigma^2_{\bW}[\ell,j]}}\sqrt{\frac{2.01\log n}{n}}\Biggr\}$

  \Return{$\hat{\bV}$, $\hat{\bW}$, $\hat{\cE}_1$, and $\hat{\cE}_2$}
        
\end{algorithm}

Analogous to the identification procedure in Algorithm \ref{alg:iden_dag}, Algorithm \ref{alg:est_order_seq} estimates causal effects in a top-down manner following the learned causal order. In particular, we initialize $\hat{\bV}$ and $\hat{\bW}$ as $\mathbf{NA}$ in line 2 to explicitly show that they have not yet been computed. In each iteration of lines 7--13, Algorithm \ref{alg:est_order_seq} constructs surrogate instruments $\hT(h)$ using the estimated $\hat{\bV}_i$ and $\hat{\bW}_i$ for $i\in \hat{\cV}^+(h)$, and then employs them to obtain the estimates $\hat{\bV}_j$ and $\hat{\bW}_j$ for $j\in \hat{\cV}(h)$ using two-stage least squares (2SLS) through \eqref{eq:2sls}. Specifically, \eqref{eq:2sls} expresses the 2SLS estimates in terms of the R language package \texttt{ivreg} for clarity, which means taking $\Y_j$ as the response, $(\Y_i)_{i\in \hat{\cV}^+(h)}$ as the endogenous regressors, $\hT(h)$ as the IVs, and $\X$ as the covariates. The remaining steps of Algorithm \ref{alg:est_order_seq} follow the same procedure as the estimation based on GMM, with both approaches using hard thresholding to estimate the causal graph. 

We note that Algorithm \ref{alg:est_order_seq} can be viewed as a sequential implementation of the GMM procedure in \eqref{eq:GMM} with a specific weighting matrix $\bOmega_{\TSLS}$. Specifically, by arranging the moment conditions in \eqref{eq:moments} for estimating the parameters $(\bV_i,\bW_i)$ of $Y_i$ according to its topological height, this weighting matrix $\bOmega_{\TSLS}$ can be rearranged as a block diagonal matrix. We rearrange the primary variables such that $\hat{h}_\cG(1)\le \hat{h}_\cG(2)\le\dots\le \hat{h}_\cG(p)$, and correspondingly rearrange $\bQ(\btheta,\O^{(k)})$ in \eqref{eq:moments} as $\tilde{\bQ}(\btheta,\Ok)=\bigl(\tilde{\bQ}_{i}(\btheta,\Ok)\bigr)_{i=0,\dots,p+1}$, where
\[
\begin{aligned}
  \tilde{\bQ}_0(\btheta,\Ok)&=\X^{(k)}-\bmu,\\
  \tilde{\bQ}_{p+1}(\btheta,\Ok) &=\Bigl(\X^{(k)}-\bmu\Bigr)\Bigl(\Y_\ell^{(k)}-\bV_\ell^T \Y^{(k)}-\bW_\ell^T \X^{(k)}\Bigr)\\
   &\quad\relph{=}{} \cdot\Bigl(\Y_j^{(k)}-\bV_j^T \Y^{(k)}-\bW_j^T \X^{(k)}\Bigr),\ for\ \ell\in\hat{\cV}(\hat{h}_\cG(j))\setminus\{j\},j=1\dots,p, 
\end{aligned}
\]
and $\tilde{\bQ}_i=(\tilde{\bQ}_{i1}^T,\tilde{\bQ}_{i2}^T)^T$ for $i=1,\dots,p$ with
\[
\begin{aligned}
  \tilde{\bQ}_{i1}(\btheta,\Ok)&=\X^{(k)}\Bigl(\Y_i^{(k)}-\bV_i^T \Y^{(k)}-\bW_i^T \X^{(k)}\Bigr),\\
  \tilde{\bQ}_{i2}(\btheta,\Ok)&=\Bigl(\X^{(k)}-\bmu\Bigr)\Bigl(\Y_i^{(k)}-\bV_i^T \Y^{(k)}-\bW_i^T \X^{(k)}\Bigr)\\
  &\quad\relph{=}{} \cdot\Bigl(\Y_j^{(k)}-\bV_j^T \Y^{(k)}-\bW_j^T \X^{(k)}\Bigr),\ for\ all\ j\in\hat{\cV}^+(\hat{h}_\cG(i)).
\end{aligned}
\]
Define $\hS(h):=\bigl(\X,\hT(h)\bigr)$ for $h=0,\dots, \hat{H}_\cG-1$, where $\hT(h)=\{\hT(h)^{(k)}\}_{k=1}^n$ is defined in line 8 of Algorithm \ref{alg:est_order_seq}. We further define
\[
\begin{aligned}
    &\bOmega(\bmu):=\bI_q,\\ &\bOmega(h):=  
\Big (\frac{1}{n}\hS(h)^T \hS(h)\Big )^{-1},\ for\ h=0,\dots, \hat{H}_\cG-1,\\
&\bOmega(\hat{H}_\cG):=\Bigl(\frac{1}{n}\X^T \X\Bigr)^{-1}.
\end{aligned}
\]
Then the block-wise diagonal of the corresponding rearranged block diagonal matrix $\bOmega_{\TSLS}$ is
\[
\bigl(\bOmega(\bmu),(\underbrace{\bOmega(h),\dots,\bOmega(h)}_{|\hat{\cV}(h)|\ times})_{h=0,\dots, \hat{H}_\cG},\bzero\bigr).
\]
In contrast to directly solving the GMM problem in \eqref{eq:GMM}, Algorithm \ref{alg:est_order_seq} proceeds sequentially through a series of lower-dimensional estimation steps, which substantially simplifies computation. In practice, this procedure can also provide a consistent starting value for efficient GMM estimation.  

\section{Estimation methodology in high-dimensional settings}\label{sec:high}

In this section, we develop methods for causal discovery and effect estimation under high-dimensional settings. In particular, our framework accommodates high-dimensional settings, including $n\ll p$ and $p\le n\ll q$, provided that $\log p+\log q=o(\sqrt n)$. For simplicity, here we assume that the variables of each observation have been centered, i.e., $\bmu=\E(\X^{(k)})=\bzero$. To accommodate large-scale DAGs, we adapt the low-dimensional estimation approach by modifying the residual estimator $\hat{\Z}$ and covariance estimator $\hat{\bzeta}_{ij}$ in Algorithm \ref{alg:est_order}, as well as the causal effect estimator $(\hat{\bV},\hat{\bW})$ in \eqref{eq:GMM}. The full estimation framework is detailed in Section \ref{ssec:est_high}, with theoretical guarantees provided in Section \ref{ssec:theo_high}. The associated proofs are presented in Sections \ref{ssec:prf_theo_high} and \ref{ssec:prf_pre_lem}. 

We now introduce some notation for further illustration. For the DAG $\cG=(\cV;\cE_1,\cE_2)$, let $\pa_\cG(i)$, $\an_\cG(i)$, and $\de_\cG(i)$ denote the index set of parents, ancestors, and descendants of $Y_i$ in the edge set $\cE_1$ of $\cG$, respectively. Let $|J|$ denote the cardinality of a set $J$. For a vector $\bnu$, let $\|\bnu\|_p$ denote its $\ell_p$-norm where $1\le p\le \infty$, $\|\bnu\|_0$ the number of its non-zero entries, and $T_s(\bnu)=\bigl(\nu_i\bone(|\nu_i|\ge s)\bigr)$ the hard-thresholding operator. For a random variable $\bX$, let $\|\bX\|_{\psi_2}$ denote its sub-gaussian norm, $\|\bX\|_{\psi_1}$ its sub-exponential norm, and $\|\bX\|_n=(\sum_{k=1}^n \|\X^{(k)}\|_2^2)^{1/2}$ denote its empirical $\ell_2$-norm. For a matrix $\bA$ and index sets $I$ and $J$, let $\|\bA\|_\infty:=\max_i\sum_j|A_{ij}|$ denote the matrix $\ell_\infty$-norm, $\sigma_{min}(\bA)$ its minimal singular value, $A_{ij}=\bA[i,j]$ its $(i,j)$th element, $\bA_{IJ}=\bA[I,J]$ the submatrix formed with the $(i,j)$th entries where $i\in I$ and $j\in J$, and $\bA_J=\bA[\cdot,J]=(\bA_j)_{j\in J}$ the submatrix formed with columns $\bA_j$ where $j\in J$. Following \cite{bickel2009simultaneous}, define the restricted eigenvalue $\kappa(\bA,s)>0$ for an $m\times m$ symmetric matrix $\bA$ and $s\le m$ by
\[
  \kappa^2(\bA,s):=\min_{
\begin{array}
{c}J\subseteq\{1,...,m\}, \\
|J|\leq s
\end{array}}\min_{
\begin{array}
{c}\bnu\ne 0,\\
\|\bnu_{J^{c}}\|_{1}\leq 3\|{\bnu}_J\|_{1}
\end{array}}\frac{\bnu^T\bA\bnu}{\|\bnu_J\|^2_{2}}.
  \]

\subsection{Estimation methodology}\label{ssec:est_high}

In this subsection, we present the estimation methodology for causal directions and effects in high-dimensional settings. Following the estimation strategy developed for low-dimensional settings, we first estimate the causal order and then recover the causal effects and directions based on the estimated order.

Algorithm \ref{alg:est_order_high} outlines the procedure for learning the causal order. It largely retains the framework of Algorithm \ref{alg:est_order}, adapting only the computation of the residual $\hat{\Z}$ and covariances $\{\hat{\bzeta}_{ij}\}$. Specifically, we rewrite the model for primary variables as
\[
  Y_i=P_\bX Y_i+P^\perp_\bX Y_i=\bA_i^T\bX+Z_i,\ i=1,\dots,p,
\]
where $\bA^T:=(\bI_p-\bV^T)^{-1}\bW^T$ and $P_\bX \bY=\bA^T \bX$. The coefficient $\hat{\bA}_i$ is then obtained via Lasso with penalty factor $\alpha_n^{(i)}$ tuned by cross-validation, and the corresponding residuals $\hat{\Z}_i$ are subsequently computed, as implemented in lines 3 and 4 of Algorithm \ref{alg:est_order_high}. To estimate the covariances $\bzeta_{ij}$, we employ the thresholding estimator of \citep{bickel2008covariance,el2008operator} with threshold $\nu_n$, as detailed in lines 8--10. To select the hyperparameter $\nu_n$, one can apply the cross-validation approach as suggested by \cite{bickel2008covariance}. Since calculating $\hat{\Z}$ and $\hat{\bA}$ becomes more complex in high-dimensional settings and they are required later for constructing surrogate IVs, Algorithm \ref{alg:est_order_high} outputs them for subsequent reuse. 

\begin{algorithm}[t]
  \caption{Empirical methods for learning causal order in high-dimensional settings.}\label{alg:est_order_high}
  \KwIn{Data matrices $\Y$ and $\X$}

  $h\gets -1$, $p\gets \mathrm{dim}_2(\Y)$, $\hcV^+(-1)\gets \{1,\dots,p\}$
  
   \For{$i=1$ to $p$}{
$\hat{\bA}_i\gets \argmin \frac{1}{n}\|\Y_i-\X\bA_i\|_2^2+\alpha_n^{(i)} \|\bA_i\|_1$

$\hat{\Z}_i\gets \Y_i-\X\hat{\bA}_i$
   }

  \For{$i=1$ to $p$}{
    \For{$j=i$ to $p$}{
      $\check{\bzeta}_{ij}\gets \frac{1}{n}\sum_{k=1}^n \X^{(k)}\hZ^{(k)}_{i}\hZ^{(k)}_{j}$

      $\hat{\bzeta}_{ij}\gets T_{\nu_n}(\check{\bzeta}_{ij})$ 
      
      $\hat{\bzeta}_{ji}\gets\hat{\bzeta}_{ij}$
    }
  }

  Execute lines 10--24 in Algorithm \ref{alg:est_order}

  \Return{$\{\hat{h}_\cG(i)\}_{i=1,\dots, p}$, $\{\hcV(h)\}_{h=0,\dots, \hat{H}_\cG}$, $\{\hcV^+(h)\}_{h=0,\dots, \hat{H}_\cG}$, $\hat{H}_\cG$, $\hat{\Z}$, and $\hat{\bA}$}
\end{algorithm}

Based on the estimated causal order in Algorithm \ref{alg:est_order_high}, we next propose the method for estimating the causal effects and directions, as presented in Algorithm \ref{alg:est_dag_high}. Since directly solving the estimating equation \eqref{eq:GMM} becomes numerically challenging in high-dimensional settings, Algorithm \ref{alg:est_dag_high} adopts the sequential top-down approach from Algorithm \ref{alg:est_order_seq} in Section \ref{sec:seq}. Following Algorithm \ref{alg:est_order_seq}, Algorithm \ref{alg:est_dag_high} initializes $\hat{\bV}$ and $\hat{\bW}$ as $\mathbf{NA}$ in line 1 to explicitly show that they have not yet been computed. Since $\bV_j=\bzero$ and $\bW_j=\bA_j$ for $j\in \cV(H_\cG)$, lines 2--5 set $\hat{\bV}_j$ to zero and $\hat{\bW}_j$ to $\hat{\bA}_j$ for $j\in \hat{\cV}(\hat{H}_\cG)$ accordingly. The sequential estimation procedure for the causal effects $(\bV,\bW)$ is then carried out in lines 6--16. At each iteration, Algorithm \ref{alg:est_dag_high} first constructs surrogate instruments $\hT(h)$ using the estimated $\hat{\bV}_i$ and $\hat{\bW}_i$ for $i\in \hat{\cV}^+(h)$, and then employs them to obtain the estimates $\hat{\bV}_j$ and $\hat{\bW}_j$ for $j\in \hat{\cV}(h)$ using an adapted 2SLS approach, as shown in lines 8--15. 

In contrast to the classic 2SLS in Algorithm \ref{alg:est_order_seq}, we incorporate $\ell_1$-penalty in both stages to overcome the difficulty of high dimensionality while simultaneously performing model selection. Specifically, since the covariates, instruments, and endogenous regressors are the same for all $j\in \cV(h)$, the first-stage model for estimating $(\bV_j,\bW_j)$ with $j\in \cV(h)$ is identical and formulated as follows:
\begin{equation}\label{eq:model_stg1}
  Y_i=\bT(h)\bB_i(h)+\bX\bF_i(h)+\delta_i(h),\ i\in \cV^+(h),
\end{equation}
where $\bT(h)\bB_i(h)+\bX\bF_i(h)=P_{(\bT(h)^T,\bX^T)^T}Y_i$ and $\delta_i(h)=P^\perp_{(\bT(h)^T,\bX^T)^T}Y_i$. Algorithm \ref{alg:est_dag_high} estimates $\bB_i(h)$ and $\bF_i(h)$ using Lasso with tuning parameter $\beta_n^{(i)}(h)$, as implemented in line 9. The first-stage prediction of $\Y_i$ is then obtained as $\hat{\Y}_i(h)\gets \hT(h)\hat{\bB}_i(h)+\X\hat{\bF}_i(h)$ in line 10. Substituting the first-stage prediction $\hat{\Y}_i(h)$ for $\Y_i$ with $i\in \hat{\cV}^+(h)$, we proceed to obtain the causal effect estimates $(\hat{\bV}_j,\hat{\bW}_j)$ using Lasso with tuning parameter $\gamma_n^{(j)}(h)$, as implemented in lines 13 and 14. Unlike edge estimation via hard thresholding in Algorithm \ref{alg:est_order_seq}, Lasso inherently performs model selection, allowing $(\cE_1,\cE_2)$ to be directly estimated based on the non-zero entries of $(\hat{\bV},\hat{\bW})$, as shown in lines 17 and 18.

\begin{algorithm}
  \caption{Empirical methods for estimating causal effects and directions in high-dimensional settings.}\label{alg:est_dag_high}
  \KwIn{Data matrices $\Y$ and $\X$, causal order $\{\hat{h}_\cG(i)\}_{i=1,\dots, p}$, $\{\hcV(h)\}_{h=0,\dots, \hat{H}_\cG}$, $\{\hcV^+(h)\}_{h=0,\dots, \hat{H}_\cG}$, $\hat{H}_\cG$, residual $\hat{\Z}$, and coefficient $\hat{\bA}$ learned from Algorithm \ref{alg:est_order_high}}

  $\hat{\bV}\gets \mathbf{NA}$, $\hat{\bW}\gets \mathbf{NA}$

  \For{$j\in \hat{\cV}(\hat{H}_\cG)$}{
    $\hat{\bV}_j\gets \bzero$

    $\hat{\bW}_j\gets \hat{\bA}_j$
  }

  \For{$h=\hat{H}_\cG-1$ to $0$}{
    $\hT^{(k)}(h)\gets\bigl(\X^{(k)}\odot(\Y_i^{(k)}-\hat{\bV}_i^T \Y^{(k)}-\hat{\bW}_i^T \X^{(k)}) \bigr)_{i\in \hat{\cV}^+(h)}$ for $k=1,\dots, n$

    \For{$i\in \hat{\cV}^+(h)$}{
      \begin{equation}\label{eq:opt_stg1}
      \hat{\bB}_i(h),\hat{\bF}_i(h)\gets \arg\min \frac{1}{n}\|\Y_i-\hT(h)\bB_i(h)-\X\bF_i(h)\|_2^2+\beta_n^{(i)}(h)\{\|\bB_i(h)\|_1+\|\bF_i(h)\|_1\}
      \end{equation}

      $\hat{\Y}_i(h)\gets \hT(h)\hat{\bB}_i(h)+\X\hat{\bF}_i(h)$

    }

    \For{$j\in \hat{\cV}(h)$}{
      $(\hat{V}_{ij})_{i\notin\hat{\cV}^+(h)}\gets \bzero$

\begin{equation}\label{eq:opt_stg2}
(\hat{V}_{ij})_{i\in \hat{\cV}^+(h)},\hat{\bW}_j\gets \argmin\frac{1}{n}\biggl\|\Y_j-\sum_{i\in\hat{\cV}^+(h)}V_{ij}\hat{\Y}_i(h)-\X\bW_j\biggr\|_2^2+\gamma_n^{(j)}(h)\Biggl\{\sum_{i\in \hat{\cV}^+(h)}|V_{ij}|+\|\bW_j\|_1\Biggr\}
\end{equation}
    }
  }
  
  $\hat{\cE}_1=\Bigl\{(i,j):\hat{h}_\cG(i)>\hat{h}_\cG(j),|\hat{V}_{ij}|\ne 0\Bigr\}$

  $\hat{\cE}_2=\Bigl\{(\ell,j):|\hat{W}_{\ell j}|\ne 0\Bigr\}$

  \Return{$\hat{\bV}$, $\hat{\bW}$, $\hat{\cE}_1$, and $\hat{\cE}_2$}
        
\end{algorithm}

Together, Algorithms \ref{alg:est_order_high} and \ref{alg:est_dag_high} form the complete estimation procedure for the causal graph and effects. In the following subsection, we establish rigorous theoretical guarantees to validate the proposed method.

\subsection{Theoretical guarantees}\label{ssec:theo_high}

In this subsection, we provide the theoretical analysis for the proposed method under the high-dimensional settings, including both cases where $p\gg n$ and $q\gg n\ge p$. For $h=-1,\dots, H_\cG-1$ and $i\in \cV^+(h)$, define $\bR_i(h)=(\bzeta_{kk})_{k\ne i,k\in\cV^+(h)}$. We first introduce some assumptions.

\begin{assumption}[Sub-gaussian residual]\label{ass:sub_Z}
  The residual $\bZ$ is sub-gaussian, i.e., there exists some positive constant $C_1$ such that $\|\bZ\|_{\psi_2}\le C_1$. 
\end{assumption}

\begin{assumption}[Restricted eigenvalue condition for $\bSigma_\bX$]\label{ass:RE}
  There exists some positive constant $C_2$, such that $\kappa(\bSigma_\bX,s_\bA)\ge C_2$ where $s_\bA=\max_{i=1,\dots, p} \|\bA_i\|_0$. 
\end{assumption}

\begin{assumption}[Bounded and sub-gaussian auxiliary variable]\label{ass:boundX}
  There exists some positive constant $C_3$ such that $ \|\bX\|_{\psi_2}\le C_3 $ and $ \|\bX\|_\infty\le C_3$ almost surely.
\end{assumption}

\begin{assumption}[Sparse graph]\label{ass:sp_graph}
  There exists some positive constant $C_4$ such that $\max_{i=1,\dots, p}\{|\an_\cG(i)|,\|\bW_i\|_0\}\le C_4$. 
\end{assumption}

\begin{assumption}[Sparse covariance]\label{ass:sp_cov}
  There exists some positive constant $C_5$ such that each row and each column of $\bPhi_{\bzeta}$ has at most $C_5$ non-zero entries. 
\end{assumption}

\begin{assumption}[Heteroscedastic strength]\label{ass:str_hete}
  There exist some sufficiently large positive constants $C_6$ and $C_7$, such that $\sigma_{min}(\bPhi_{\bzeta})\ge C_6$ and
\[
\min_{i\in\cV^+(0)}\ \min_{h=-1,\dots,h_\cG(i)-2}\ \max_{j\ne i,j\in\cV^+(h)} \bzeta_{ij}^T P^\perp_{\bR_{i}(h)}\bzeta_{ij} \ge C_7\frac{\log p+\log q}{\sqrt{n}}.
\]
\end{assumption}

Assumptions \ref{ass:sub_Z}--\ref{ass:boundX} are standard regularity conditions in high-dimensional settings \citep{bickel2009simultaneous,buhlmann2011statistics}. In particular, the boundedness condition in Assumption \ref{ass:boundX} is mild and commonly imposed for high-dimensional theoretical analysis, especially when the covariates are naturally bounded or uniformly bounded after appropriate preprocessing. Furthermore, Assumptions \ref{ass:sp_graph} and \ref{ass:sp_cov} enforce sparsity on the DAG and covariances, respectively, aligning with the frameworks of \cite{buhlmann2014cam,frot2019robust} and \cite{el2008operator,bickel2008covariance}. To overcome the challenges of high-dimensional estimation, Assumption \ref{ass:str_hete} requires that the heteroscedastic strength is bounded away from zero and is stricter than Assumption \ref{ass:hete}.

The following lemma shows that, under Assumptions \ref{ass:uncorr}, \ref{ass:hete}, and \ref{ass:sub_Z}--\ref{ass:str_hete}, Algorithm \ref{alg:est_order_high} yields a consistent estimator of the causal order in the high-dimensional setting.

\begin{lemma}\label{lem:con_order_high}
  Suppose that Assumptions \ref{ass:uncorr}, \ref{ass:hete}, and \ref{ass:sub_Z}--\ref{ass:str_hete} hold. Let $\alpha_n^{(i)}=C_\alpha\sqrt{(\log p+\log q)/n}$ for $i=1,\dots, p$ and $\nu_n=C_\nu\sqrt{(\log p+\log q)/n}$ with some positive constants $C_\alpha$ and $C_\nu$. Then for sufficiently large $n$, 
  \[
  \P(\hat{h}_\cG(i)=h_\cG(i)\ for\ all\ i=1,\dots, p)\ge 1-Cn^{-1},
  \]
where $C$ is some positive constant. 
\end{lemma}

Although the tuning parameters $\alpha_n^{(i)}$ with $i=1,\dots, p$ are set to a common value for simplicity in Lemma \ref{lem:con_order_high}, they are permitted to differ as long as $\alpha_n^{(i)}= O(\sqrt{(\log p+\log q)/n})$ for all $i=1,\dots, p$. 

To establish consistency of causal discovery and effect estimation, we introduce the following regularity conditions. Let $\bS(h)=(\bT(h)^T,\bX^T)^T$ for $h=0,\dots,H_\cG-1$ represent the first-stage regressors and $\bSigma_{\bS(h)}=\var(\bS(h))$ the corresponding covariance. For $i\in \cV^+(h)$ and $h=0,\dots, H_\cG-1$, define $\Y_i(h):=\TT(h)\bB_i(h)+\X\bF_i(h)$, where the coefficients $\bB_i(h)$ and $\bF_i(h)$ are given in \eqref{eq:model_stg1}. Define $\Q(h):=((\Y_i(h))_{i\in\cV^+(h)},\X)$ for $i\in\cV^+(h)$ with $h=0,\dots, H_\cG-1$. Define further $s(h):=\max_{i\in\cV^+(h)} \{\|\bB_i(h)\|_0+\|\bF_i(h)\|_0\}$, $\tbSigma(h):=\Q(h)^T \Q(h)/n$ for $h=0,\dots, H_\cG-1$, and $\tbSigma(H_\cG):=\X^T \X/n$. Let $J_i$ denote the support set of $((V_{ji})_{j\in\cV^+(h_\cG(i))},\bW_i)$ for $i=1,\dots, p$.

\begin{assumption}[Sub-gaussian errors]\label{ass:sub_errors}
    The errors $\bve$ and $(\delta_i(h))_{i\in\cV^+(h)}$ with $h=0,\dots,H_\cG-1$ are sub-gaussian, i.e., there exists some positive constants $C_8$ and $C_9$ such that $\|\bve\|_{\psi_2}\le C_8$ and $\|(\delta_i(h))_{i\in\cV^+(h)}\|_{\psi_2}\le C_9$ for $h=0,\dots,H_\cG-1$. 
\end{assumption}

\begin{assumption}[Restricted eigenvalue condition for the first stage]\label{ass:stg1}
   There exists some positive constants $C_{10}$ and $C_{11}$, such that $\kappa(\bSigma_{\bS(h)},s(h))\ge C_{10}$ and $s(h)\le C_{11}$ for $h=0,\dots,H_\cG-1$.
\end{assumption}

\begin{assumption}[Irrepresentablity condition for the second stage]\label{ass:stg2}
  The following event holds with probability at least $1-C_{12}n^{-1}$ for sufficiently large $n$,
    \[
  \bigl\|\tbSigma(h)_{J^c_i J_i}\{\tbSigma(h)_{J_iJ_i}\}^{-1}\bigr\|_\infty\le 1-C_{13}\quad and\quad\|\{\tbSigma(h)_{J_iJ_i}\}^{-1}\|_\infty\le C_{14} ,
  \]
  for all $i\in\cV(h)$ and $h=0,\dots, H_\cG$, where $C_{12},C_{13},C_{14}$ are some positive constants with $C_{13}<1$. 
\end{assumption}

\begin{assumption}[$\ell_1$-sparsity for both stages]\label{ass:both_stg}
There exists some positive constant $C_{15}$ and $C_{16}$ such that $\max_{i=1,\dots,p}\{\|\bV_i\|_1,\|\bW_i\|_1\}\le C_{15}$ and $\max_{h=0,\dots,H_\cG-1}\max_{i\in\cV^+(h)} \{\|\bB_i(h)\|_1,\|\bF_i(h)\|_1\}\le C_{16}$.   
\end{assumption}

\begin{assumption}[Signal strength]\label{ass:str_sig}
  There exist some sufficiently large positive constants $C_{17}$ and $C_{18}$, such that 
  \[
  \min_{V_{ij}\ne 0}|V_{ij}|\ge C_{17} \sqrt{\frac{\log p+\log q}{n}}\quad and\quad \min_{W_{\ell j}\ne 0}|W_{\ell j}|\ge C_{18} \sqrt{\frac{\log p+\log q}{n}}. 
  \]
\end{assumption}

Assumption \ref{ass:stg1} is a restricted eigenvalue condition \citep{bickel2009simultaneous}, and Assumption \ref{ass:stg2} is in the same spirit as the irrepresentablity condition \citep{zhao2006model}, both of which are standard in high-dimensional estimation. Assumption \ref{ass:both_stg} imposes $\ell_1$-sparsity on coefficients in both stages of the 2SLS, which is common in the recovery of large linear SEM \citep{chen2018two,chen2024discovery}. Finally, Assumption \ref{ass:str_sig} imposes restrictions on the minimal signal strengths of $\bV$ and $\bW$ so that the DAG $\cG$ can be consistently recovered, which is similar to the beta-min condition in the high-dimensional linear regression literature \citep{meinshausen2006high}. 

Building on Lemma \ref{lem:con_order_high}, the following theorem establishes the consistency of the proposed method in estimating both the causal effects and directions. 

\begin{theorem}\label{thm:con_graph_high}
  Suppose that conditions of Lemma \ref{lem:con_order_high} and Assumptions \ref{ass:sub_errors}--\ref{ass:str_sig} hold. Let $\beta_n^{(i)}(h)=C_\beta\sqrt{(\log p+\log q)/n}$ and $\gamma_n^{(j)}(h)=C_\gamma\sqrt{(\log p+\log q)/n}$ for $i\in \hat{\cV}^+(h)$, $j\in \hat{\cV}(h)$, and $h=0,\dots,\hat{H}_\cG-1$, where $C_\beta$ and $C_\gamma$ are some positive constants. Then when $n$ is sufficiently large, with probability at least $1-Cn^{-1}$, the following holds for all $i,j=1,\dots,p$, and $\ell=1,\dots,q$, 
  \[
  \begin{aligned}
  &|\hat{V}_{ij}-V_{ij}|\le C^\prime\sqrt{\frac{\log p+\log q}{n}}\quad ,\quad |\hat{W}_{\ell j}-W_{\ell j}|\le C^{\prime\prime}\sqrt{\frac{\log p+\log q}{n}},\quad and\\
  &\sgn(\hat{V}_{ij})=\sgn(V_{ij})\quad,\quad \sgn(\hat{W}_{\ell j})=\sgn(W_{\ell j}), 
  \end{aligned}
  \]
  where $C,C^\prime,C^{\prime\prime}$ are some positive constants. Consequently, for sufficiently large $n$,
  \[
  \P(\hat{\cE}_1=\cE_1\ and\ \hat{\cE}_2=\cE_2)\ge 1-Cn^{-1}.
  \]
\end{theorem}

For the tuning parameters $\beta_n^{(i)}(h)$ with $i\in \hat{\cV}^+(h)$ and $h=0,\dots, \hat{H}_\cG-1$, while Theorem \ref{thm:con_graph_high} sets them to a common value for simplicity, they can be different as long as $\beta_n^{(i)}(h)= O(\sqrt{(\log p+\log q)/n})$, similar to the setup of tuning parameters $\alpha_n^{(i)}$ in Lemma \ref{lem:con_order_high}. This also applies to the tuning parameters $\gamma_n^{(j)}(h)$ with $j\in\hat{\cV}(h)$ and $h=0,\dots, \hat{H}_\cG-1$.

\subsection{Proofs for Section \ref{ssec:theo_high}}

\begin{proof}[Proof of Lemma \ref{lem:con_order_high}]\label{ssec:prf_theo_high}

  The proof proceeds by mathematical induction and is conditional on the results of Lemmas \ref{lem:A_est_high}--\ref{lem:project_high} in Section \ref{ssec:prf_pre_lem}. 

  We first note that the population covariance vectors are uniformly bounded in
$\ell_2$-norm. Since the variables are centered, we have
\[
\bzeta_{ij}
=
\operatorname{cov}(\bX,Z_iZ_j)
=
E(\bX Z_iZ_j),
\qquad i,j=1,\ldots,p.
\]
Therefore, by the dual representation of the $\ell_2$-norm and the Cauchy--Schwarz inequality,
\[
\begin{aligned}
\|\bzeta_{ij}\|_2
&=
\sup_{\|\bu\|_2=1}
\left|
\E\bigl\{(\bu^T\bX)Z_iZ_j\bigr\}
\right| \\
&\le
\sup_{\|\bu\|_2=1}
\left\{\E(\bu^T\bX)^2\right\}^{1/2}
\left\{\E(Z_i^2Z_j^2)\right\}^{1/2}.
\end{aligned}
\]
By Assumption~\ref{ass:boundX}, $\|\bX\|_{\psi_2}\le C_1$, and hence
\[
\sup_{\|\bu\|_2=1}
\left\{\E(\bu^T\bX)^2\right\}^{1/2}
\le
C_2\sup_{\|\bu\|_2=1}\|\bu^T\bX\|_{\psi_2}
\le C_3.
\]
Moreover, by Assumption \ref{ass:sub_Z},
\[
\left\{\E(Z_i^2Z_j^2)\right\}^{1/2}
\le
\left\{\E(Z_i^4)\E(Z_j^4)\right\}^{1/4}
\le
C_4\|Z_i\|_{\psi_2}\|Z_j\|_{\psi_2}
\le C_5.
\]
Consequently,
\[
\max_{i,j=1,\ldots,p}\|\bzeta_{ij}\|_2
\le C_3C_5
=O(1).
\]
Thus, throughout the proof, the factors
$(\|\bzeta_{ij}\|_2+1)$ and
$(\|\bzeta_{ij}\|_2+1)^2$
can be absorbed into generic positive constants.

We then show that, under the results of Lemmas \ref{lem:A_est_high}--\ref{lem:project_high}, $\widehat{\leaf}(\cG)=\leaf(\cG)$ holds. For any $i\notin \leaf(\cG)$, Proposition \ref{prop:leaf} illustrates that there exists some $\ell\ne i$ such that $P_{\bR_i(-1)}^\perp \bzeta_{i\ell}\ne \bzero$ where $\bR_i(-1)=(\bzeta_{kk})_{k\ne i}$.  Define $j=\arg\max_{\ell\ne i}\|P_{\bR_i(-1)}^\perp \bzeta_{i\ell}\|_2$. Thus, Lemma \ref{lem:project_high} implies
\[
  \Bigl|\hat{\bzeta}_{ij}^T P_{\hat{\bR}_i(-1)}^\perp\hat{\bzeta}_{ij}-\bzeta_{ij}^T P^\perp_{\bR_i(-1)}\bzeta_{ij}\Bigr|\le C_6(\|\bzeta_{ij}\|_2+1)^2\sqrt{\frac{\log p+\log q }{n}}.
\]
  Moreover, since $P_{\bPhi_{\bzeta}}^\perp \bzeta_{ij} = \bzero$ according to Proposition \ref{prop:full}, Lemma \ref{lem:project_high} leads to
  \[
  \Bigl|\hat{\bzeta}_{ij}^T P_{\hat{\bPhi}_{\bzeta}}^\perp\hat{\bzeta}_{ij}-\bzeta_{ij}^T P^\perp_{\bPhi_{\bzeta}}\bzeta_{ij}\Bigr|\le C_7(\|\bzeta_{ij}\|_2+1)^2\frac{\log p+\log q }{n}.
  \] 
  Therefore, 
  \[
  \begin{aligned}
      \hat{F}_i(-1)&= \max_{\ell\ne i}\frac{\hat{\bzeta}_{i\ell}^T\Bigl(P_{\hat{\bPhi}_{\bzeta}(-1)}-P_{\hat{\bR}_i(-1)}\Bigr)\hat{\bzeta}_{i\ell}}{\hat{\bzeta}_{i\ell}^T\Bigl(\bI_q-P_{\hat{\bPhi}_{\bzeta}(-1)}\Bigr)\hat{\bzeta}_{i\ell}+C_0/n}\\
      &\ge \frac{\bzeta_{ij}^T P_{\bR_i(-1)}^\perp\bzeta_{ij}-C_8(\|\bzeta_{ij}\|_2+1)^2\sqrt{(\log p+\log q) /n}}{C_9(\|\bzeta_{ij}\|_2+1)^2(\log p+\log q) /n+C_0/n}\\
      &\ge C_{10}\sqrt{n},
  \end{aligned}
  \]
  where the last line holds due to Assumption \ref{ass:str_hete}. Conversely, if $i\in \leaf(\cG)$, then we have $\hat{F}_i(-1)=o(\sqrt{\log n})$ by repeating the preceding analysis. Consequently, for all $i\notin \leaf(\cG)$, we have $\hat{F}_i(-1)\gg \sqrt{2.01\log n}$, whereas for all $i\in \leaf(\cG)$, we have $\hat{F}_i(-1)\ll \sqrt{2.01\log n}$. Hence, for sufficiently large $n$, $\widehat{\leaf}(\cG)=\leaf(\cG)$ conditional on the results of Lemmas \ref{lem:A_est_high}--\ref{lem:project_high}.  

  The following analysis is still conducted conditional on the events established in Lemmas \ref{lem:A_est_high}--\ref{lem:project_high}. Now suppose that $\hat{\cV}(k)=\cV(k)$ for $k=0,\dots, h$ with some $h=0,\dots, H_\cG-1$. We next prove that $\hat{\cV}(h+1)=\cV(h+1)$. On the event $\cA:=\{\hat{\cV}(k)=\cV(k)\ for\ k=0,\dots,h\}$, we have $\hat{\cV}^+(h)=\cV^+(h)$. Thus, the estimated leaves $\hat{\cV}(h+1)$ can be viewed as being derived from the true subgraph $\cG(h)$ on $\cA$. Since Assumptions \ref{ass:uncorr}, \ref{ass:hete}, and \ref{ass:sub_Z}--\ref{ass:str_hete} also hold in $\cG(h)$, using the result that $\widehat{\leaf}(\cdot)=\leaf(\cdot)$ for the subgraph $\cG(h)$, we have $\hat{\cV}(h+1)=\cV(h+1)$. The proof is then finished by induction. 
\end{proof}

\begin{proof}[Proof of Theorem \ref{thm:con_graph_high}] 
  The proof is conditional on the results in Lemma \ref{lem:con_order_high}. Let $\hat{\ve}_j^{(k)}=\Y_j^{(k)}-\hat{\bV}_j^T \Y^{(k)}-\hat{\bW}_j^T \X^{(k)}$ denote the estimated structural error associated with the $j$th primary variable for the $k$th observation. Define the following event for $h=-1,\dots,H_\cG-1$,
 \[
  \begin{aligned}
  \cA_h&:=\Biggl\{|\hat{V}_{ij}-V_{ij}|\le C_1\sqrt{\frac{\log p+\log q}{n}}\quad ,\quad |\hat{W}_{\ell j}-W_{\ell j}|\le C_2\sqrt{\frac{\log p+\log q}{n}},\\
  &\relph{=}{}\sgn(\hat{V}_{ij})=\sgn(V_{ij})\quad,\quad \sgn(\hat{W}_{\ell j})=\sgn(W_{\ell j}), \quad and\\
  & \relph{=}{}\|\hat{\ve}_j-\ve_j\|_n\le C_3\sqrt{\log p+\log q},\ for\ all\ i=1,\dots,p,j\in\cV^+(h),\ell=1,\dots,q\Biggr\}. 
  \end{aligned}
  \]
  Lemma \ref{lem:stg2_high} implies that $\P(\cA_{h-1}\mid \cA_h)\ge 1-c_1n^{-1}$ for all $h=0,\dots,H_\cG-1$ when $n$ is sufficiently large. Since $H_\cG=O(1)$ by Assumption \ref{ass:sp_graph}, it suffices to show that $\P(\cA_{H_\cG-1})\ge 1-c_3n^{-1}$ in order to establish $\P(\cA_{-1})\ge 1-c_2n^{-1}$. Conditional on the results in Lemma \ref{lem:con_order_high}, we have$\hat{\bV}_j=\bV_j=\bzero$ for $j\in\cV(H_\cG)$. Moreover, the estimation of $\bW_j$ for $j\in\cV(H_\cG)$ can be decomposed into $|\cV(H_\cG)|$ penalized least squares problems. Therefore, applying standard Lasso sign-consistency theory to each problem and taking a union bound over $j\in\cV(H_\cG)$ yields $\P(\cA_{H_\cG-1})\ge 1-c_3n^{-1}$. As a result, we have $\P(\cA_{-1})\ge 1-c_2n^{-1}$ for sufficiently large $n$, which completes the proof. 
\end{proof}

\subsection{Preliminary lemmas}\label{ssec:prf_pre_lem}

\begin{lemma}\label{lem:A_est_high}
  Suppose that Assumptions \ref{ass:sub_Z}--\ref{ass:sp_graph} hold. Then for sufficiently large $n$, it follows that with probability at least $1-C_1n^{-1}$, 
  \[
  \begin{aligned}
    &\max_{i=1,\dots,p} \|\hat{\bA}_i-\bA_i\|_1\le \frac{C_2 s_\bA}{\kappa^2(\bSigma_\bX,s_\bA)}\sqrt{\frac{\log p+\log q}{n}},\quad and\\
    &\max_{i=1,\dots,p} \|\hat{\Z}_i-\Z_i\|_2^2\le \frac{C_3  s_\bA}{\kappa^2(\bSigma_\bX,s_\bA)}(\log p+\log q),
  \end{aligned}
  \]
  where $C_1,C_2$, and $C_3$ are some positive constants. 
\end{lemma}

\begin{proof}
  By Theorem 1.6 in \cite{zhou2009restricted}, $\P(\kappa(\X^T \X/n,s_\bA)\ge C_1)\ge 1-C_2n^{-1}$ for sufficiently large $n$ under Assumptions \ref{ass:RE} and \ref{ass:boundX}. Since the problem can be decomposed into $p$ penalized least squares problems, the result follows as a straightforward extension of Theorem 7.2 in \cite{bickel2009simultaneous} using the union bound. 
\end{proof}

\begin{lemma}\label{lem:zeta_high}
  Under the conditions of Lemma \ref{lem:con_order_high}, the following holds with probability at least $1-Cn^{-1}$ for sufficiently large $n$,
    \[\|\hat{\bPhi}_{\bzeta}-\bPhi_{\bzeta}\|_2 \le C_1\sqrt{\frac{\log p+\log q}{n}}\quad and\quad \max_{i,j=1,\dots, p}\|\hat{\bzeta}_{ij}-\bzeta_{ij}\|_2\le C_2\sqrt{\frac{\log p+\log q}{n}},
  \]
  where $\hat{\bPhi}_{\bzeta}=(\hat{\bzeta}_{ii})_{i=1,\dots,p}$ and $C,C_1,C_2$ are some positive constants. 
\end{lemma}

\begin{proof}
  The following proof is conditional on the results in Lemma \ref{lem:A_est_high}. We first obtain the upper bound of $\max_{i,j}\|\check{\bzeta}_{ij}-\bzeta_{ij}\|_\infty$. Note that 
  \[
  \begin{aligned}
      \max_{i,j=1,\dots,p}\|\check{\bzeta}_{ij}-\bzeta_{ij}\|_\infty&\le \underbrace{\max_{i,j=1,\dots,p}\Biggl\|\frac{1}{n}\X^T(\Z_i\odot \Z_j)-\bzeta_{ij}\Biggr\|_\infty}_{\tT_1}+2\underbrace{\max_{i,j=1,\dots,p}\Biggl\|\frac{1}{n}\X^T\{(\hat{\Z}_i-\Z_i)\odot \Z_j\}\Biggr\|_\infty}_{\tT_2}\\
      &\relph{=}{}+\underbrace{\max_{i,j=1,\dots,p}\Biggl\|\frac{1}{n}\X^T\{(\hat{\Z}_i-\Z_i)\odot(\Z_j-\hat{\Z}_j)\}\Biggr\|_\infty}_{\tT_3}.
  \end{aligned}
  \]
  
  For the term $\tT_1$, note that  $\E(\bX Z_iZ_j)=\bzeta_{ij}$, and $\|X_\ell Z_iZ_j\|_{\psi_1}\le \|X_\ell Z_i\|_{\psi_2}\|Z_j\|_{\psi_2}\le C_1$ under Assumptions \ref{ass:sub_Z} and \ref{ass:boundX}. Thus, for all $i,j=1,\dots,p$ and $\ell=1,\dots,q$, 
  \[
  \P\biggl\{\biggl|\frac{1}{n}\sum_{k=1}^n \X_\ell^{(k)}\Z_i^{(k)} \Z_j^{(k)}-\E(\X_\ell^{(k)}\Z_i^{(k)} \Z_j^{(k)})\biggr|\ge C_2\sqrt{\frac{\log p+\log q}{n}}\biggr\}\le \exp(-C_3 (\log p+\log q)). 
  \]
  Therefore,  
  \[
  \P\biggl\{\max_{i,j,\ell}\biggl|\frac{1}{n}\sum_{k=1}^n \X_\ell^{(k)}\Z_i^{(k)} \Z_j^{(k)}-\E(\X_\ell^{(k)}\Z_i^{(k)} \Z_j^{(k)})\biggr|\ge C_2\sqrt{\frac{\log p+\log q}{n}}\biggr\}\le p^2 q\exp(-C_3 (\log p+\log q))\le C_4n^{-1}, 
  \]
  for sufficiently large $n$. As a result, $\P(\tT_1\le C_4 \sqrt{(\log p+\log q)/n})\ge 1-C_5n^{-1}$.

  For the term $\tT_2$, let
\[
\bDelta_i=\hat{\bA}_i-\bA_i,\qquad i=1,\ldots,p.
\]
Since $\hat{\Z}_i-\Z_i=-\X\bDelta_i$, Assumption \ref{ass:boundX} gives
\[
\begin{aligned}
\tT_2
&=
\max_{i,j=1,\ldots,p}
\left\|
\frac{1}{n}
\X^T\{(\widehat{\Z}_i-\Z_i)\odot\Z_j\}
\right\|_\infty\\
&\le
\max_{i,j=1,\ldots,p}
\max_{\ell=1,\ldots,q}
\frac{1}{n}
\sum_{k=1}^n
|X_\ell^{(k)}|
\left|
\X^{(k)T}\bDelta_i
\right|
|\Z_j^{(k)}|\\
&\le
C_5
\max_{i=1,\ldots,p}\|\bDelta_i\|_1
\max_{j=1,\ldots,p}
\frac{1}{n}\sum_{k=1}^n|\Z_j^{(k)}|\\
&\le
C_5
\max_{i=1,\ldots,p}\|\bDelta_i\|_1
\max_{j=1,\ldots,p}
\frac{\|\Z_j\|_2}{\sqrt n}.
\end{aligned}
\]
By Assumption \ref{ass:sub_Z}, for sufficiently large $n$,
\[
\P\left(
\max_{j=1,\ldots,p}\|\Z_j\|_2^2
\le C_6n
\right)
\ge 1-C_7n^{-1}.
\]
Combining this result with Lemma \ref{lem:A_est_high} yields
\[
\P\left\{
\tT_2
\le
C_8
\sqrt{\frac{\log p+\log q}{n}}
\right\}
\ge
1-C_9n^{-1}.
\]

Similarly, for the term $\widetilde T_3$, recalling that
$\bDelta_i=\hat{\bA}_i-\bA_i$, Assumption \ref{ass:boundX} gives
\[
\begin{aligned}
\widetilde T_3
&=
\max_{i,j=1,\ldots,p}
\left\|
\frac{1}{n}
\X^T
\{(\widehat{\Z}_i-\Z_i)
\odot
(\Z_j-\hat{\Z}_j)\}
\right\|_\infty\\
&\le
\max_{i,j=1,\ldots,p}
\max_{\ell=1,\ldots,q}
\frac{1}{n}
\sum_{k=1}^n
|\X_\ell^{(k)}|
\left|\X^{(k)T}\bDelta_i\right|
\left|\X^{(k)T}\bDelta_j\right|\\
&\le
C_{10}
\left(
\max_{i=1,\ldots,p}
\|\bDelta_i\|_1
\right)^2.
\end{aligned}
\]
Therefore, by Lemma \ref{lem:A_est_high},
\[
\P\left\{
\widetilde T_3
\le
C_{11}\frac{\log p+\log q}{n}
\right\}
\ge
1-C_{12}n^{-1}.
\]

  Combining these results, we have 
  \[
  \P\Biggl\{\max_{i,j=1,\dots,p}\|\check{\bzeta}_{ij}-\bzeta_{ij}\|_\infty \le C_{13} \sqrt{\frac{\log p+\log q}{n}}\Biggr\}\ge 1-C_{14} n^{-1}.
  \] 
  Define the event $\cA_1:=\{\max_{i,j=1,\dots,p}\|\check{\bzeta}_{ij}-\bzeta_{ij}\|_\infty\le \nu_n/2\}$. Let $\nu_n=2C_{13} \sqrt{(\log p+\log q)/n}$. It follows that $\P(\cA_1)\ge 1-C_{15} n^{-1}$. The proof below is additionally conditional on the event $\cA_1$. Let $\check{\bPhi}_{\bzeta}:=(\check{\bzeta}_{ii})_{i=1,\dots,p}$ denote the $q\times p$ matrix whose columns are given by $\check{\bzeta}_{ii}$ for all $i=1,\dots p$. For any $i,j=1,\dots,p$ and the $\ell$th element $\bzeta_{ij}[\ell]$ of $\bzeta_{ij}$, if $\bzeta_{ij}[\ell]=0$, then $T_{\nu_n}(\check{\bzeta}_{ij})[\ell]=0$, otherwise, 
  \[
  |T_{\nu_n}(\check{\bzeta}_{ij}[\ell])-\bzeta_{ij}[\ell]| \le |T_{\nu_n}(\check{\bzeta}_{ij}[\ell])-\check{\bzeta}_{ij}[\ell]|+|\check{\bzeta}_{ij}[\ell]-\bzeta_{ij}[\ell]|\le 2\nu_n. 
  \]
  As a result, we have $|T_{\nu_n}(\check{\bPhi}_{\bzeta})-\bPhi_{\bzeta}|\le 2\nu_n \tbPhi_{\bzeta}$ entrywise, where the $(\ell,k)$th element of $\tbPhi_{\bzeta}$ is $\bone(\bPhi_{\bzeta}[\ell,k]\ne 0)$. Since $|T_{\nu_n}(\check{\bPhi}_{\bzeta})-\bPhi_{\bzeta}|\ge 0$ and $2\nu_n \tbPhi_{\bzeta}\ge 0$ entrywise, 
  \[
    \|T_{\nu_n}(\check{\bPhi}_{\bzeta})-\bPhi_{\bzeta}\|_2 \le 2\nu_n \|\tbPhi_{\bzeta}\|_2\le C_{16} \sqrt{\frac{\log p+\log q}{n}}, 
  \]
  where the last inequality holds due to Assumption \ref{ass:sp_cov}. 
  
  Moreover, for all $i,j=1,\dots, p$, we have $|T_{\nu_n}(\check{\bzeta}_{ij})-\bzeta_{ij}|\le 2\nu_n \tbzeta_{ij}$ entrywise, where the $\ell$th element of $\tbzeta_{ij}$ is $\bone(\bzeta_{ij}[\ell]\ne 0)$ with $\bzeta_{ij}[\ell]$ denoting the $\ell$th element of $\bzeta_{ij}$. Noting that $\bzeta_{ij}$ is the linear combination of $\{\bzeta_{kk}\}_{k\in \an_\cG^+(i)\cup \an_\cG^+(j)}$ from \eqref{eq:zeta_ij_repre}, it follows that $\max_{i,j}\|\bzeta_{ij}\|_0\le C_{17}$ under Assumptions \ref{ass:sp_graph} and \ref{ass:sp_cov}. Therefore, for all $i,j=1,\dots,p$, 
  \[
  \|T_{\nu_n}(\check{\bzeta}_{ij})-\bzeta_{ij}\|_2 \le 2\nu_n \|\tbzeta_{ij}\|_2 \le C_{17}\sqrt{\frac{\log p+\log q}{n}}.
  \] 

\end{proof}

\begin{lemma}\label{lem:project_high}
  Under the conditions of Lemma \ref{lem:con_order_high}, for sufficiently large $n$, the following holds with probability at least $1-Cn^{-1}$ for all nonempty $J\subseteq \{1,\dots, p\}$ and $i\ne j$,
  \[
  \begin{aligned}
  &\Bigl|\hat{\bzeta}_{ij}^T P_{\hPhiJ}^\perp\hat{\bzeta}_{ij}-\bzeta_{ij}^T P^\perp_{\PhiJ}\bzeta_{ij}\Bigr|\le C_1(\|\bzeta_{ij}\|_2+1)^2\sqrt{\frac{\log p+\log q }{n}},\ if\ P_{\PhiJ}^\perp \bzeta_{ij}\ne \bzero,\quad and\\
  &\bigl|\hat{\bzeta}_{ij}^T P_{\hPhiJ}^\perp\hat{\bzeta}_{ij}-\bzeta_{ij}^T P^\perp_{\PhiJ}\bzeta_{ij}\Bigr|\le C_2(\|\bzeta_{ij}\|_2+1)^2\frac{\log p+\log q }{n},\ if\ P_{\PhiJ}^\perp \bzeta_{ij}= \bzero, 
  \end{aligned}
  \]
  where $C,C_1,C_2$ are some positive constants. 
\end{lemma}
\begin{proof}
  The proof below is conditional on the results in Lemmas \ref{lem:A_est_high} and \ref{lem:zeta_high}. Note that for any given nonempty $J\subseteq \{1,\dots, p\}$ and $i\ne j$,
  \begin{align}
      &\|P^\perp_{\hPhiJ}\hat{\bzeta}_{ij}-P^\perp_{\PhiJ}\bzeta_{ij}\|_2\notag\\
      &\quad \le \|\hat{\bzeta}_{ij}-\bzeta_{ij}\|_2+  \|P_{\hPhiJ}\hat{\bzeta}_{ij}-P_{\PhiJ}\bzeta_{ij}\|_2\notag\\
      &\quad \le C_1\sqrt{\frac{\log p+\log q}{n}}+  \|P_{\hPhiJ}\hat{\bzeta}_{ij}-P_{\PhiJ}\bzeta_{ij}\|_2,
      \label{eq:residual_high_initial}
  \end{align}
  where the last line comes from Lemma \ref{lem:zeta_high}. 
  Define $\balpha_{ij}=(\PhiJ^T \PhiJ)^{-1} \PhiJ^T \bzeta_{ij}$. It then follows that $P_{\PhiJ}\bzeta_{ij}=\PhiJ\balpha_{ij}$ and 
  \begin{align}
    & \|P_{\hPhiJ}\hat{\bzeta}_{ij}-P_{\PhiJ}\bzeta_{ij}\|_2\notag\\
    &\quad = \|P_{\hPhiJ}\{\hPhiJ\balpha_{ij}+(\PhiJ-\hPhiJ)\balpha_{ij}+\hat{\bzeta}_{ij}-\bzeta_{ij}+P^\perp_{\PhiJ}\bzeta_{ij}\}-\PhiJ\balpha_{ij}\|_2\notag\\
    &\quad \le \|(\hPhiJ-\PhiJ)\balpha_{ij}\|_2+\|P_{\hPhiJ}(\PhiJ-\hPhiJ)\balpha_{ij}\|_2\notag\\
    &\quad \relph{=}{}+\|P_{\hPhiJ}(\hat{\bzeta}_{ij}-\bzeta_{ij})\|_2+ \|P_{\hPhiJ} P^\perp_{\PhiJ}\bzeta_{ij}\|_2\notag\\
    &\quad \le  \|\hPhiJ-\PhiJ\|_2\|\balpha_{ij}\|_2+ \|P_{\hPhiJ}\|_2\|\PhiJ-\hPhiJ\|_2\|\balpha_{ij}\|_2\notag\\
    &\quad \relph{=}{}+\|P_{\hPhiJ}\|_2\|\hat{\bzeta}_{ij}-\bzeta_{ij}\|_2+ \|P_{\hPhiJ} P^\perp_{\PhiJ}\bzeta_{ij}\|_2\notag\\
    &\quad\le 2\|\hat{\bPhi}_{\bzeta}-\bPhi_{\bzeta}\|_2\|\balpha_{ij}\|_2+ \|\hat{\bzeta}_{ij}-\bzeta_{ij}\|_2+ \|P_{\hPhiJ} P^\perp_{\PhiJ}\bzeta_{ij}\|_2\notag\\
    &\quad \le C_2\sqrt{\frac{\log p+\log q}{n}}(\|\balpha_{ij}\|_2+1)+ \|P_{\hPhiJ} P^\perp_{\PhiJ}\bzeta_{ij}\|_2,\label{eq:residual_high_med}
  \end{align}
  where the last line holds by Lemma \ref{lem:zeta_high}. Besides, under Assumption \ref{ass:str_hete}, for all $J\subseteq\{1,\dots,p\}$, 
  \[
  \sigma_{min}(\hPhiJ)\ge \sigma_{min}(\PhiJ)-\|\hPhiJ-\PhiJ\|_2\ge \sigma_{min}(\bPhi_{\bzeta})-\|\hat{\bPhi}_{\bzeta}-\bPhi_{\bzeta}\|_2\ge C_3>0,\] 
  where the bound $\sigma_{\min}(\bPhi_{\bzeta})-\|\hat{\bPhi}_{\bzeta}-\bPhi_{\bzeta}\|_2\ge C_3$ follows from Lemma \ref{lem:zeta_high} for sufficiently large $n$. Therefore,
  \[
  \|\balpha_{ij}\|_2\le \sigma_{min}^{-1}(\PhiJ)\|\bzeta_{ij}\|_2\le  \|\bzeta_{ij}\|_2/C_3.
  \]
  We further have the following result combining Lemma \ref{lem:zeta_high},
  \[
  \begin{aligned}
    &\|P_{\hPhiJ} P^\perp_{\PhiJ}\bzeta_{ij}\|_2\\
    &\quad =\|\hPhiJ(\hPhiJ^T \hPhiJ)^{-1} (\hPhiJ-\PhiJ)^T P^\perp_{\PhiJ} \bzeta_{ij}\|_2\\
    &\quad \le \|\hPhiJ(\hPhiJ^T \hPhiJ)^{-1}\|_2 \|\hPhiJ-\PhiJ\|_2 \|P^\perp_{\PhiJ}\|_2 \|\bzeta_{ij}\|_2\\
    &\quad \le \sigma_{min}^{-1}(\hPhiJ) \|\hat{\bPhi}_{\bzeta}-\bPhi_{\bzeta}\|_2 \|\bzeta_{ij}\|_2\\
    &\quad \le C_4 \sqrt{\frac{\log p+\log q}{n}} \|\bzeta_{ij}\|_2. 
  \end{aligned}
  \]
  Substituting the above bounds into \eqref{eq:residual_high_med}, we have
  \begin{equation}\label{eq:projection_high}
   \|P_{\hPhiJ}\hat{\bzeta}_{ij}-P_{\PhiJ}\bzeta_{ij}\|_2 \le C_5 (\|\bzeta_{ij}\|_2+1)\sqrt{\frac{\log p+\log q}{n}}. 
  \end{equation}
  Combining \eqref{eq:projection_high} with \eqref{eq:residual_high_initial} yields
  \begin{equation}\label{eq:residual_high}
  \|P^\perp_{\hPhiJ}\hat{\bzeta}_{ij}-P^\perp_{\PhiJ}\bzeta_{ij}\|_2\le C_6 (\|\bzeta_{ij}\|_2+1)\sqrt{\frac{\log p+\log q}{n}}. 
  \end{equation}

  Note that for vectors $\bnu$ and $\hbnu$ of the same dimension,
  \[
  \begin{aligned}
  |\bnu^T\bnu-\hbnu^T\hbnu|&=|\bnu^T\bnu-(\hbnu-\bnu)^T\hbnu-\bnu^T\hbnu|\\
  &=|\bnu^T\bnu-(\hbnu-\bnu)^T(\hbnu-\bnu)-\bnu^T(\hbnu-\bnu)-\bnu^T\hbnu|\\
  &=|\bnu^T\bnu-(\hbnu-\bnu)^T(\hbnu-\bnu)-\bnu^T(\hbnu-\bnu)-\bnu^T(\hbnu-\bnu)-\bnu^T\bnu|\\
  &=|(\hbnu-\bnu)^T(\hbnu-\bnu)+2\bnu^T(\hbnu-\bnu)|\\
  &\le \|\hbnu-\bnu\|_2^2+2\|\bnu\|_2\|\hbnu-\bnu\|_2. 
  \end{aligned}
  \] 
  Therefore, 
  \begin{equation}\label{eq:rss_high}
    \Bigl|\hat{\bzeta}_{ij}^T P_{\hPhiJ}^\perp\hat{\bzeta}_{ij}-\bzeta_{ij}^T P^\perp_{\PhiJ}\bzeta_{ij}\Bigr|\le \|P^\perp_{\hPhiJ}\hat{\bzeta}_{ij}-P^\perp_{\PhiJ}\bzeta_{ij}\|_2^2+2\|P_{\PhiJ}^\perp \bzeta_{ij}\|_2 \|P^\perp_{\hPhiJ}\hat{\bzeta}_{ij}-P^\perp_{\PhiJ}\bzeta_{ij}\|_2.
  \end{equation}
  If $P_{\PhiJ}^\perp \bzeta_{ij}=\bzero$, then \eqref{eq:residual_high} and \eqref{eq:rss_high} imply
  \[
  \Bigl|\hat{\bzeta}_{ij}^T P_{\hPhiJ}^\perp\hat{\bzeta}_{ij}-\bzeta_{ij}^T P^\perp_{\PhiJ}\bzeta_{ij}\Bigr|\le C_7(\|\bzeta_{ij}\|_2+1)^2\frac{\log p+\log q}{n}.
  \]
  If $P_{\PhiJ}^\perp \bzeta_{ij}\ne \bzero$, then \eqref{eq:residual_high} and \eqref{eq:rss_high} imply that for sufficiently large $n$,
  \[
  \begin{aligned}
  \Bigl|\hat{\bzeta}_{ij}^T P_{\hPhiJ}^\perp\hat{\bzeta}_{ij}-\bzeta_{ij}^T P^\perp_{\PhiJ}\bzeta_{ij}\Bigr|&\le C_8\|\bzeta_{ij}\|_2(\|\bzeta_{ij}\|_2+1)\sqrt{\frac{\log p+\log q}{n}}\\
  &\le C_8(\|\bzeta_{ij}\|_2+1)^2\sqrt{\frac{\log p+\log q}{n}}.
  \end{aligned}
  \]
\end{proof}

\begin{lemma}\label{lem:stg1_high}
  Suppose that the conditions of Theorem \ref{thm:con_graph_high} hold. Suppose further the following holds for all $i=1,\dots, p$, $j\in\cV^+(h)$, and $\ell=1,\dots,q$ with some $h=0,\dots,H_\cG-1$, 
  \[
  \begin{aligned}
    &|\hat{V}_{ij}-V_{ij}|\le C_1\sqrt{\frac{\log p+\log q}{n}}\quad ,\quad |\hat{W}_{\ell j}-W_{\ell j}|\le C_2\sqrt{\frac{\log p+\log q}{n}},\\
  &\sgn(\hat{V}_{ij})=\sgn(V_{ij})\quad,\quad \sgn(\hat{W}_{\ell j})=\sgn(W_{\ell j}), \quad and\\
  & \|\hat{\ve}_j-\ve_j\|_n\le C_3\sqrt{\log p+\log q}.
  \end{aligned}
  \]
  Then for sufficiently large $n$, the following holds with probability at least $1-Cn^{-1}$ for all $k\in\cV^+(h)$,
  \[
  \begin{aligned}
    &\|\hat{\bB}_k(h)-\bB_k(h)\|_1+\|\hat{\bF}_k(h)-\bF_k(h)\|_1\le C_4\sqrt{\frac{\log p+\log q}{n}},\quad and\\
    & \|\hat{\TT}(h)(\hat{\bB}_i(h)-\bB_i(h))+\X(\hat{\bF}_i(h)-\bF_i(h))\|_2\le C_5 \sqrt{\log p+\log q}.
  \end{aligned}
  \]
\end{lemma}

\begin{proof}
  The following proof is conditional on the results in Lemma \ref{lem:con_order_high}. Define $\bGamma_i(h):=(\bB_i(h)^T,\bF_i(h)^T)^T$, $\hat{\bGamma}_i(h):=(\hat{\bB}_i(h)^T,\hat{\bF}_i(h)^T)^T$ for all $i\in\cV^+(h)$, and $\hS(h):=(\hT(h),\X)$. Let $J_i(h)$ be the support of $\bGamma_i(h)$ and thus $|J_i(h)|\le s(h)$. Let $d(h)=\dim(\bS(h))=(|\cV^+(h)|+1)q$ be the dimension of $\bS(h)$. Define the event
  \[
  \cA_1:=\Biggl\{\kappa\Biggl(\frac{\SS(h)^T\SS(h)}{n},s(h)\Biggr)\ge \frac{1}{2}\kappa\bigl(\bSigma_{\bS(h)},s(h)\bigr),\ for\ h=0,\dots,H_\cG-1\Biggr\}. 
  \]
By Assumptions \ref{ass:boundX} and \ref{ass:sub_errors}, every coordinate of $\bS(h)$ is uniformly sub-Gaussian. 
Indeed, each coordinate of $\bS(h)$ is either $X_\ell$ or $X_\ell\ve_j$ for some 
$\ell=1,\ldots,q$ and $j\in\mathcal V^+(h)$. Hence,
\[
\max_{h=0,\ldots,H_{\cG}-1}
\max_{k=1,\ldots,d(h)}
\|\bS(h)_k\|_{\psi_2}
\le C_1,
\]
where we have used $
\|X_\ell\ve_j\|_{\psi_2}
\le
\|X_\ell\|_\infty\|\ve_j\|_{\psi_2}$
under Assumptions \ref{ass:boundX} and \ref{ass:sub_errors}.
Consequently, for any $k_1,k_2=1,\ldots,d(h)$,
\[
\left\|
\bS(h)_{k_1}\bS(h)_{k_2}
-
\E\{\bS(h)_{k_1}\bS(h)_{k_2}\}
\right\|_{\psi_1}
\le C_2.
\]
By Bernstein's inequality and a union bound, for sufficiently large $n$,
\[
\P\left\{
\max_{h=0,\ldots,H_{\cG}-1}
\max_{k_1,k_2=1,\ldots,d(h)}
\left|
\frac{1}{n}\SS_{k_1}(h)^T\SS_{k_2}(h)
-
\bSigma_{\bS(h)}[k_1,k_2]
\right|
\le
C_3\sqrt{\frac{\log p+\log q}{n}}
\right\}
\ge 1-C_4n^{-1}.
\]
Therefore, applying Lemma \ref{lem:perturb_RE} under Assumption \ref{ass:stg1} gives $\P(\cA_1)\ge 1-C_4n^{-1}$.
Moreover, the uniform sub-Gaussian bound on the coordinates of $\SS(h)$ implies that, for
\[
\cA_2
:=
\left\{
\max_{h=0,\ldots,H_{\cG}-1}
\max_{k=1,\ldots,d(h)}
\|\SS(h)_k\|_2
\le C_5\sqrt n
\right\},
\]
we have $\P(\cA_2)\ge 1-C_5^\prime n^{-1}$ for sufficiently large $n$. 
The following proof is further conditional on the event $\cA_1\cap\cA_2$.

   Define
\[
\Delta:=\max_{i,j=1,\dots,d(h)}\Delta_{ij}
=\max_{i,j=1,\dots,d(h)}
\frac{\bigl|\hS(h)_i^T\hS(h)_j-\SS(h)_i^T\SS(h)_j\bigr|}{n}.
\]
If $i,j>|\cV^+(h)|q$, then $\hS(h)_i=\SS(h)_i$ and $\hS(h)_j=\SS(h)_j$, implying that $\Delta_{ij}=0$. Otherwise, suppose without loss of generality that $i>|\cV^+(h)|q$ and $j\le|\cV^+(h)|q$. Then $\hS(h)_i=\SS(h)_i$, while $\SS(h)_j=X_{\ell_1}\odot\ve_{\ell_2}$ for some $\ell_1\in\{1,\dots,q\}$ and $\ell_2\in\cV^+(h)$. Consequently, we have
  \begin{equation}\label{eq:prf_stg1_mid}
  \begin{aligned}
    \Delta_{ij}^2&=\biggl|\frac{1}{n}\SS(h)_i^T(\hS(h)_j-\SS(h)_j)\biggr|^2\\
    &\le \frac{1}{n^2}\|\SS(h)_i\|^2_2\|\hS(h)_j-\SS(h)_j\|^2_2\\\
    &\le C_6\frac{1}{n} \|(\hat{\ve}_{\ell_2}-\ve_{\ell_2})\odot X_{\ell_1}\|_n^2\\
    &= C_6\frac{1}{n}\sum_{k=1}^n (\hat{\ve}^{(k)}_{\ell_2}-\ve_{\ell_2}^{(k)})^2 \X_{\ell_1}^{(k)2}\\
    &\le C_7 \frac{\log p+\log q}{n}, 
  \end{aligned}
\end{equation}
  where the last line holds by using Assumption \ref{ass:boundX} and the condition that $\|\hat{\ve}_{\ell_2}-\ve_{\ell_2}\|_n\le C\sqrt{\log p+\log q}$. The above analysis also shows that 
  \begin{equation}\label{eq:S_est_err}
    \max_{j=1,\dots, d(h)}\|\hS(h)_j-\SS(h)_j\|_2\le C_8\sqrt{\log p+\log q}.
  \end{equation} 
  Finally, if $i,j\le |\cV^+(h)|q$, we can rewrite $\SS_i(h)=\X_{\ell_1}\odot \ve_{\ell_2}$ and $\SS(h)_j=\X_{\ell_3}\odot \ve_{\ell_4}$ for some $\ell_1,\ell_3\in\{1,\dots,q\}$ and $\ell_2,\ell_4\in\cV^+(h)$. Then we have 
  \[
  \begin{aligned}
    \Delta_{ij}&=\frac{1}{n}\|\hS(h)_i^T\hS(h)_j-\SS(h)_i^T\SS(h)_j\|_2\\
    &=\frac{1}{n}\|(\hS(h)_i-\SS(h)_i)^T\SS(h)_j+(\hS(h)_j-\SS(h)_j)^T \SS(h)_i+ (\hS(h)_i-\SS(h)_i)^T(\hS(h)_j-\SS(h)_j)\|_2\\
    &\le \frac{1}{n}\{\|\hS(h)_i-\SS(h)_i\|_2\|\SS(h)_j\|_2+\|\hS(h)_j-\SS(h)_j\|_2 \|\SS(h)_i\|_2+ \|\hS(h)_i-\SS(h)_i\|_2\|\hS(h)_j-\SS(h)_j\|_2\}\\
    & \le C_9 \sqrt{\frac{\log p+\log q}{n}},
  \end{aligned}
  \]
  when $n$ is sufficiently large. Therefore, $\Delta\le C_{10}\sqrt{(\log p+\log q)/n}$. Under Assumption \ref{ass:stg1}, substituting $\hat{\bC}=\hS(h)^T\hS(h)/n$, $\bC=\SS(h)^T\SS(h)/n$, and $s=s(h)$ into Lemma \ref{lem:perturb_RE} yields, 
  \begin{equation}\label{eq:RE_perturb}
  \kappa\Biggl(\frac{\hS(h)^T\hS(h)}{n},\,s(h)\Biggr)\ge C_{11},
  \end{equation}
  for some positive constant $C_{11}$ and sufficiently large $n$,

  By the optimality of $\hat{\bGamma}_i(h)$ for the optimization problem \eqref{eq:opt_stg1}, we have 
  \[
  \frac{1}{n}\|\Y_i-\hS(h)\hat{\bGamma}_i(h)\|_2^2+\beta_n^{(i)}(h)\|\hat{\bGamma}_i(h)\|_1 \le \frac{1}{n}\|\Y_i-\hS(h)\bGamma_i(h)\|_2^2+\beta_n^{(i)}(h)\|\bGamma_i(h)\|_1.
  \]
  Substituting $\Y_i=\SS(h)\bGamma_i(h)+\delta_i(h)$ yields 
  \begin{equation}\label{eq:upper_basic}
    \begin{aligned}
    &\frac{1}{n}\|\hS(h) (\hat{\bGamma}_i(h)-\bGamma_i(h))\|_2^2\\
    &\quad\le \beta_n^{(i)}(h) (\|\bGamma_i(h)\|_1-\|\hat{\bGamma}_i(h)\|_1)+\frac{2}{n}\delta_i(h)^T \hS(h)(\hat{\bGamma}_i(h)-\bGamma_i(h))-\frac{2}{n}\bGamma_i(h)^T (\hS(h)\\
    & \quad\relph{=}{}-\SS(h))^T \hS(h)(\hat{\bGamma}_i(h)-\bGamma_i(h)) \\
    &\quad\le \beta_n^{(i)}(h) (\|\bGamma_i(h)\|_1-\|\hat{\bGamma}_i(h)\|_1) + \frac{2}{n}\|\hS(h)^T \delta_i(h)-\hS(h)^T (\hS(h)-\SS(h)) \bGamma_i(h) \|_\infty \|\hat{\bGamma}_i(h)-\bGamma_i(h)\|_1.
    \end{aligned}
  \end{equation}
  Note that for every $i\in\cV^+(h)$,
  \begin{equation}\label{eq:upper_perturb_basic}
    \begin{aligned}
      &\frac{1}{n}\|\hS(h)^T \delta_i(h)-\hS(h)^T (\hS(h)-\SS(h)) \bGamma_i(h) \|_\infty\\
      & \quad = \frac{1}{n} \|\delta_i(h)^T \SS(h)+\delta_i(h)^T (\hS(h)-\SS(h))-\SS(h)^T (\hS(h)-\SS(h))\bGamma_i(h)-(\hS(h)-\SS(h))^T (\hS(h)-\SS(h))\bGamma_i(h)\|_\infty\\
      &\quad \le \underbrace{\max_{i\in\cV^+(h)} \frac{1}{n} \|\delta_i(h)^T \SS(h)\|_\infty}_{T_1}+\underbrace{ \max_{i\in\cV^+(h)} \frac{1}{n}\|\delta_i(h)^T (\hS(h)-\SS(h))\|_\infty}_{T_2}\\
      &\quad \relph{=}{} +\underbrace{\max_{i\in\cV^+(h)} \frac{1}{n}\|\SS(h)^T (\hS(h)-\SS(h))\bGamma_i(h)\|_\infty}_{T_3}+\underbrace{\max_{i\in\cV^+(h)} \frac{1}{n}\|(\hS(h)-\SS(h))^T (\hS(h)-\SS(h))\bGamma_i(h)\|_\infty}_{T_4}. 
    \end{aligned}
  \end{equation}
  We next bound $T_1,\dots, T_4$, respectively.

  For the term $T_1$, since $\max_{i\in \cV^+(h),j=1,\dots, d(h)}\|\delta_i(h)S(h)_j\|_{\psi_1}\le \|\delta_i(h)\|_{\psi_2}\|S(h)_j\|_{\psi_2}\le C_{12}$ by Assumptions \ref{ass:boundX} and \ref{ass:sub_errors}, we have 
  \[
  \begin{aligned}
      \P\Biggl(T_1\le C_{13}\sqrt{\frac{\log p+\log q}{n}}\Biggr)&\ge 1-|\cV^+(h)|d(h) \exp(-C_{14} (\log p+\log q))\\
      &\ge 1-p(p+1)q \exp(-C_{14} (\log p+\log q))\\
      &\ge 1-C_{15} n^{-1},
  \end{aligned}
  \]
  when $n$ is sufficiently large. For the term $T_2$, noting that $\P(\max_{i\in\cV^+(h)} \|\delta_i(h)\|_n^2\ge C_{16} n)\le C_{17}n^{-1}$ by Assumption \ref{ass:sub_errors}, the following holds with probability at least $1-C_{17} n^{-1}$:
  \[
  \begin{aligned}
    T_2&=\max_{i\in \cV^+(h)} \frac{1}{n} \|\delta_i(h)^T (\hS(h)-\SS(h))\|_\infty\\
    &\le \max_{i\in \cV^+(h)} \frac{1}{n} \|\delta_i(h)\|_n\max_{j=1,\dots, d(h)}\|\hS(h)_j-\SS(h)_j\|_2\\
    &\le \sqrt{\frac{C_{16}}{n}} \max_{j=1,\dots, d(h)}\|\hS(h)_j-\SS(h)_j\|_2\\
    &\le C_{18}\sqrt{\frac{\log p +\log q}{n}}, 
  \end{aligned}
  \]
  where the last line holds by \eqref{eq:S_est_err}. For the term $T_3$, by Assumption \ref{ass:both_stg}, we have 
  \[
  \begin{aligned}
    T_3&=\max_{i\in\cV^+(h)} \frac{1}{n}\|\SS(h)^T (\hS(h)-\SS(h))\bGamma_i(h)\|_\infty \\
    &\le \frac{1}{n} \max_{i\in\cV^+(h)} \|\bGamma_i(h)\|_1 \max_{k,\ell=1,\dots, d(h)} |\SS(h)_k^T (\hS(h)_\ell-\SS(h)_\ell)|\\
    &\le \frac{1}{n} C_{19} \max_{k,\ell=1,\dots, d(h)} \|\SS(h)_k\|_2 \|\hS(h)_\ell-\SS(h)_\ell\|_2\\
    & \le C_{20} \sqrt{\frac{\log p+\log q}{n}},
  \end{aligned}
  \]
  where the last line holds on the event $\cA_2$ and by \eqref{eq:S_est_err}. Similarly, for the term $T_4$, by Assumption \ref{ass:both_stg} and \eqref{eq:S_est_err}, we have
  \[
  \begin{aligned}
      T_4&=\max_{i\in\cV^+(h)} \frac{1}{n}\|(\hS(h)-\SS(h))^T (\hS(h)-\SS(h))\bGamma_i(h)\|_\infty\\
      &\le \max_{i\in\cV^+(h)} \frac{1}{n} \|\bGamma_i(h)\|_1 \max_{k,\ell=1,\dots, d(h)} \|\hS(h)_k-\SS(h)_k\|_2 \|\hS(h)_\ell-\SS(h)_\ell\|_2\\
      &\le C_{21} \frac{\log p+\log q}{n}. 
  \end{aligned}
  \]
  Substituting the upper bounds of $T_1,\dots,T_4$ into \eqref{eq:upper_perturb_basic}, the following holds with probability at least $1-C_{22}n^{-1}$ for sufficiently large $n$,
  \[
  \max_{i\in \cV^+(h)}\frac{1}{n}\|\hS(h)^T \delta_i(h)-\hS(h)^T (\hS(h)-\SS(h)) \bGamma_i(h) \|_\infty\le C_{23}\sqrt{\frac{\log p+\log q}{n}}. 
  \]
  Define $\cA_3=\{\max_{i\in \cV^+(h)}\frac{1}{n}\|\hS(h)^T \delta_i(h)-\hS(h)^T (\hS(h)-\SS(h)) \bGamma_i(h) \|_\infty\le C_{23}\sqrt{(\log p+\log q)/n}\}$. The proof below is further conditional on $\cA_3$. 

  On the event $\cA_3$, \eqref{eq:upper_basic} implies
  \[
  \frac{1}{n}\|\hS(h) (\bGamma_i(h)-\hat{\bGamma}_i(h))\|_2^2\le \beta_n^{(i)}(h) (\|\bGamma_i(h)\|_1-\|\hat{\bGamma}_i(h)\|_1) + 2C_{23}\sqrt{\frac{\log p+\log q}{n}} \|\hat{\bGamma}_i(h)-\bGamma_i(h)\|_1.
  \]
  Let $\beta_n^{(i)}(h) =4C_{23}\sqrt{(\log p+\log q)/n}$ for $i\in \cV^+(h)$. Then  
  \[
  \frac{1}{n}\|\hS(h) (\hat{\bGamma}_i(h)-\bGamma_i(h))\|_2^2\le \beta_n^{(i)}(h) (\|\bGamma_i(h)\|_1-\|\hat{\bGamma}_i(h)\|_1) + \frac{1}{2}\beta_n^{(i)}(h)\|\hat{\bGamma}_i(h)-\bGamma_i(h)\|_1.
  \]
  Adding $\frac{1}{2}\beta_n^{(i)}(h) \|\hat{\bGamma}_i(h)-\bGamma_i(h)\|_1$ on the both sides yields
  \[
  \begin{aligned}
      \frac{1}{n}\|\hS(h) (\hat{\bGamma}_i(h)-\bGamma_i(h))\|_2^2 +\frac{1}{2}\beta_n^{(i)}(h) \|\hat{\bGamma}_i(h)-\bGamma_i(h)\|_1&\le \beta_n^{(i)}(h) (\|\bGamma_i(h)\|_1-\|\hat{\bGamma}_i(h)\|_1+\|\hat{\bGamma}_i(h)-\bGamma_i(h)\|_1)\\
      & \le \beta_n^{(i)}(h) \bigl(\|\bGamma_i(h)[J_i(h)]\|_1-\|\hat{\bGamma}_i(h)[J_i(h)]\|_1\\
      &\relph{=}{}+\|\hat{\bGamma}_i(h)[J_i(h)]-\bGamma_i(h)[J_i(h)]\|_1\bigr)\\
      &\le 2 \beta_n^{(i)}(h) \|\hat{\bGamma}_i(h)[J_i(h)]-\bGamma_i(h)[J_i(h)]\|_1. 
  \end{aligned}
  \]
  This implies
\[
\frac{1}{n}\|\hS(h)(\hat{\bGamma}_i(h)-\bGamma_i(h))\|_2^2\le2\beta_n^{(i)}(h)\|\hat{\bGamma}_i(h)[J_i(h)]-\bGamma_i(h)[J_i(h)]\|_1,
\]
and
\[
\|\hat{\bGamma}_i(h)[J_i(h)^c]-\bGamma_i(h)[J_i(h)^c]\|_1\le 3\|\hat{\bGamma}_i(h)[J_i(h)]-\bGamma_i(h)[J_i(h)]\|_1.
\]
 Using \eqref{eq:RE_perturb}, we then have 
 \[
 \|\hat{\bGamma}_i(h)[J_i(h)]-\bGamma_i(h)[J_i(h)]\|_2 \le C_{24} \sqrt{s(h)}\beta_n^{(i)}(h),\ for\ all\ i\in \cV^+(h).
 \]
 This further implies that 
 \[
 \|\hat{\bGamma}_i(h)[J_i(h)]-\bGamma_i(h)[J_i(h)]\|_1 \le C_{24} s(h)\beta_n^{(i)}(h).
 \] 
 Under Assumption \ref{ass:stg1}, combining these pieces leads to 
 \[
 \frac{1}{n}\|\hS(h)(\hat{\bGamma}_i(h)-\bGamma_i(h))\|_2^2\le C_{25} \frac{\log p+\log q}{n},  
 \] 
 and 
 \[
 \|\hat{\bGamma}_i(h)-\bGamma_i(h)\|_1\le 4\|\hat{\bGamma}_i(h)[J_i(h)]-\bGamma_i(h)[J_i(h)]\|_1\le C_{26} \sqrt{\frac{\log p+\log q}{n}}.
 \]

\end{proof}

\begin{lemma}\label{lem:stg2_high}
  Suppose that the conditions of Theorem \ref{thm:con_graph_high} hold. Suppose further the following holds for all $i=1,\dots, p$, $j\in\cV^+(h)$, and $\ell=1,\dots,q$ with some $h=0,\dots,H_\cG-1$, 
  \[
  \begin{aligned}
    &|\hat{V}_{ij}-V_{ij}|\le C_1\sqrt{\frac{\log p+\log q}{n}}\quad ,\quad |\hat{W}_{\ell j}-W_{\ell j}|\le C_2\sqrt{\frac{\log p+\log q}{n}},\\
  &\sgn(\hat{V}_{ij})=\sgn(V_{ij})\quad,\quad \sgn(\hat{W}_{\ell j})=\sgn(W_{\ell j}), \quad and\\
  & \|\hat{\ve}_j-\ve_j\|_n\le C_3\sqrt{\log p+\log q}.
  \end{aligned}
  \]
  Then for sufficiently large $n$, the following holds with probability at least $1-Cn^{-1}$ for all $\ell=1,\dots,q$, $i=1,\dots,p$, and $j\in\cV(h)$,
  \[
  \begin{aligned}
    &|\hat{V}_{ij}-V_{ij}|\le C_4\sqrt{\frac{\log p+\log q}{n}}\quad ,\quad |\hat{W}_{\ell j}-W_{\ell j}|\le C_5\sqrt{\frac{\log p+\log q}{n}},\\
  &\sgn(\hat{V}_{ij})=\sgn(V_{ij})\quad,\quad \sgn(\hat{W}_{\ell j})=\sgn(W_{\ell j}), \quad and\\
  & \|\hat{\ve}_j-\ve_j\|_n\le C_6\sqrt{\log p+\log q}.
  \end{aligned}
  \]
\end{lemma}

\begin{proof}
The following proof is conditional on the results in Lemma \ref{lem:con_order_high} and \ref{lem:stg1_high}. Define $\bGamma_i(h):=(\bB_i(h)^T,\bF_i(h)^T)^T$ and $\hat{\bGamma}_i(h):=(\hat{\bB}_i(h)^T,\hat{\bF}_i(h)^T)^T$ for all $i\in\cV^+(h)$. Let $d(h)=\dim(\bS(h))=(|\cV^+(h)|+1)q$ be the dimension of $\bS(h)$ and define the event $\cA_1:=\bigl\{\max_{k=1,\dots, d(h)}\|\SS(h)_k\|_2\le c\sqrt{n}\bigr\}$. 
By Assumptions \ref{ass:boundX} and \ref{ass:sub_errors}, every coordinate of $\bS(h)$ is uniformly sub-Gaussian. Specifically, each coordinate of $\bS(h)$ is either $X_\ell$ or $X_\ell\ve_i$ for some $\ell=1,\ldots,q$ and $i\in\mathcal V^+(h)$. Since $\|X_\ell\|_{\psi_2}\le c_1$ and $|X_\ell|\le C_3$ almost surely under Assumption \ref{ass:boundX}, we have $\max_{k=1,\ldots,d(h)}
\|\bS(h)_k\|_{\psi_2}
\le c_2$, where we have used $
\|X_\ell\ve_i\|_{\psi_2}
\le
\|X_\ell\|_\infty\|\ve_i\|_{\psi_2}$
under Assumption \ref{ass:sub_errors}. Consequently, $\P(\cA_1)\ge 1-c_3n^{-1}$ for sufficiently large $n$. 
Moreover, define
\[
\Delta_j:=\ve_j+\sum_{i\in\cV^+(h)}V_{ij}\delta_i(h),
\qquad j\in\cV(h).
\]
Under Assumptions \ref{ass:sp_graph}, \ref{ass:sub_errors}, and \ref{ass:both_stg}, we have $\max_{j\in\cV(h)}\|\Delta_j\|_{\psi_2}\le c_4$ and consequently, $\P(\cA_2)\ge 1-c_5n^{-1}$, where $\cA_2:=\bigl\{\max_{j\in\cV(h)}\|\Delta_j\|_n\le c_6\sqrt{n}\bigr\}$. Define $Y_i(h)=\bB_i(h)^T\bT(h)+\bF_i(h)^T\bX$, $\bQ(h):=((Y_i(h))_{i\in\cV^+(h)},\bX^T)^T$, $d^\prime(h)=\dim(\bQ(h))=|\cV^+(h)|+q$. Define further $\hQ(h):=((\hat{\Y}_i(h))_{i\in\cV^+(h)},\X)$, $\bL_j=((V_{ij})_{i\in\cV^+(h)}^T, \bW_j^T)^T$, and $\hat{\bL}_j=((\hat{V}_{ij})_{i\in\cV^+(h)}^T, \hat{\bW}_j^T)^T$ for $j\in\cV(h)$. By Assumptions \ref{ass:boundX}, \ref{ass:sp_graph}, \ref{ass:sub_errors}, and \ref{ass:both_stg}, we have $\max_{k=1,\dots, d^\prime(h)}\|\bQ(h)_k\|_{\psi_2}\le c_7$. Thus, $\P(\cA_3)\ge 1-c_8n^{-1}$ for sufficiently large $n$, where $\cA_3:=\{\max_{k=1,\dots, d^\prime(h)}\|\bQ(h)_k\|_n\le c_9\sqrt{n}\}$. The following proof is further conditional on the event $\cap_{i=1}^3\cA_i$. 

  We first prove that the irrepresentablity condition holds for the design matrix $\hQ(h)$ with high probability. To this end, we bound the entrywise difference $|\hQ(h)_i^T \hQ(h)_j- \Q(h)_i^T \Q(h)_j|$. If $i,j> |\cV^+(h)|$, then $\hQ(h)_i=\Q(h)_i$ and $\hQ(h)_j=\Q(h)_j$, which leads to $|\hQ(h)_i^T \hQ(h)_j- \Q(h)_i^T \Q(h)_j|=0$. If $i\le |\cV^+(h)|$ and $j>|\cV^+(h)|$, then $\Q(h)_i=\Y(h)_{\ell_1}$ and $\hQ(h)_j=\Q(h)_j=\X_{\ell_2}$ and  for some $\ell_1\in\cV^+(h)$ and $\ell_2=1,\dots, q$. Consequently, by Assumption \ref{ass:boundX}, we have 
  \begin{equation}\label{eq:stg2_irr_perturb1}
  \begin{aligned}
    &\max_{i\in\cV^+(h),j>|\cV^+(h)|} |\hQ(h)_i^T \hQ(h)_j- \Q(h)_i^T \Q(h)_j|\\
    & \quad \le \max_{\ell_1\in\cV^+(h),\ell_2=1,\dots,q} |(\hat{\Y}(h)_{\ell_1}-\Y(h)_{\ell_1})^T \X_{\ell_2}|\\
    &\quad \le \max_{\ell=1,\dots,q}\|\X_{\ell}\|_2 \max_{i\in\cV^+(h)}\|(\hS(h)\hat{\bGamma}_i(h)-\SS(h)\bGamma_i(h)\|_2 \\
    &\quad \le \max_{\ell=1,\dots,q}\|\X_{\ell}\|_2 \max_{i\in\cV^+(h)}\|(\hS(h)(\hat{\bGamma}_i(h)-\bGamma_i(h))\|_2+\|(\hS(h)-\SS(h))\bGamma_i(h)\|_2\\
    &\quad \le C \sqrt{n} (T_1+T_2),
  \end{aligned}
\end{equation}
  where $T_1:=\max_{i\in\cV^+(h)} \|\hS(h) (\hat{\bGamma}_i(h)-\bGamma_i(h))\|_2$ and $T_2:=\max_{i\in\cV^+(h)}\|(\hS(h)-\SS(h))\bGamma_i(h)\|_2$. 
  Finally, if $i,j\le |\cV^+(h)|$, then 
  \begin{equation}\label{eq:stg2_irr_perturb2}
    \begin{aligned}
      & \max_{i,j\le |\cV^+(h)|} |\hQ(h)_i^T \hQ(h)_j- \Q(h)_i^T \Q(h)_j|\\
      & \quad  = \max_{i,j\in\cV^+(h)}|(\hS(h) \hat{\bGamma}_i(h))^T \hS(h) \hat{\bGamma}_j(h)- (\SS(h)\bGamma_i(h))^T \SS(h)\bGamma_j(h)|\\
      & \quad  \le \max_{i,j\in\cV^+(h)} \bigl|\{\hS(h) (\hat{\bGamma}_i(h)-\bGamma_i(h))+(\hS(h)-\SS(h))\bGamma_i(h)+\SS(h)\bGamma_i(h)\}^T \\
      &\quad \relph{=}{} \cdot\{\hS(h) (\hat{\bGamma}_j(h)-\bGamma_j(h))+(\hS(h)-\SS(h))\bGamma_j(h)+\SS(h)\bGamma_j(h)\}- (\SS(h) \bGamma_i(h))^T \SS(h) \bGamma_j(h)\bigr|\\
      & \quad \le (T_1+T_2)(T_1+T_2+T_3)+T_3(T_1+T_2),  
    \end{aligned}
  \end{equation}
  where $T_3:=\max_{i\in\cV^+(h)} \|\SS(h)\bGamma_i(h)\|_2$. We then bound $T_1,T_2,T_3$ respectively. 

  For the term $T_1$, we have $T_1\le C_1\sqrt{\log p+\log q}$ according to Lemma \ref{lem:stg1_high}. For the term $T_2$, by employing \eqref{eq:S_est_err} and Assumption \ref{ass:both_stg}, we have 
  \[
    \begin{aligned}
    T_2&=\max_{i\in\cV^+(h)} \|(\hS(h)-\SS(h))\bGamma_i(h)\|_2\\
      &\le\max_{i\in\cV^+(h)}\|\bGamma_i(h)\|_1\max_{k,\ell=1\dots, d(h)} |(\hS(h)_k-\SS(h)_k)^T (\hS(h)_\ell-\SS(h)_\ell)|^{1/2}\\
      &\le \max_{i\in\cV^+(h)}\|\bGamma_i(h)\|_1\max_{k=1\dots, d(h)} |\hS(h)_k-\SS(h)_k\|_2\\
      & \le C_2 \sqrt{\log p+\log q}.  
    \end{aligned}
  \]
 For the term $T_3$, by Assumption \ref{ass:both_stg}, we have, on the event $\cA_1$,
  \[
  \begin{aligned}
    T_3&=\max_{i\in\cV^+(h)} \|\SS(h)\bGamma_i(h)\|_2\\
      &\le\max_{i\in\cV^+(h)}\|\bGamma_i(h)\|_1\max_{k,\ell=1\dots, d(h)} |\SS(h)_k^T \SS(h)_\ell|^{1/2}\\
      &\le\max_{i\in\cV^+(h)}\|\bGamma_i(h)\|_1\max_{k=1\dots, d(h)} \|\SS(h)_k\|_2\\
      & \le C_3 \sqrt{n}.  
    \end{aligned}
  \]
  Substituting the upper bounds for $T_1,T_2,T_3$ into \eqref{eq:stg2_irr_perturb1} and \eqref{eq:stg2_irr_perturb2}, the following holds for sufficiently large $n$,
  \[
  \max_{i,j=1,\dots,d^\prime(h)} |\hQ(h)_i^T \hQ(h)_j- \Q(h)_i^T \Q(h)_j| \le C_4 \sqrt{(\log p+\log q)n}. 
  \]

  Consequently, by taking $\bC=\Q(h)^T \Q(h)/n$, $\hat{\bC}=\hQ(h)^T \hQ(h)/n$, $S=J_j$ for $j\in \cV(h)$, and $\Delta=C_4 \sqrt{(\log p+\log q)/n}$ in Lemma \ref{lem:perturb_irr}, we have the following by Assumptions \ref{ass:sp_graph} and \ref{ass:stg2}
  \begin{equation}\label{eq:irr_con}
    \begin{aligned}
      &\Bigl\|\Bigl(\frac{1}{n}\hQ(h)^T \hQ(h)\Bigr)_{J_j^cJ_j} \Bigl\{\Bigl(\frac{1}{n}\hQ(h)^T \hQ(h)\Bigr)_{J_jJ_j}\Bigr\}^{-1}\Bigr\|_\infty\le C_5<1 \quad and \\
      & \Bigl\|\Bigl\{\Bigl(\frac{1}{n}\hQ(h)^T \hQ(h)\Bigr)_{J_jJ_j}\Bigr\}^{-1}\Bigr\|_\infty\le C_6,\ for\ j\in \cV(h).
    \end{aligned}
  \end{equation}
  
  Similar to the analysis in Lemma \ref{lem:stg1_high}, by the optimality of $\hat{\bL}_j$ for \eqref{eq:opt_stg2}, we have 
  \begin{equation}\label{eq:stg2_upper_basic}
      \frac{1}{n} \|\hQ(h)(\hat{\bL}_j-\bL_j)\|_2^2 \le \gamma_n^{(j)}(h) (\|\bL_j\|_1-\|\hat{\bL}_j\|_1)+\frac{2}{n}\|\hQ(h)^T \Delta_j-\hQ(h)^T (\hQ(h)-\Q(h))\bL_j\|_\infty\|\hat{\bL}_j-\bL_j\|_1. 
  \end{equation}
  Note that for all $j\in \cV(h)$,
  \[
  \begin{aligned}
  & \frac{1}{n}\|\hQ(h)^T \Delta_j-\hQ(h)^T (\hQ(h)-\Q(h))\bL_j\|_\infty\\
  &\quad = \frac{1}{n}\|\Q(h)^T \Delta_j+(\hQ(h)-\Q(h))^T \Delta_j-\Q(h)^T (\hQ(h)-\Q(h))\bL_j+(\Q(h)-\hQ(h))^T (\hQ(h)-\Q(h))\bL_j\|_\infty\\
  & \quad \le \underbrace{\max_{j\in\cV(h)} \frac{1}{n} \|\Q(h)^T \Delta_j\|_\infty}_{\tT_1} + \underbrace{\max_{j\in\cV(h)} \frac{1}{n} \|(\hQ(h)-\Q(h))^T \Delta_j\|_\infty}_{\tT_2}\\
  & \quad \relph{=}{}+\underbrace{\max_{j\in\cV(h)} \frac{1}{n} \|\Q(h)^T (\hQ(h)-\Q(h))\bL_j\|_\infty}_{\tT_3}+\underbrace{\max_{j\in\cV(h)} \frac{1}{n} \|(\hQ(h)-\Q(h))^T (\hQ(h)-\Q(h))\bL_j\|_\infty}_{\tT_4}. 
  \end{aligned}
  \]
  We next bound $\tT_1,\dots, \tT_4$, respectively. 

  For the term $\tT_1$, since $\max_{k\in\cV(h),\ell=1,\dots,d^\prime(h)}\|\Delta_k \Q(h)_\ell\|_{\psi_1}\le \max_{k,\ell}\|\Delta_k\|_{\psi_2} \|\Q(h)_\ell\|_{\psi_2}\le C_9$, we have the following for sufficiently large $n$,
  \[
  P\Biggl(\tT_1\le C_{10} \sqrt{\frac{\log p+\log q}{n}}\Biggr)\ge 1-C_{11} n^{-1}.
  \]
  For the term $\tT_2$, we first note that 
  \begin{equation}\label{eq:Q_est_err}
  \begin{aligned}
    &\max_{k=1,\dots,d^\prime(h)} \|\hQ(h)_k-\Q(h)_k \|_2\\
    &\quad =\max_{k=1,\dots,|\cV^+(h)|} \|\hQ(h)_k-\Q(h)_k \|_2\\
    &\quad =\max_{k\in\cV^+(h)} \|\hS(h)\hat{\bGamma}_k(h)-\SS(h)\bGamma_k(h) \|_2\\
    &\quad \le \max_{k\in\cV^+(h)} \|\hS(h)(\hat{\bGamma}_k(h)-\bGamma_k(h)) \|_2+\|(\hS(h)-\SS(h))\bGamma_k\|_2\\
    &\quad \le \max_{k\in\cV^+(h)} \|\hS(h)(\hat{\bGamma}_k(h)-\bGamma_k(h)) \|_2+T_2\\
    &\quad \le C_{12} \sqrt{\log p+\log q},  
  \end{aligned}
\end{equation}
  where the last line comes from Lemma \ref{lem:stg1_high} and the upper bound of $T_2$. Therefore, on the event $\cA_2$, 
  \[
  \begin{aligned}
    \tT_2&=\max_{j\in\cV(h)} \frac{1}{n} \|(\hQ(h)-\Q(h))^T \Delta_j\|_\infty\\
    &\le \max_{j\in\cV(h),k=1,\dots,d^\prime(h)} \frac{1}{n} \|\Delta_j\|_n \|\hQ(h)_k-\Q(h)_k \|_2\\
    &\le C_{13}\sqrt{\frac{\log p+\log q}{n}}.
  \end{aligned}
  \]
  For the term $\tT_3$, by \eqref{eq:Q_est_err}, we have 
  \[
  \begin{aligned}
  \tT_3&=\max_{j\in\cV(h)} \frac{1}{n} \|\Q(h)^T (\hQ(h)-\Q(h))\bL_j\|_\infty\\
  &\le \max_{j\in\cV(h)} \frac{1}{n}\|\bL_j\|_1 \max_{k,\ell=1,\dots, d^\prime(h)} |\Q(h)_k^T (\hQ(h)_\ell-\Q(h)_\ell)|\\
  &\le \max_{j\in\cV(h)} \frac{1}{n}\|\bL_j\|_1 \max_{k,\ell=1,\dots, d^\prime(h)} \|\Q(h)_k\|_2 \|\hQ(h)_\ell-\Q(h)_\ell\|_2\\
  &\le C_{14}\sqrt{\frac{\log p+\log q}{n}},
  \end{aligned}
  \]
  where the last line holds on the event $\cA_3$ and by Assumption \ref{ass:both_stg}. Finally, for the term $\tT_4$, we have 
  \[
  \tT_4\le \frac{1}{n}\max_{j\in\cV(h)} \|\bL_j\|_1 \max_{k,\ell} |(\hQ(h)_k-\Q(h)_k)^T (\hQ(h)_\ell-\Q(h)_\ell)| \le C_{15} \frac{\log p+\log q}{n}. 
  \]
  Combining these pieces yields 
  \begin{equation}\label{eq:upper_pert_basic_stg2}
  \frac{1}{n}\|\hQ(h)^T \Delta_j-\hQ(h)^T (\hQ(h)-\Q(h))\bL_j\|_\infty\le C_{16}\sqrt{\frac{\log p+\log q}{n}},  
  \end{equation}
  with probability at least $1-C_{11}n^{-1}$ for sufficiently large $n$.

Combining \eqref{eq:irr_con} and \eqref{eq:upper_pert_basic_stg2}, let
$\rho\in(0,1)$ denote the constant satisfying
\[
\left\|
\left(
\frac{1}{n}\hQ(h)_{J_j^c}^{T}\hQ(h)_{J_j}
\right)
\left\{
\frac{1}{n}\hQ(h)_{J_j}^{T}\hQ(h)_{J_j}
\right\}^{-1}
\right\|_\infty
\le \rho,
\]
and recall that
\[
\left\|
\left\{
\frac{1}{n}\hQ(h)_{J_j}^{T}\hQ(h)_{J_j}
\right\}^{-1}
\right\|_\infty
\le C_6.
\]
For notational simplicity, define
\[
r_n:=\sqrt{\frac{\log p+\log q}{n}}
\]
and
\[
g_j
:=
\frac{1}{n}\hQ(h)^{T}
\left\{
\Delta_j-
\bigl(\hQ(h)-\Q(h)\bigr)\bL_j
\right\}.
\]
By \eqref{eq:upper_pert_basic_stg2}, we have the following with probability at least $1-Cn^{-1}$ for sufficiently large $n$,
\[
\max_{j\in\cV(h)}\|g_j\|_\infty
\le C_{16}r_n.
\]

We choose
\begin{equation}
\gamma_n^{(j)}(h)
=
\frac{4(1+\rho)C_{16}}{1-\rho}r_n,
\qquad j\in\cV(h).
\label{eq:gamma_stg2}
\end{equation}
To establish support recovery, we use the Karush--Kuhn--Tucker conditions of the
Lasso problem \eqref{eq:opt_stg2}. Specifically, it suffices to construct
a vector $\widetilde{\bL}_j$ with
$\widetilde{\bL}_j[J_j^c]=0$ that satisfies
\begin{align}
\frac{2}{n}\hQ(h)_{J_j}^{T}
\left\{
\Y_j-\hQ(h)\widetilde{\bL}_j
\right\}
&=
\gamma_n^{(j)}(h)
\sgn\bigl(\widetilde{\bL}_j[J_j]\bigr)
\quad and,\label{eq:KKT_nonzero}\\
\left\|
\frac{2}{n}\hQ(h)_{J_j^c}^{T}
\left\{
\Y_j-\hQ(h)\widetilde{\bL}_j
\right\}
\right\|_\infty
&\le
\gamma_n^{(j)}(h).
\label{eq:KKT_zero}
\end{align}
We construct such a primal-dual witness as follows. For each
$j\in\cV(h)$, set $\widetilde{\bL}_j[J_j^c]=0$ and define
$\widetilde{\bL}_j[J_j]$ by
\begin{equation}
\widetilde{\bL}_j[J_j]-\bL_j[J_j]
=
\left\{
\frac{1}{n}
\hQ(h)_{J_j}^{T}
\hQ(h)_{J_j}
\right\}^{-1}
\left\{
g_j[J_j]
-
\frac{1}{2}\gamma_n^{(j)}(h)
\sgn\bigl(\bL_j[J_j]\bigr)
\right\}.
\label{eq:pdw_active_stg2}
\end{equation}
It follows from \eqref{eq:irr_con},
\eqref{eq:upper_pert_basic_stg2}, and \eqref{eq:gamma_stg2} that
\[
\begin{aligned}
\|\widetilde{\bL}_j[J_j]-\bL_j[J_j]\|_\infty
&\le
C_6
\left\{
C_{16}r_n+\frac{1}{2}\gamma_n^{(j)}(h)
\right\}\\
&\le
C_{17}r_n.
\end{aligned}
\]
 Therefore, by Assumption
\ref{ass:str_sig}, 
\[
\sgn\bigl(\widetilde{\bL}_j[J_j]\bigr)
=
\sgn\bigl(\bL_j[J_j]\bigr).
\]

We first verify the Karush--Kuhn--Tucker condition on the active set $J_j$. By the
definition of $\Delta_j$, $\Y_j
=
\Q(h)\bL_j+\Delta_j$.
Since $\bL_j[J_j^c]=\widetilde{\bL}_j[J_j^c]=0$, we have
\begin{equation}
\begin{aligned}
\Y_j-\hQ(h)\widetilde{\bL}_j
=
\Delta_j
-
\bigl(
\hQ(h)_{J_j}-\Q(h)_{J_j}
\bigr)\bL_j[J_j]-
\hQ(h)_{J_j}
\bigl\{
\widetilde{\bL}_j[J_j]-\bL_j[J_j]
\bigr\}.
\end{aligned}
\label{eq:residual_decomp_stg2}
\end{equation}
Substituting \eqref{eq:pdw_active_stg2} into
\eqref{eq:residual_decomp_stg2} gives
\[
\begin{aligned}
\frac{2}{n}
\hQ(h)_{J_j}^{T}
\left\{
\Y_j-\hQ(h)\widetilde{\bL}_j
\right\}
=
\gamma_n^{(j)}(h)
\sgn\bigl(\bL_j[J_j]\bigr)=
\gamma_n^{(j)}(h)
\sgn\bigl(\widetilde{\bL}_j[J_j]\bigr).
\end{aligned}
\]
Thus, the active Karush--Kuhn--Tucker condition \eqref{eq:KKT_nonzero} is satisfied.

We next verify the Karush--Kuhn--Tucker condition on the inactive set $J_j^c$, namely,
the dual feasibility condition. Using \eqref{eq:pdw_active_stg2} and
\eqref{eq:residual_decomp_stg2}, we obtain
\[
\begin{aligned}
&\left\|
\frac{2}{n}
\hQ(h)_{J_j^c}^{T}
\left\{
\Y_j-\hQ(h)\widetilde{\bL}_j
\right\}
\right\|_\infty\\
&\quad=
\Bigg\|
2g_j[J_j^c]
-
2
\left(
\frac{1}{n}
\hQ(h)_{J_j^c}^{T}
\hQ(h)_{J_j}
\right)
\left\{
\frac{1}{n}
\hQ(h)_{J_j}^{T}
\hQ(h)_{J_j}
\right\}^{-1}
g_j[J_j]\\
&\quad \relph{=}{}
+
\gamma_n^{(j)}(h)
\left(
\frac{1}{n}
\hQ(h)_{J_j^c}^{T}
\hQ(h)_{J_j}
\right)
\left\{
\frac{1}{n}
\hQ(h)_{J_j}^{T}
\hQ(h)_{J_j}
\right\}^{-1}
\sgn\bigl(\bL_j[J_j]\bigr)
\Bigg\|_\infty\\
&\quad\le
2\|g_j[J_j^c]\|_\infty
+
2
\left\|
\left(
\frac{1}{n}
\hQ(h)_{J_j^c}^{T}
\hQ(h)_{J_j}
\right)
\left\{
\frac{1}{n}
\hQ(h)_{J_j}^{T}
\hQ(h)_{J_j}
\right\}^{-1}
\right\|_\infty
\|g_j[J_j]\|_\infty\\
&\quad \relph{=}{}
+
\gamma_n^{(j)}(h)
\left\|
\left(
\frac{1}{n}
\hQ(h)_{J_j^c}^{T}
\hQ(h)_{J_j}
\right)
\left\{
\frac{1}{n}
\hQ(h)_{J_j}^{T}
\hQ(h)_{J_j}
\right\}^{-1}
\right\|_\infty\\
&\quad\le
2(1+\rho)C_{16}r_n
+
\rho\gamma_n^{(j)}(h)\\
&\quad=
\frac{1-\rho}{2}\gamma_n^{(j)}(h)
+
\rho\gamma_n^{(j)}(h)\\
&\quad=
\frac{1+\rho}{2}\gamma_n^{(j)}(h)\\
&\quad<\gamma_n^{(j)}(h),
\end{aligned}
\]
where the penultimate line follows from \eqref{eq:gamma_stg2}.
Hence, the strict dual feasibility condition
\eqref{eq:KKT_zero} holds.

Therefore, $\widetilde{\bL}_j$ satisfies the Karush--Kuhn--Tucker conditions of
\eqref{eq:opt_stg2}. Since \eqref{eq:opt_stg2} is a convex optimization
problem, the Karush--Kuhn--Tucker conditions are sufficient for global optimality.
Hence, $\widetilde{\bL}_j$ is a solution to \eqref{eq:opt_stg2}.
Moreover, since $\hQ(h)_{J_j}^{T}
\hQ(h)_{J_j}/n$
is invertible by \eqref{eq:irr_con} and the dual feasibility condition
is strict, the solution is unique. Consequently,
\[
\hat{\bL}_j
=
\widetilde{\bL}_j\quad 
and \quad
\hat J_j
=
J_j.
\]
Thus, for every $j\in\cV(h)$,
\[
\|\hat{\bL}_j-\bL_j\|_\infty
\le
C_{17}\sqrt{\frac{\log p+\log q}{n}}
\quad and \quad
\sgn(\hat{\bL}_j)
=
\sgn(\bL_j).
\]

  Besides, noting that $\hat{V}_{ij}=V_{ij}=0$ for all $i\notin\cV^+(h)$ and $j\in\cV(h)$, the following holds for all $\ell=1,\dots,q$, $i=1,\dots,p$, and $j\in\cV(h)$,
  \[
  \begin{aligned}
    &|\hat{V}_{ij}-V_{ij}|\le C_{18}\sqrt{\frac{\log p+\log q}{n}}\quad ,\quad |\hat{W}_{\ell j}-W_{\ell j}|\le C_{19}\sqrt{\frac{\log p+\log q}{n}},\quad and \\
  &\sgn(\hat{V}_{ij})=\sgn(V_{ij})\quad,\quad \sgn(\hat{W}_{\ell j})=\sgn(W_{\ell j}). 
  \end{aligned}
  \]

Finally, we bound the estimation error of the structural residuals. 
Recall that
\[
\hat{\ve}_j
=
\Y_j-\Y\hat{\bV}_j-\X\hat{\bW}_j\quad 
and \quad
\ve_j
=
\Y_j-\Y \bV_j-\X  \bW_j.
\]
Hence, for every $j\in\mathcal V(h)$,
\[
\hat{\ve}_j-\ve_j
=
-\Y(\hat{\bV}_j-\bV_j)
-\X(\hat{\bW}_j-\bW_j).
\]
Moreover, since $\hat{V}_{ij}=V_{ij}=0$ for all
$i\notin\mathcal V^+(h)$ and $j\in\mathcal V(h)$, we have
\[
\begin{aligned}
\|\hat{\ve}_j-\ve_j\|_n
&\le
\sum_{i\in\mathcal V^+(h)}
|\hat{V}_{ij}-V_{ij}|\|Y_i\|_n
+
\sum_{\ell=1}^q
|\hat{W}_{\ell j}-W_{\ell j}|\|X_\ell\|_n\\
&\le
\max_{i\in\mathcal V^+(h)}\|Y_i\|_n
\|\hat{\bV}_j-\bV_j\|_1
+
\max_{\ell=1,\ldots,q}\|X_\ell\|_n
\|\hat{\bW}_j-\bW_j\|_1.
\end{aligned}
\]

We first control the empirical norms of the regressors. For
$i\in\mathcal V^+(h)$, recall that
\[
\Y_i=\Y_i(h)+\delta_i(h)\quad 
and \quad
\Y_i(h)=\TT(h)\bB_i(h)+\X\bF_i(h).
\]
By Assumptions \ref{ass:boundX}, \ref{ass:sub_errors}, and \ref{ass:both_stg},
$Y_i(h)$ and $\delta_i(h)$ have uniformly bounded
sub-Gaussian norms. Therefore, for sufficiently large $n$, with
probability at least $1-C_{20}n^{-1}$, $\max_{i\in\mathcal V^+(h)}\|Y_i\|_n
\le C_{21}\sqrt n$. 
In addition, Assumption \ref{ass:boundX} implies deterministically that $\max_{\ell=1,\ldots,q}\|X_\ell\|_n
\le C_{22}\sqrt n$. By the sign consistency established above,
\[
\operatorname{supp}(\hat{\bV}_j)
=
\operatorname{supp}(\bV_j)\quad 
and \quad
\operatorname{supp}(\hat{\bW}_j)
=
\operatorname{supp}(\bW_j).
\]
Together with Assumption \ref{ass:sp_graph} and the previously established
$\ell_\infty$ error bounds, this yields
\[
\begin{aligned}
\|\hat{\bV}_j-\bV_j\|_1
&\le
|\operatorname{supp}(\bV_j)|
\|\hat{\bV}_j-\bV_j\|_\infty
\le
C_{23}\sqrt{\frac{\log p+\log q}{n}},\ and\\
\|\hat{\bW}_j-\bW_j\|_1
&\le
|\operatorname{supp}(\bW_j)|
\|\hat{\bW}_j-\bW_j\|_\infty
\le
C_{24}\sqrt{\frac{\log p+\log q}{n}},
\end{aligned}
\]
where we have used
$|\operatorname{supp}(\bV_j)|
\le |\operatorname{an}_{\mathcal G}(j)|=O(1)$
and $\|\bW_j\|_0=O(1)$ under Assumption \ref{ass:sp_graph}.

Combining the above bounds, for all $j\in\mathcal V(h)$,
\[
\begin{aligned}
\|\hat{\ve}_j-\ve_j\|_n
&\le
C_{25}\sqrt n\,
\|\hat{\bV}_j-\bV_j\|_1
+
C_{26}\sqrt n\,
\|\hat{\bW}_j-\bW_j\|_1\\
&\le
C_{27}\sqrt{\log p+\log q}.
\end{aligned}
\]
Thus, with probability at least $1-C_{28}n^{-1}$,
\[
\max_{j\in\mathcal V(h)}
\|\hat{\ve}_j-\ve_j\|_n
\le
C_{27}\sqrt{\log p+\log q},
\]
which completes the proof.

\end{proof}

\begin{lemma}\label{lem:perturb_RE}
  For two symmetric matrices $\bC,\hat{\bC}\in\R^{m\times m}$ and $s\le m$, if $\max_{i,j}| \hat{C}_{ij}-C_{ij}|\le \Delta$ and $\kappa^2(\bC,s)\ge 32s\Delta$, then $\kappa(\hat{\bC},s)\ge \kappa(\bC,s)/2$. 
\end{lemma}

\begin{proof}
  For any set $J\subseteq\{1,\dots,m\}$ satisfying $|J|\le s$ and nonzero $\bnu\in\R^{m}$ satisfying $\|\bnu_{J^c}\|_1\le 3\|\bnu_J\|_1$, we have 
  \[
  \frac{|\bnu^T( \hat{\bC}-\bC)\bnu|}{\|\bnu_J\|_2^2}\le \frac{\|\bnu\|_1^2 \max_{i,j}|\hat{C}_{ij}-C_{ij}|}{\|\bnu_J\|_2^2}\le 16 s\Delta.
  \]
  By the definition of restricted eigenvalue, $\kappa^2(\hat{\bC},s)\ge \kappa^2(\bC,s)-16s\Delta\ge \kappa^2(\bC,s)/4$, which leads to the conclusion. 
\end{proof}

\begin{lemma}\label{lem:perturb_irr}
  For two square matrices $\bC,\hat{\bC}\in\R^{m\times m}$ and $S\subseteq\{1,\dots,m\}$ with $|S|=s$, suppose that $\|\bC_{S^cS}(\bC_{SS})^{-1}\|_\infty\le C_1$ for some $C_1\in [0,1)$ and $\|(\bC_{SS})^{-1}\|_\infty\le C_2$. If $\max_{i,j}|\hat{C}_{ij}-C_{ij}|\le \Delta\le 1/(2C_2s)$ and $C_1+2C_2\Delta s<1$, then 
  \[
  \|\hat{\bC}_{S^cS}(\hat{\bC}_{SS})^{-1}\|_\infty\le \frac{C_1+C_2\Delta s}{1-C_2\Delta s}<1\quad and \quad \|(\hat{\bC}_{SS})^{-1}\|_\infty\le \frac{C_2}{1-C_2\Delta s}. 
  \]
\end{lemma}

\begin{proof}
  We first prove that $\|(\hat{\bC}_{SS})^{-1}\|_\infty\le C_2/(1-C_2\Delta s)$. Note that $\|(\bC_{SS})^{-1}\|_\infty\|\hat{\bC}_{SS}-\bC_{SS}\|_\infty\le C_2\Delta s<1$, and similarly, $\|(\bC_{SS})^{-1}\|_\infty\|\hat{\bC}_{S^cS}-\bC_{S^cS}\|_\infty\le C_2\Delta s<1$. By an error bound for matrix inversion \citep[p. 336]{horn1985matrix}, we have 
  \[
  \|(\hat{\bC}_{SS})^{-1}-(\bC_{SS})^{-1}\|_\infty\leq\frac{\|(\bC_{SS})^{-1}\|_\infty\|\hat{\bC}_{SS}-\bC_{SS}\|_\infty}{1-\|(\bC_{SS})^{-1}\|_\infty\|\hat{\bC}_{SS}-\bC_{SS}\|_\infty} \|(\bC_{SS})^{-1}\|_\infty\le \frac{C_2^2\Delta s}{1-C_2\Delta s}. 
  \]
  Therefore, 
  \[
  \|(\hat{\bC}_{SS})^{-1}\|_\infty\le \|(\hat{\bC}_{SS})^{-1}-(\bC_{SS})^{-1}\|_\infty+\|(\bC_{SS})^{-1}\|_\infty\le \frac{C_2}{1-C_2\Delta s}. 
  \]
  For the term $\|\hat{\bC}_{S^cS}(\hat{\bC}_{SS})^{-1}\|_\infty$, note that 
  \[
  \begin{aligned}
    &\|\hat{\bC}_{S^cS}(\hat{\bC}_{SS})^{-1}-\bC_{S^cS}(\bC_{SS})^{-1}\|_\infty\\
    \quad &=\|(\hat{\bC}_{S^cS}-\bC_{S^cS})(\hat{\bC}_{SS})^{-1}-\bC_{S^cS}(\bC_{SS})^{-1}(\hat{\bC}_{SS}-\bC_{SS})(\hat{\bC}_{SS})^{-1}\|_\infty\\
    \quad &\le \|\hat{\bC}_{S^cS}-\bC_{S^cS}\|_\infty \|(\hat{\bC}_{SS})^{-1}\|_\infty+\|\bC_{S^cS}(\bC_{SS})^{-1}\|_\infty\|\hat{\bC}_{SS}-\bC_{SS}\|_\infty\|(\hat{\bC}_{SS})^{-1}\|_\infty\\
    \quad & \le \frac{C_2}{1-C_2\Delta s}\Delta s+C_1\Delta s\frac{C_2}{1-C_2\Delta s}\\
    \quad &=\frac{(1+C_1) C_2\Delta s}{1-C_2\Delta s}.
  \end{aligned}
  \]
  The proof is then completed by using $\|\hat{\bC}_{S^cS}(\hat{\bC}_{SS})^{-1}\|_\infty\le \|\bC_{S^cS}(\bC_{SS})^{-1}\|_\infty+\|\hat{\bC}_{S^cS}(\hat{\bC}_{SS})^{-1}-\bC_{S^cS}(\bC_{SS})^{-1}\|_\infty$. 
\end{proof}

\end{document}